\documentclass{aa}  
\usepackage{graphicx}
\usepackage[varg]{txfonts}
\pdfoutput=1
\usepackage{natbib,twoopt}
\bibpunct{(}{)}{;}{a}{}{,}
\usepackage{subfigure}
\newcommand{\Halpha}{H$\alpha$}
\newcommand{\kms}{km\,s$^{-1}$}

\begin{document}
%

   \title{The magnetic field topology of the weak-lined T Tauri star V410\,Tauri}
   \subtitle{New strategies for Zeeman-Doppler imaging}

   \author{T. A. Carroll\inst{1}, K. G. Strassmeier\inst{1}, J. B. Rice \inst{2} \and
      A. K\"unstler\inst{1}}

   \offprints{T. A. Carroll}

\institute{Leibniz-Institute for Astrophysics Potsdam, An der
Sternwarte 16, D-14482 Potsdam, Germany
\email{[tcarroll,kstrassmeier,akuenstler]@aip.de} \and Department
of Physics, Brandon University, Brandon, Manitoba R7A 6A9, Canada
\email{rice@BrandonU.ca} }

\date{Received: 13 August 2012 ; Accepted: 25 October 2012 }

\abstract  
  {} 
  {
   In a follow-up investigation we present Zeeman-Doppler maps of the weak-lined T Tauri star (WTTS) V410~Tau.
   As a rapid rotating star and a typical WTTS the stellar surface of V410 Tau is accessible to surface imaging 
   techniques and allows us to detect and reconstruct the major magnetic surface features on 
   this pre-main sequence star.
  }
  {
   The polarized signals we are measuring are of the order of $10^{-4}$ 
   to $10^{-3} $ and are hidden well below the noise level of a single observation.  
   A new line profile reconstruction technique based on a singular value decomposition (SVD) allows 
   us to extract the weak polarized line profiles (Stokes~V) as well as the intensity profiles (Stokes~I). 
   One of the key features of the line profile reconstruction
   is that the SVD line profiles are amenable to radiative transfer modeling within our 
   Zeeman-Doppler Imaging code \emph{iMap}. The code also utilizes a new iterative regularization
   scheme which is independent of any additional surface constraints.
   To provide more stability a vital part of our inversion strategy is to invert both 
   Stokes~I and Stokes~V profiles to simultaneously reconstruct the temperature and magnetic field surface 
   distribution of V410~Tau. A new image-shear analysis is also implemented to allow the 
   search for image and line profile distortions induced by a differential rotation of the star.
  }  
  {
   The magnetic field structure we obtain for V410~Tau shows a good spatial correlation with the surface
   temperature and is dominated by a strong field within the cool polar spot. 
   The Zeeman-Doppler maps exhibit a large-scale organization of both polarities around the 
   polar cap in the form of a twisted bipolar structure.
   The magnetic field reaches a value of almost 2~kG within the polar region but smaller fields are also 
   present down to lower latitudes. 
   The pronounced non-axisymmetric field structure and the non-detection of a differential rotation for V410~Tau 
   supports the idea of an underlying $\alpha^2$-type dynamo, which is predicted for weak-lined T Tauri stars. 
  }
  {} 


\authorrunning{Carroll et al.}

\maketitle


\section{Introduction}

T~Tauri stars are very young stars on their way towards the zero-age main
sequence (ZAMS). With the ongoing contraction they prepare the
conditions in their interiors to start the first nuclear process.
In classical T~Tauri stars, the outside is still dominated by the
accretion of matter from the disk while in the weak-lined T~Tauri
stars (WTTS), or ``naked'' T~Tauri stars, the accretion disk is
basically gone and the star is soon to arrive on the ZAMS. Little
is known about the magnetic field generating processes during
both of these evolutionary phases 
\citep[see, e.g. the review by][]{Gregory11}. 
How do WTTS stars produce,
maintain, and reorganize their internal fields at this
rapidly changing evolutionary stage? An even more
relevant question from a general point of view is how do these
fields manifest themselves on the surface of the star and whether their
appearance, i.e. the topology, is directly linked to the
generating process in the interior? Or do convection and other 
surface flows alter the surface appearance in such a way that 
they will prevent us from making any direct conclusions about the 
the internal dynamo processes? We are still far from answering all these
questions which is, unfortunately, not only true for WTTS stars
but also for any other type of star \citep{Schrijver00}. Even for
the Sun there is no clear answer yet of how strong sunspots and active
regions are still connected deep down into the tachocline region
\citep{Rempel09,Rempel11} where the
internal dynamo is believed to operate
\citep{Schrijver99,Schuessler05}. We are still in the phase of
characterization and it may still take some time before we
can connect observations with theory (and vice versa) 
in a conclusive way.

However, the characterization of surface magnetic fields with
techniques like Doppler imaging (DI) \citep[e.g.][]{Rice02},
which tracks the magnetically induced cool spots on the surface of
stars, or more directly with Zeeman-Doppler Imaging (ZDI)
\citep[see, e.g.][]{Donati01}, has provided us in the last two
decades with a wealth of new information about the surface spots
and magnetic fields on many different types of stars, see
\cite{Strassmeier09} and \cite{Donati09} for comprehensive
overviews.

In this work, we concentrate on DI and ZDI of V410~Tau, a rapidly
rotating WTTS that has fully stripped away its surrounding disk.
Many WTTS like V410~Tau show strong signatures of photospheric
variability attributed to atmospheric magnetic activity rather
than to a left-over from external accretion processes.
Indirect signs of magnetic activity are known for a long time from
photometric light curve variations \citep[see, e.g.][]{Rydgren83,Bouvier89}. 
More direct evidence of photospheric
magnetic activity could be obtained by Doppler images, in particular for
V410~Tau \citep{Joncour94, Strassmeier94, Hatzes95, Rice96, Rice11}. 
Skelly et al. (2010) presented the first Zeeman-Doppler map of V410~Tau.
Surprisingly, the reconstructed large-scale field exhibit a rather complex
structure dispersed over different latitudes with strongly inclined fields 
that are only weakly correlated with the spots shown in their brightness maps.

Signs of chromospheric and coronal activity on V410~Tau are also
known for many years and strong ultraviolet and optical emission
lines and X-rays emission were observed \citep{Hatzes95,Stelzer01,Stelzer03}. 
An intriguing fact here is that the modulation of the chromospheric
and coronal activity tracers seem not to be strongly correlated
with the photospheric spot locations \citep{Stelzer03}.

To shed more light on the magnetic field distribution on stellar
surfaces, in particular for V410~Tau, we conducted a simultaneous
investigation of rotationally modulated spectral line profiles
(Stokes~I) and spectropolarimetric line profiles (Stokes~V) in
terms of a rigorous Doppler imaging and Zeeman-Doppler
imaging analysis. In this respect, the present paper is a
continuation of the work by 
\cite{Rice11}, which we hereby call Paper~I. The present paper is
organized as follows. In Sect.~2, we briefly describe the
acquisition of the data. In Sect.~3, we introduce our improved
polarized-line profile extraction and reconstruction technique based on a 
singular-value decomposition (SVD). Sect.~4 is devoted to a recap
of the basic outline and assumptions in our ZDI code \emph{iMap} and its
recent improvements.
In Sect.~5, we present the results of our simultaneous DI and ZDI
approach as well as the results of our search for a differential
rotation on V410~Tau. Sect.~6 summarizes our findings and conclusions.


\section{Observations}
\label{Sect:2}

High-resolution spectropolarimetric observations were obtained
with the ESPADONS echelle spectrograph and polarimeter \citep{Espadons}
in Stokes~I and Stokes~V at the 3.6m Canada-France-Hawaii
telescope (CFHT) on Mauna Kea, Hawaii. Data were obtained in queue
mode during one observing block over four nights from 2008 October
15 to 19, and one over 13 nights distributed between the dates of
2008, December~5 to 2009, January~14. A spectral resolution of
60,000 with a wavelength coverage of 390--900 nm was obtained. All
integrations on V410 Tau ($V \approx 11^m.0$) were set to an
exposure time of 4 $\times$ 600 s. This allowed for a total of 17
spectra with an average signal-to-noise (S/N) ratio of around
160:1 per single exposure per pixel. All spectra were reduced and
extracted using the Libre-ESPRIT package provided by CFHT and
executed automatically. Details of this reduction procedure are
given in \citep{Donati97}.


\section{SVD based signal extraction and reconstruction}\label{Sect:3}

\subsection{Method}
\label{Sect:3.1}

For most active cool late-type stars the typical signal amplitude of 
a single polarized line profile is of the order of a few times  $10^{-4}$
relative to the normalized continuum.
These values are far below the signal-to-noise ratio of a typical
spectropolarimetric observation, which hardly exceeds a few times
$10^{3}$. So the individual polarimetric line profiles are buried
well below the noise level by typically one order of magnitude.
This requires techniques that are able to recover and extract the
line profile signals from the measurement noise prior to the actual 
ZDI analysis.
Most of the techniques for circularized, i.e. Stokes~V, line
profiles take advantage of the fact that the polarized Stokes~V
patterns for almost all Zeeman sensitive atomic spectral lines are
similar except for a non-linear scaling (in wavelength and
amplitude) caused by different atomic line parameters, broadening
mechanisms, and magnetic sensitivities. For rapidly rotating
stars, when the rotational broadening dominates as well as under the weak magnetic field
approximation \citep{Stenflo94}, line-profile shapes can be considered 
as equal and amplitudes are only \emph{linearly} dependent 
upon their individual Land\'{e} factors and line strengths. 
One popular and very successful extraction
technique which follows this line of arguments is the so called
least-squares deconvolution \citep[LSD,][]{Donati97, Kochukhov10}. Another
method is the principal component analysis \citep[PCA,][]{Carroll07b,Marian08,Carroll09,Paletou12}
or the simple but very effective coherent addition of line profiles in the
velocity or logarithmic wavelength domain \citep{Semel09,Ramirez10}.

For our application to V410~Tau, we present here an improved
version of the PCA technique that is more robust in the case of
weak polarimetric signals and which can be described in terms of an 
eigenvalue decomposition (EVD) of the observation covariance matrix. 
An EVD of the signal covariance matrix (or a SVD of the observation matrix)
allows us to obtain an orthogonal decomposition of the signal or line
profiles in terms of the covariance eigenprofiles. The basic idea
is similar as in the previous PCA-based approach where the
similarity of the individual Stokes~V profiles allows one to
describe the most coherent and systematic features 
present in all spectral line profiles as a projection onto a small number
of eigenprofiles. In parallel, incoherent features like noise,
and line blends etc. will be dispersed along many dimensions in the
transformed eigenspace. In other words the similarity and
coherence of all individual Stokes~V patterns allows one to capture
the signal pattern by a low-rank representation of the observation
covariance matrix. The respective eigenvalues of the observation
covariance matrix, together with an estimate of the noise level in
the data, then provide a means of separating the low-dimensional
signal subspace from the noise subspace. 
Once the reduced rank of the signal subspace is determined a 
projection of the observed line profiles onto the signal subspace
and a subsequent subspace-averaging is performed to eventually provide 
the signal boosting effect for the polarimetric line profile.
What is described in the following for Stokes~V profiles also holds
for Stokes~I but for the sake of brevity we will use a notation
that makes use of Stokes~V only.

The problem setup is as follows: An individual, observed, line profile
$\tilde{\vec{V}}(\lambda)$ is the result of the true signal vector
$\vec{V}(\lambda)$ and a noise vector $\vec{N}$, which we consider
as additive and independent of wavelength and time and of zero mean.
Although our idealize assumption are not met (and we will discuss the consequence later in
this section), let us proceed
with this assumption to describe the basic mechanism of the subspace method.

The observed line profile can be written in a discrete form as,
\begin{equation}
\tilde{V}_{ij} \; = \;  V_{ij} \; + \; N \; ,
\label{Eq:3_1}
\end{equation}
where $i$ and $j$ denote the wavelength and the spectral
line index, respectively. In vector form, we may write Eq.~\ref{Eq:3_1}
as
\begin{equation}
\tilde{\vec{V}}_j \; = \;  \vec{V}_j \; + \; \vec{N} \; .
\label{Eq:3_2}
\end{equation} 
We further assume that the net circular polarization of each
individual Stokes~V line profile is zero such that their integral
over the wavelength range is also zero. This implies that there
are no gradients in magnetic field and velocity present in the
atmosphere \citep{Carroll07a} which is also the usual starting
point for ZDI to keep the problem tractable.

Arranging now the individual, observed, Stokes profiles in a $m
\times n$ observation matrix $\tilde{\vec{V}}$, where $m$ denotes
the number of wavelength pixels and $n$ the number of observed
individual spectral lines, we can calculate the non-normalized
covariance matrix of the observations as
\begin{equation}
\vec{C}_{\tilde{V}} \: = \; \tilde{\vec{V}} \tilde{\vec{V}}^T \: ,
\label{Eq:3_3}
\end{equation}
where the superscript $T$ denotes the transpose. The covariance
matrix $\vec{C}_{\tilde{V}}$ is a $m \times m$ real symmetric
matrix and has $m$ linear independent eigenvectors, which can be
used to decompose $\vec{C}_{\tilde{V}}$ by a spectral
decomposition, i.e. an eigenvalue decomposition, such as
\begin{equation}
\vec{C}_{\tilde{V}} \: = \: \vec{Q} \tilde{\vec{\Lambda}} \vec{Q}^T \: ,
\label{Eq:3_4}
\end{equation}
with $\tilde{\vec{\Lambda}}$ being a diagonal matrix containing
the eigenvalues $\tilde{\lambda}_i$, and $\vec{Q}$ an $m \times m$
orthogonal matrix containing in each column the eigenvectors $\vec{\tilde{q}}_i$. By
assumption, signal and noise are uncorrelated and we may therefore
write the covariance matrix $\vec{C}_{\tilde{V}}$ as the sum of
the two matrices $\vec{C}_{S}$ and $\vec{C}_{N}$, i.e. the clean
signal and the noise covariance matrix:
\begin{equation}
\vec{C}_{\tilde{V}} \: = \: \vec{C}_{S} \: + \: \vec{C}_{N} \: .
\label{Eq:3_5}
\end{equation}
Based on our assumption that the clean signals span only a lower
dimensional subspace of the observed space (i.e. signals plus
noise), which is equivalent of assuming that the signal matrix is
rank deficient, we can use the orthogonal matrix $\vec{Q}$ of the
observed signal space for the eigenvalue decomposition of the clean signal
covariance matrix $\vec{C}_{S}$ and the noise covariance matrix
$\vec{C}_{N}$, to write
\begin{equation}
\vec{C}_{S} \: = \: \vec{Q} \: \vec{\Lambda}_S \: \vec{Q}^T \: ,
\label{Eq:3_6}
\end{equation}
and
\begin{equation}
\vec{C}_{N} \: = \: \vec{Q} \: (\sigma^2\vec{I}) \:  \vec{Q}^T \: , 
\label{Eq:3_7}
\end{equation}
which directly follows from Eq.~(\ref{Eq:3_5}).
Here, $\sigma^2$ is the eigenvalue of the noise and corresponds to
the noise variance along a direction in subspace, the matrix $\vec{I}$ is the
identity matrix and $\vec{\Lambda}_S$ is the diagonal matrix of the
eigenvalues $\lambda$ of the clean signal matrix. 
Note, because the noise is assumed to be isotropic and featureless it will be homogeneously
distributed across the eigenspace with all eigenvalues of similar magnitude.
Note also that by definition of the covariance matrix each eigenvalue
provide the total energy (variance) of the observation along its
corresponding dimension (eigenprofile). 
The noise eigenvalue is related to the estimated observation error $\sigma_{obs}$ by
\begin{equation}
\sigma^2 \; = \; m \; \sigma^2_{obs} \: .
\label{Eq:3_8}
\end{equation}
Using the definition of Eqs.~(\ref{Eq:3_6}) and (\ref{Eq:3_7}) the
eigenvalue decomposition of the covariance matrix
$\vec{C}_{\tilde{V}}$ can now be written as
\begin{equation}
\vec{C}_{\tilde{V}^*} \: = \: \vec{Q} \: (\vec{\Lambda}_S +
\sigma^2\vec{I}) \: \vec{Q}^T \: . 
\label{Eq:3_9}
\end{equation}
If the clean signals are confined to an $s$-dimensional subspace
with $s < m$, then the eigenvalues of $\tilde{\vec{\Lambda}}$ can be
expressed as
\begin{equation}
\tilde{\lambda}_i \: = \: \left \{
\begin{array}{ll}
\lambda_i + \sigma^2 \hspace{0.5cm}  i = 1...s \\
\sigma^2  \hspace{1.1cm}  i = s+1...m \: .
\end{array}
\right .
\label{Eq:3_10}
\end{equation}
Partitioning $\vec{Q}$ in an orthogonal $m \times s$ matrix
$\vec{Q}_S$ for the signal subspace and a $m \times (m-s)$ matrix
$\vec{Q}_N$ for the noise subspace allows us to write for the
observational covariance matrix
\begin{equation}
\vec{C}_{\tilde{V}^*} \: = \: \left [ \vec{Q}_S \vec{Q}_N \right ] \:
\left ( \:
\left [
\begin{array}{cc}
\vec{\Lambda}_s & 0 \\
0 & 0
\end{array}
\right ]
+ \sigma^2
\left [
\begin{array}{cc}
\vec{I}_s & 0 \\
0 & \vec{I}_{m-s}
\end{array}
\right ] \:
\right )
\: \left [ \vec{Q}_S \vec{Q}_N \right ]^T \: .
\label{Eq:3_11}
\end{equation}
From this we can define a simple but effective
criterion for the separation of the signal subspace. All those
eigenvalues $\lambda_i$, i.e. the combination of clean signal and
noise eigenvalues, which are larger than the noise eigenvalues
alone carry more energy than a pure noise component and may
therefore exhibit significant structural information. The
condition for the separation of the signal space then reads
\begin{equation}
\tilde{\lambda}_i \: > \: \sigma^2  \: . \label{Eq:3_12}
\end{equation}
It provides an estimate of the number $s$ of the signal space,
i.e. the reduced rank. The above described eigenvalue
decomposition can efficiently calculated by utilizing a singular value
decomposition (SVD) of the original observation matrix $\tilde{\vec{V}}$.

Let us now briefly discuss the restriction and consequences of our idealized 
assumption, i.e. that we deal with homogeneous and stationary noise. 
The real noise that we are dealing with in our spectral data is certainly not 
stationary and may have a dependence on wavelength. The noise can therefore 
introduce a systematic feature and the noise matrix will have off-diagonal elements.
If the correlation of this feature is strong enough among the contributing lines
it would give rise to a structure in signal space that is different
from pure random noise. In that case this systematic will therefore be detected as 
a significant signal feature in our framework as it follows our basic notion of 
being systematic and correlated among all contributing lines.  
So without giving up the derivation of the rank estimation we need
to change the meaning of the signal space. The signal space is not exclusively 
given by the true underlying Stokes signal but rather includes all systematic effects
that possess a certain correlation within the observation matrix.

In principle, we could now retrieve each individual Stokes~V
profile by projecting the observed line profiles $\vec{V}_j$
onto the restricted set of $s$ signal
eigenprofiles from $\vec{Q}_S$. This could be done for any
line profile of interest. Unfortunately, this approach alone is limited to
relative noise levels above unity. The situation changes when the original
profiles are noise dominated such that the true
signal patterns are completely hidden by noise and any projection operations
is increasingly affected by the noise contamination of the original signal.
For this case we use a two-stage strategy where the reduced rank estimation of 
the covariance matrix is used to perform a coherent addition of the projection coefficients 
in signal subspace.
Consequently our second stage of the signal extraction consists
of a projection of each individual line profile onto the eigenprofiles $\vec{Q}_S$  
in order to obtain a low-dimensional representation of each observation vector. 
This projection and the respective matrix of projection coefficients $\vec{V}_P$  can be
written as
\begin{equation}
\vec{V}_P \; = \;\vec{Q}^T_S \; \tilde{\vec{V}} \; .
\label{Eq:3_13}
\end{equation}
We can use the projection matrix to obtain the subspace averaged projection vector $\bar{\vec{V}}_P$
by multiplying an $n \times 1$ sum vector $\vec{Z}$ to the projection matrix 
$\vec{V}_P$. Each entry of the vector $\vec{Z}$ may hold a specific weighting for 
each individual spectral line or may be given by a simple equal weighting scheme 
(e.g. 1/n for the sample mean).
\begin{equation}
\bar{\vec{V}}_P \; = \; \vec{V}_P \; \vec{Z} \: .
\label{Eq:3_14}
\end{equation}
We can then use the averaged projection vector $\bar{\vec{V}}_P$  
together with the signal space eigenprofiles 
to finally obtain the signal reconstruction $\vec{V}_R$,
\begin{equation}
\vec{V}_R\; = \; \vec{Q}_S \; \bar{\vec{V}}_P \: .
\label{Eq:3_15}
\end{equation}
Now we have a reconstructed (weighted) mean line profile 
that is comparable to the mean Zeeman feature obtained by a 
conventional LSD approach. 
However, the reconstruction here is
performed over the estimated signal subspace, which carries the
majority of the significant information of the profile pattern.
This is in contrast to the LSD method which operates in the
original data domain where the observed signals are still a
mixture of coherent and non-coherent features. Much of the noise
and incoherent structures are already reduced in the first
processing step while the main signal-boosting is provided
by Eq.~(\ref{Eq:3_14}). The second crucial difference lies in the
fact that we use this subspace-averaged line profile not as a
proxy for the general line shape in the ZDI inversion but rather
as the real approximate \emph{mean} line profile that can be
synthetically calculated by using the same lines that have been used
for the observation matrix $\tilde{\vec{V}}$. Even though it is a
formidable computational task to calculate the mean line profile
from several hundreds or even thousands of individual spectral 
lines in each iteration cycle of the inversion process it has 
been shown that artificial neural networks \citep{Bishop95}
provide a fast and accurate approximation of the polarized
radiative transfer equation to calculate thousands of synthetic
Stokes profiles with a minimum amount of computational time \citep{Carroll08}.

The net gain of the SVD method can be readily estimated by using
our starting assumption of incoherent white noise that is
homogeneously distributed in eigenspace. 
The total energy (variance) of a
line profile is given by the following expectation
\begin{equation}
\sigma^2_{m} \; = \; E\{\bar{\vec{V}}^T\bar{\vec{V}}\} \; = \;
\sum_{i=1}^{m} \; \tilde{\lambda}_i \: . 
\label{Eq:3_16}
\end{equation}
We know that the reconstructed signal only 
uses eigenprofiles of the signal subspace so that we can write according to 
Eq.~(\ref{Eq:3_11}) their total energy as,
\begin{equation}
\sigma^2_S \; = \; \sum_{i=1}^{s} \lambda_i \; + \; s \; \sigma^2  \: .
\label{Eq:3_17}
\end{equation}
The relative signal-to-noise ratio (SNR) of the reconstructed signal can then be expressed as
\begin{equation}
SNR_S \; = \;  \frac{\sum_{i=1}^{s} \lambda_i } {\; s \; \sigma^2}  \; .
\label{Eq:3_18}
\end{equation}
On the other hand using again Eq.~(\ref{Eq:3_10}) as well as Eq.~(\ref{Eq:3_16}) 
the energy of the original signal  which results from the signal subspace eigenvalues plus the 
noise contribution from the entire m-dimensional domain is given by,
\begin{equation}
\sigma^2_{m} \; = \; \sum_{i=1}^{s} \lambda_i \; + \; m \; \sigma^2 \: ,
\label{Eq:3_20}
\end{equation}
and the relative signal-to-noise ratio in this case is
\begin{equation}
SNR_{m} \; = \;  \frac{ \sum_{i=1}^{s} \lambda_i } {\; m \; \sigma^2} \; . 
\label{Eq:3_21}
\end{equation}
This finally allows us to write the gain factor $g$ for the first stage of the SVD reconstruction,
\begin{equation}
g \; = \; \frac{ SNR_{S} } {SNR_m} \; = \;  \frac{m}{s} \; .
\label{Eq:3_22}
\end{equation}
We see that due to the projection of the observations onto the
signal subspace we already obtain a signal improvement of the order of
$m/s$, i.e. the noise level scales with $\sqrt{m/s}$. 
The sample size also plays an import role in the reconstruction process 
and is relevant for the noise level within the signal eigenprofiles  
as well as for the second stage of the reconstruction process. 
Using Eq.~(\ref{Eq:3_15}) together with the mean weighting scheme in the second stage 
and moreover realizing that each subspace dimension contributes a total noise variance which 
is equal to the eigenvalue $\sigma^2$, the variance of the average of all subspace dimensions 
then scales with $\sigma^2/n$.

However, let us again emphasize that our entire derivation above is based
on an idealized noise behavior (i.e.wavelength independent  white noise) 
which is might be not fulfilled in real high-resolution spectroscopic observations. 
For a more reliable rank estimation as
well as an estimation of the noise level for the reconstructed
Stokes profiles, we will rely on a randomized
bootstrap procedure (see next section).

\begin{figure}[!t]
\begin{minipage}{9cm}
\centering
\includegraphics[width=8cm]{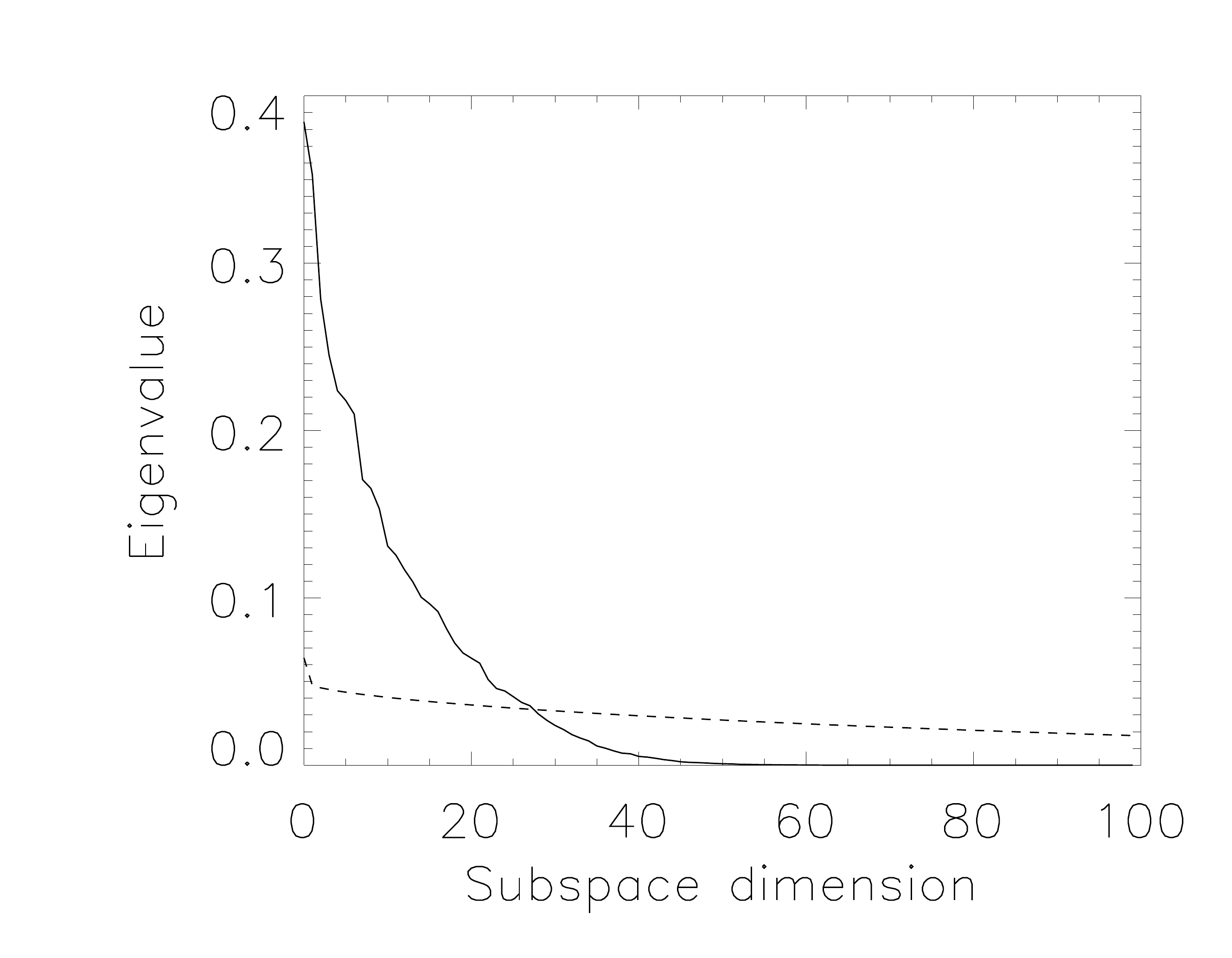}
\caption{Stokes~V eigenvalue spectrum of a V410-Tau observation
(solid line) and the mean eigenvalue spectrum obtained from a
bootstrap randomization process (dashed line) for the first 100
dimensions of rotation phase 0.014.  The magnitude of the
eigenvalues of the corresponding observation covariance matrix and
the mean eigenvalues of all randomized matrices are plotted over
the corresponding subspace dimension index. The crossing point at dimension
27 marks the dimension from which on the individual eigenprofiles
contain no more significant information.} 
\label{Fig:1}
\end{minipage}
\begin{minipage}{9cm}
\centering
\includegraphics[width=8cm]{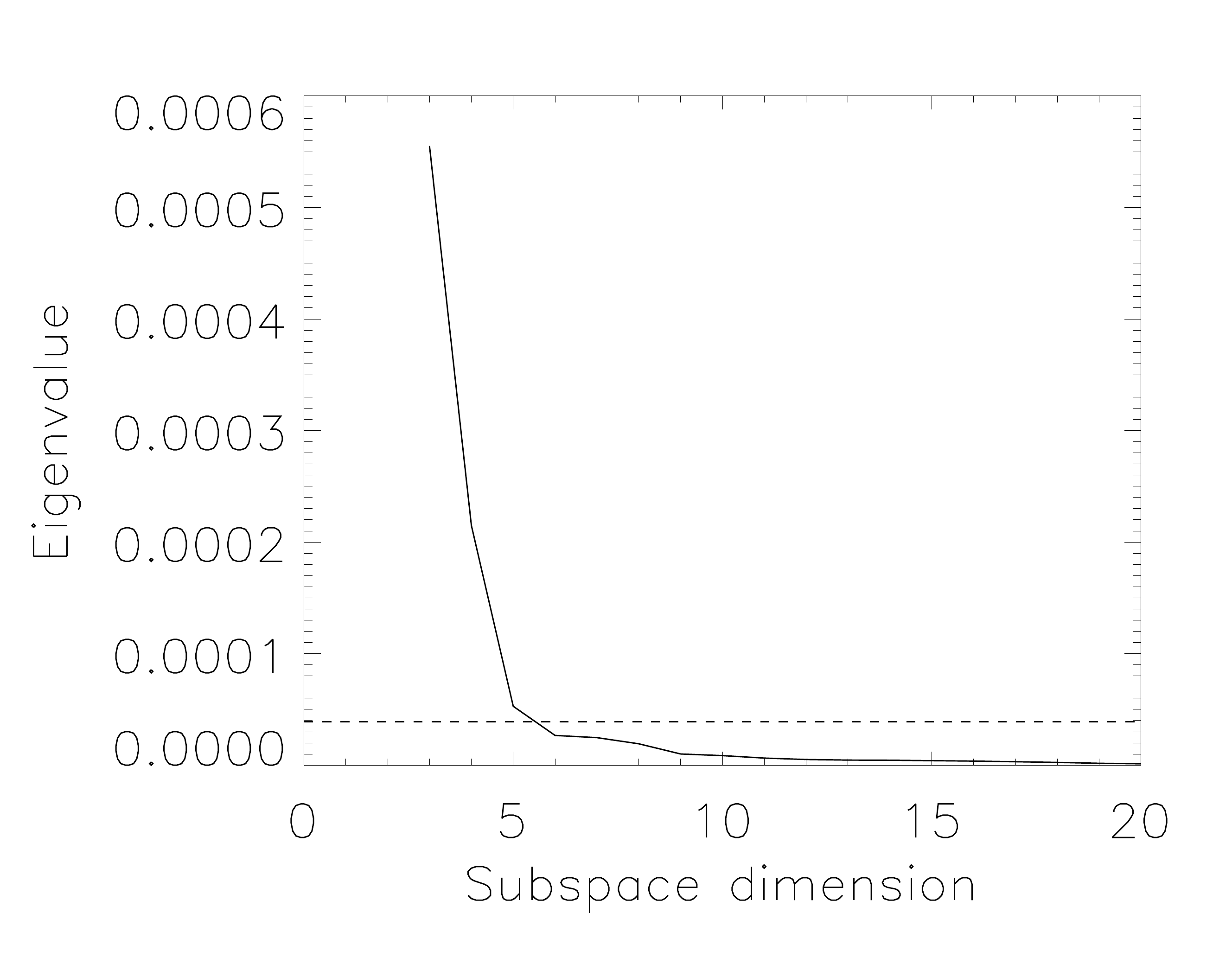}
\caption{Normalized Stokes~I eigenvalue spectrum of the
observation (solid line) and the noise variance (dashed line) for
the first 20 subspace dimensions of rotation phase 0.014.  For the purpose
of a better illustration, the first three eigenvalues are not
shown since they are significantly larger. The crossing point
determines the signal subspace dimension to be 5.} \label{Fig:2}
\end{minipage}
\end{figure}

\subsection{Application to V410~Tau}
\label{Sect:3.2}

To prepare the observation matrix $\tilde{\vec{V}}$, we first
have to select a suitable number of Zeeman-sensitive spectral lines. 
This is done by calculating a synthetic spectral atlas to judge which spectral 
line profiles are strong enough and provide enough information.

By utilizing the forward module of \emph{iMap} \citep{Carroll08} we
synthesize Stokes~I and Stokes~V spectra for the observed
wavelength range of our data (480--740nm). The individual line
parameters are taken from the VALD line list \citep{Piskunov95,Kupka99}. 
The astrophysical parameters for the
synthetic calculations are those summarized in
Sect.~\ref{Sect:Parameters}, except for the rotational velocity
which is set to zero. A uniform temperature star with a
homogeneous and purely radial surface magnetic field of 500~G is
taken for the model spectra. The model atmospheres used throughout the
analysis are Kurucz/Atlas-9 models \citep{Castelli04}.

The resulting synthetic spectral atlas (line list) in Stokes~I and Stokes~V is
then used to select those spectral lines that fulfill the following
criteria: Stokes~I line depth $\geq 0.6$ and Land\'{e} factor $\geq
1.5$. 

This procedure allowed us to select a list of 929 spectral lines.
The majority of these line are neutral Fe, Ca, Ti, Na, and C
lines. For Stokes~I, we only select a short list of the strongest
56 iron lines because the intensity profiles are two to three 
orders of magnitude larger than the polarized spectra. 
In this case there is no need for a full
\emph{reconstruction} but instead a \emph{denoising} suffices,
which requires a significantly smaller number of spectral lines.
In a first step, the observed spectra are transformed into velocity
domain with a step size of 1.5~\kms, which corresponds approximately 
to the lowest wavelength sampling in the observed spectra.
Then the line list is used to extract the corresponding Stokes~I and
Stokes~V profiles within a range of $\pm$150~\kms\ around their 
respective line centers.
Using the extracted line profiles, we then build the Stokes~V observation matrix
$\tilde{\vec{V}}$ and the Stokes~I observation matrix
$\tilde{\vec{I}}$ as described in Sect.~\ref{Sect:3}.

In total, 12 rotational phases were obtained within a time span of
four weeks during December 2008 and January 2009. For each of these
12 rotation phases, indexed with $p$, we create an observation
matrix $\tilde{\vec{V}}_{p}$ and $\tilde{\vec{I}}_{p}$ with 201
variables (intensities along the velocity domain) and a sample size of 
929 for $\tilde{\vec{V}}_{p}$ and 56 for $\tilde{\vec{I}}_{p}$. 
As described in Sect.~\ref{Sect:3}, a
singular-value decomposition provides us with the eigenvalues and
eigenprofiles of the observation covariance matrix from which we
derive for each phase a reconstructed average SVD profile.

Because the original raw Stokes~V spectra show no apparent sign of a
detectable line profile polarization, i.e the polarimetric signals are 
much weaker than the noise level, we adopt as an initial hypothesis
that the observed Stokes~V spectra contain \emph{no} signal
information except pure noise (not necessarily white). If this
hypothesis is true then a random permutation of the 
intensity values of each line across the velocity domain should
not change the outcome of our SVD analysis. In other words, if
there are no systematic and coherent features present among the 
spectral line profiles in the observation matrix, a random scrambling 
of their individual wavelength or 
velocity binned intensity values would give rise to the same
series of eigenvalues. This series of ordered eigenvalues 
(ordered according their magnitude) is called the eigenvalue spectrum.
We therefore build a second observation matrix for
Stokes~V where each extracted line profile is subject to
randomization across its velocity domain. 
If there are some sort of systematic and coherent structures
present in all observations the randomization process will
eliminate them such that the randomized covariance matrix provides 
essentially a flat spectrum of eigenvalues, i.e. all 
eigenvalues $\lambda_i^*$ have the same value 
$\lambda_1^* = \lambda_2^* =...=\lambda_m^*$. 
Hence, in the case where no significant systematic structures are present in the original 
observation matrix the eigenvalue spectrum of the observation and randomized
covariance matrix will result in a flat eigenvalue spectrum within some error margin. 
On the other hand if there are correlated and systematic features present we expect a 
distinct difference in the eigenvalue spectrum. 

However, from random matrix
theory it is known that a pure random matrix does not necessarily
provide a flat eigenvalue spectrum
and depend on the number of variables (wavelength) and measurements (spectral lines) 
\citep[see e.g.][]{Johnstone01}. For this reason, we perform a Monte-Carlo
simulation to provide a large sample (1,000) of randomized
covariance matrices from which we can deduce the mean eigenvalue
spectrum and robust error margins for each eigenvalue. The errors
of the eigenvalues are estimated by a bootstrap
procedure \citep{Efron94} where we used again a sample size of 1,000. 
After calculating the SVD of all the randomized matrices, we use a
parallel analysis \citep{Jolliffe02} to compare the significance of the 
eigenvalues of the observation covariance matrix relative to the mean
eigenvalues (averaged over the respective subspace dimension) 
of all randomized covariance matrices.


\begin{figure*}[!t]
\begin{minipage}{\textwidth}
\centering
\includegraphics[width=4.25cm]{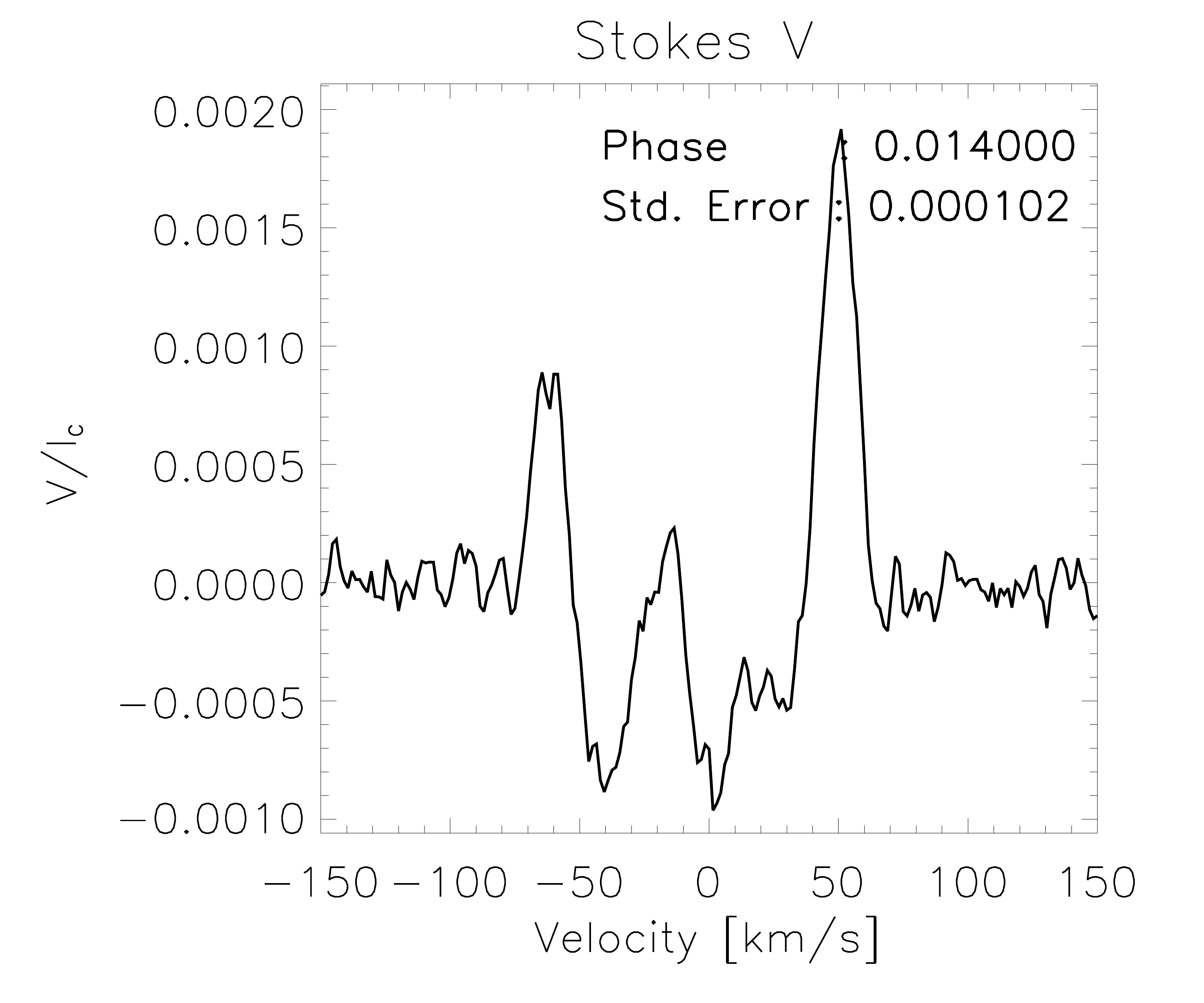}
\includegraphics[width=4.25cm]{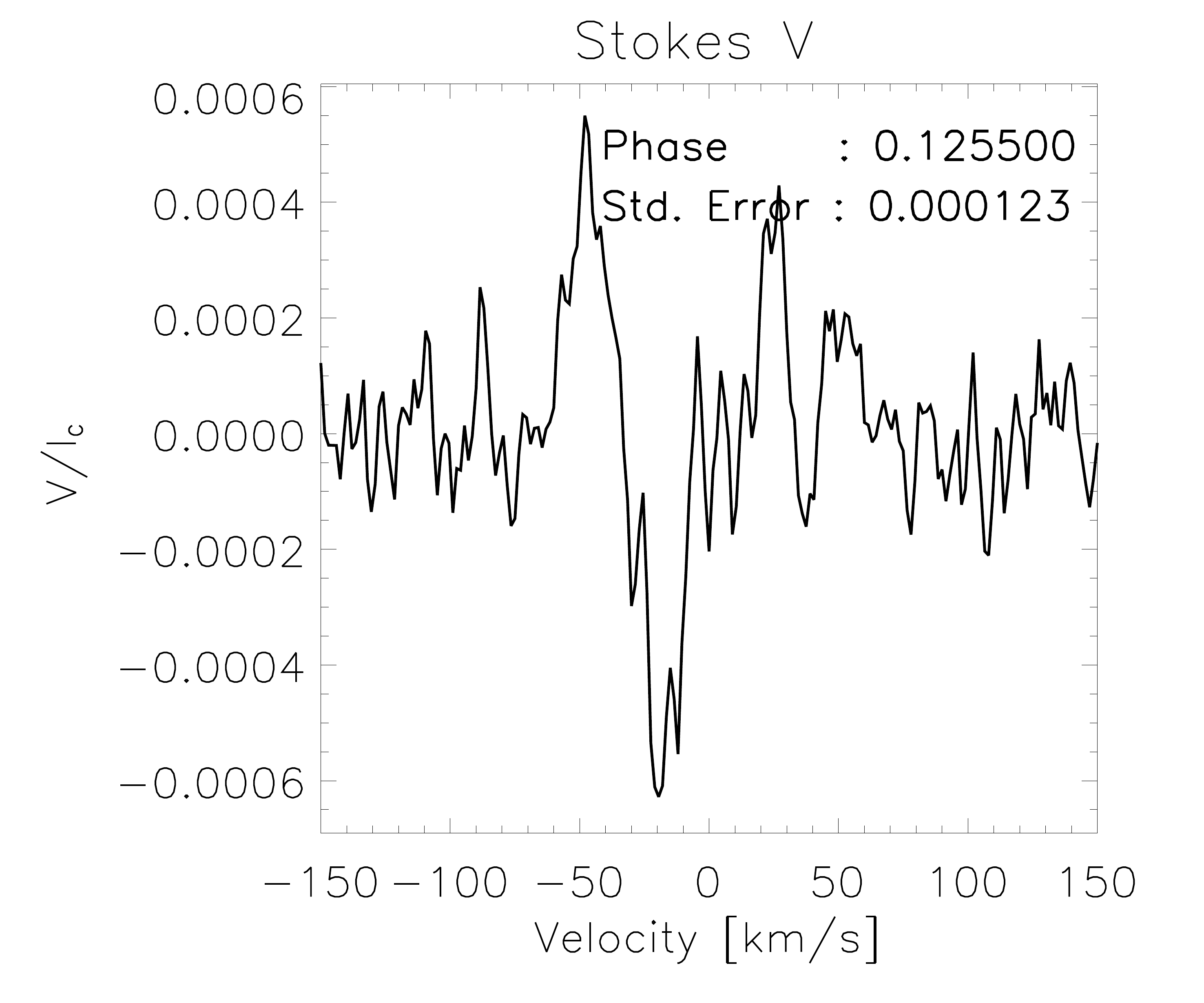}
\includegraphics[width=4.25cm]{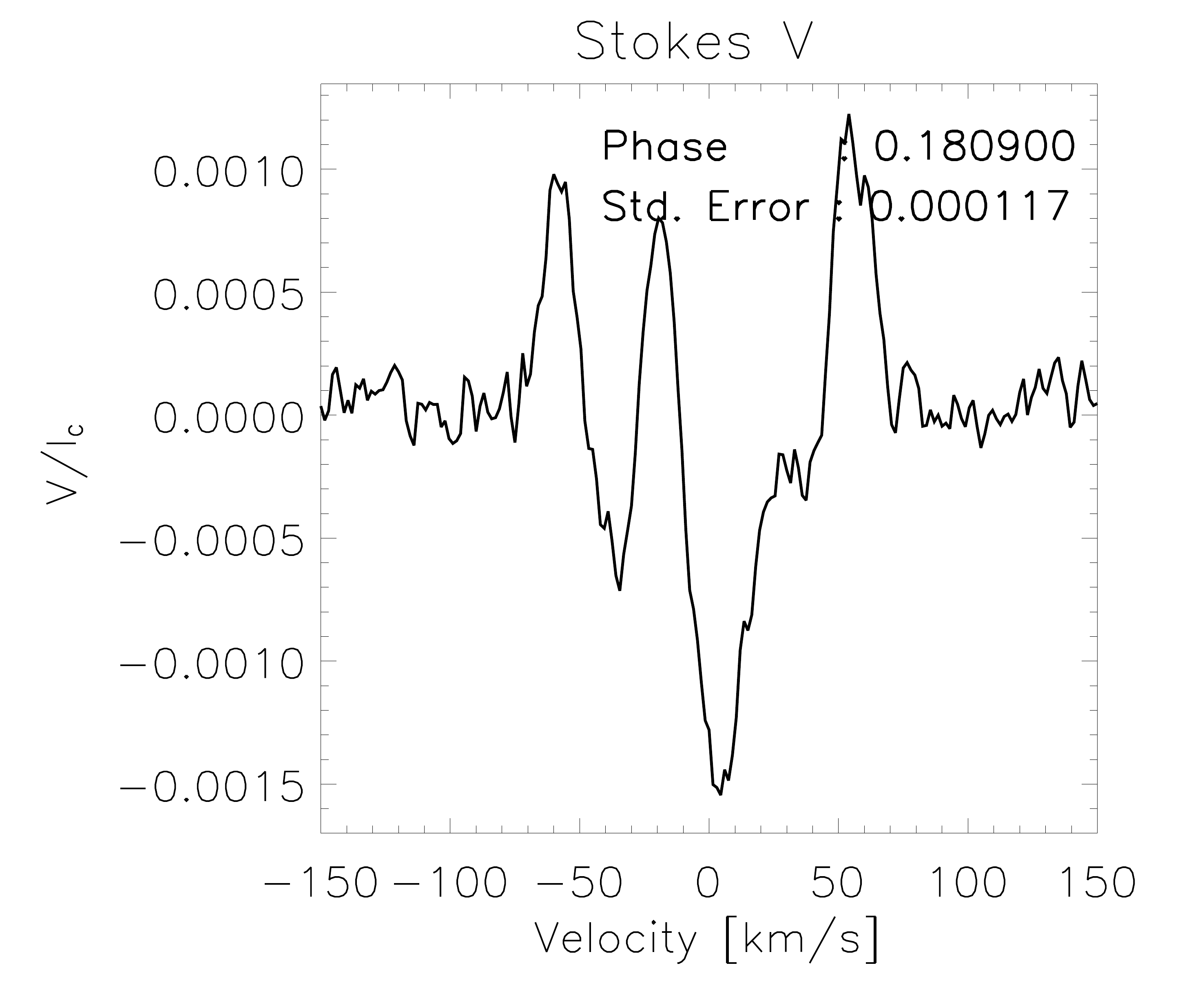}
\includegraphics[width=4.25cm]{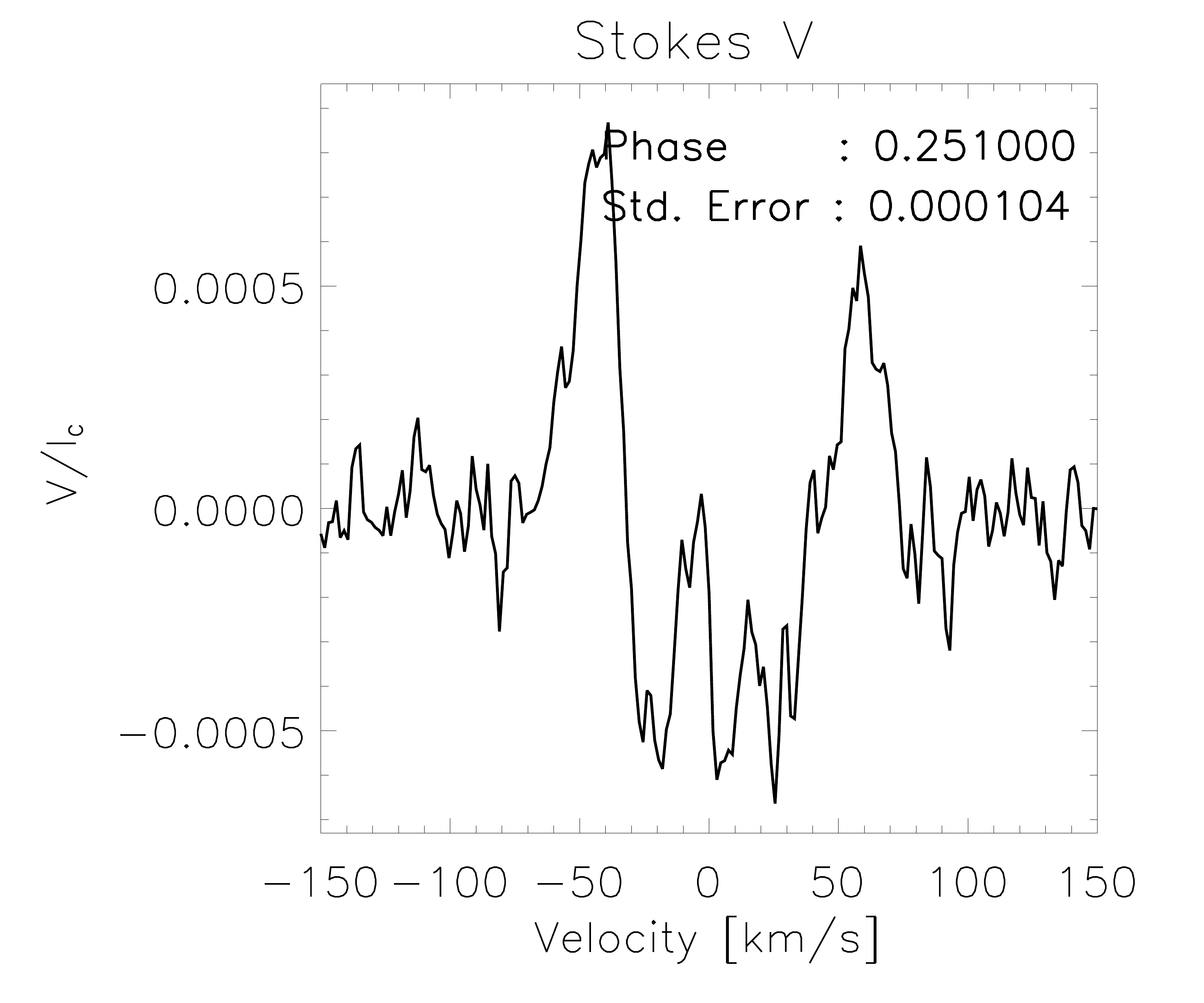}
\includegraphics[width=4.25cm]{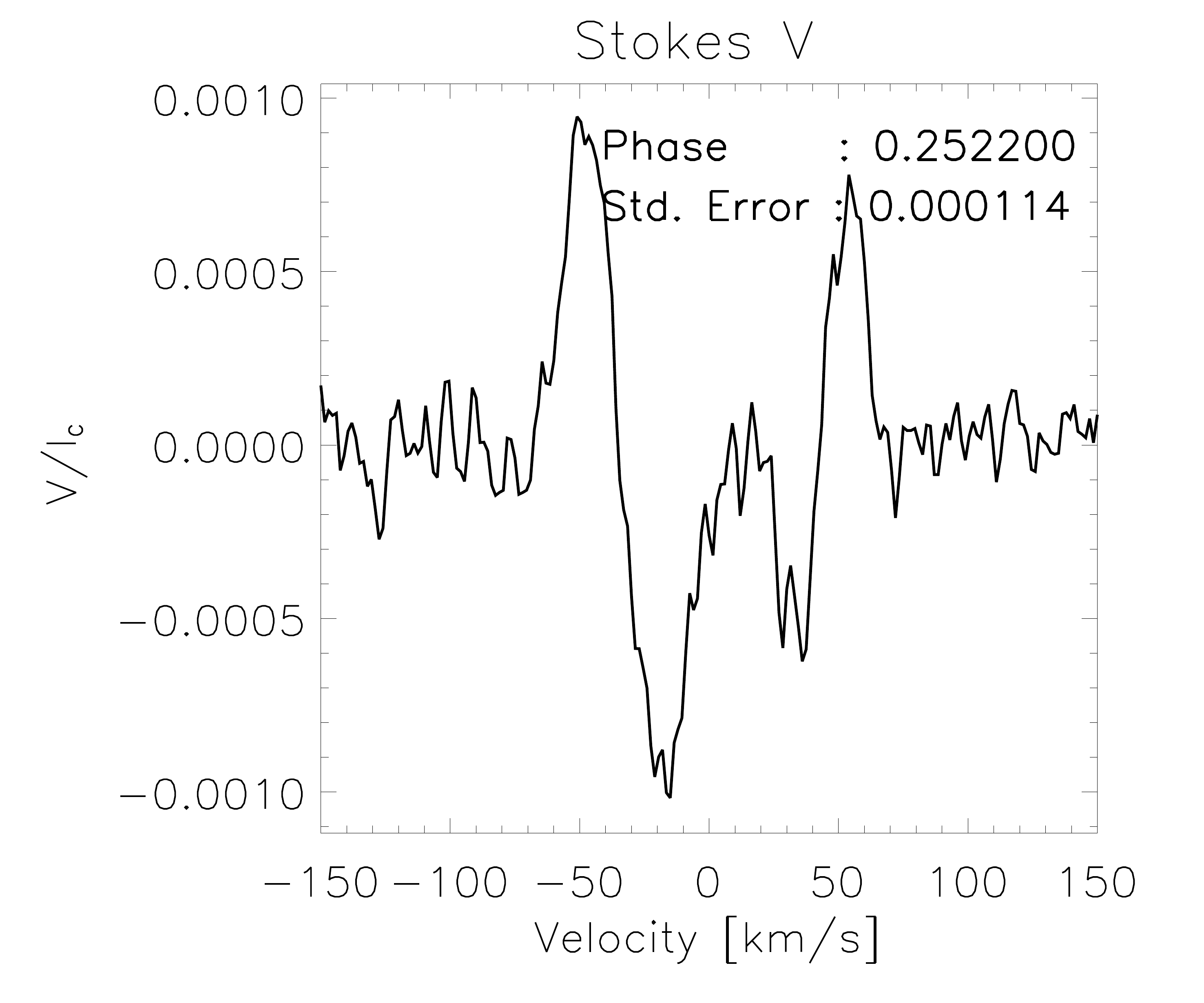}
\includegraphics[width=4.25cm]{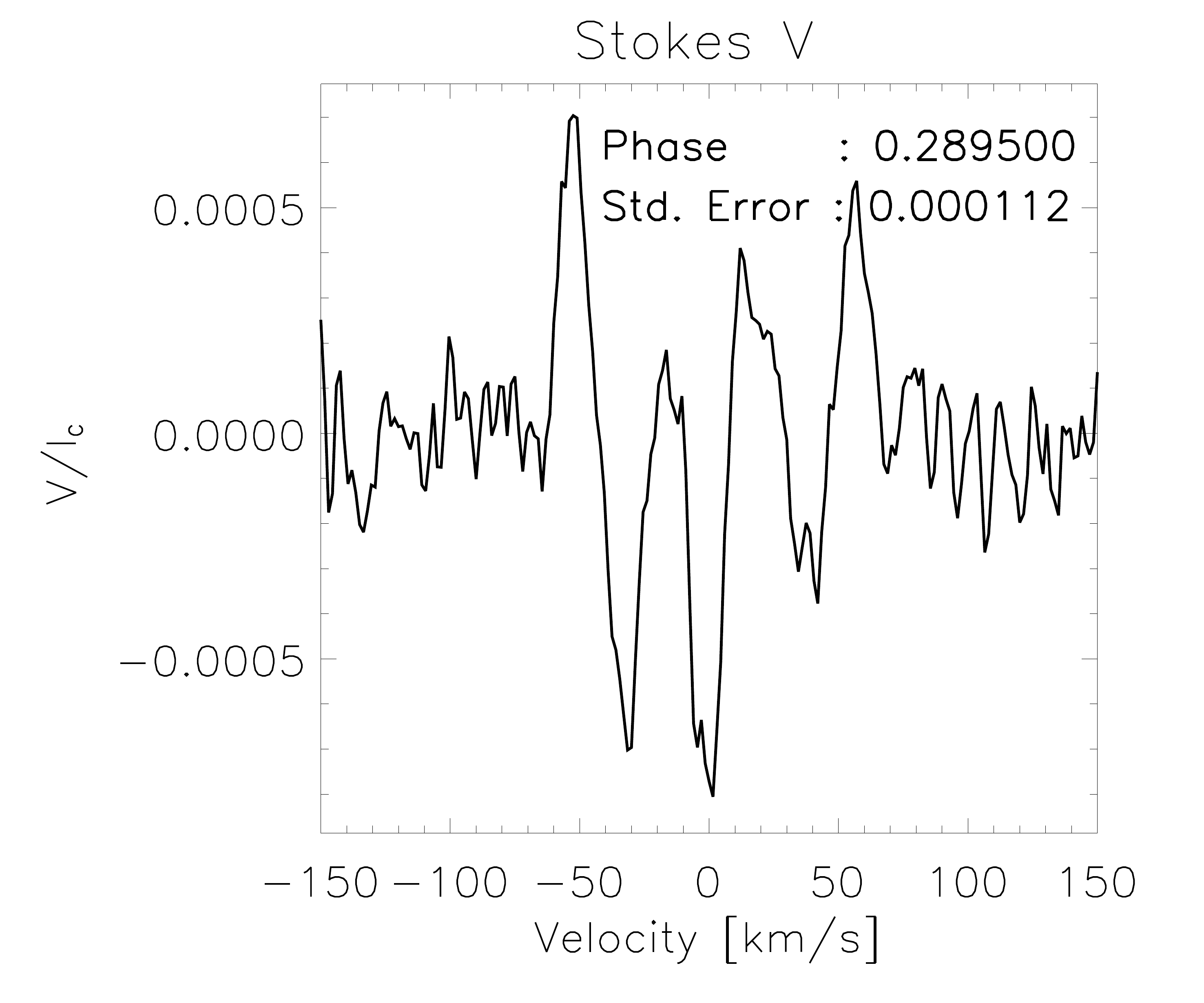}
\includegraphics[width=4.25cm]{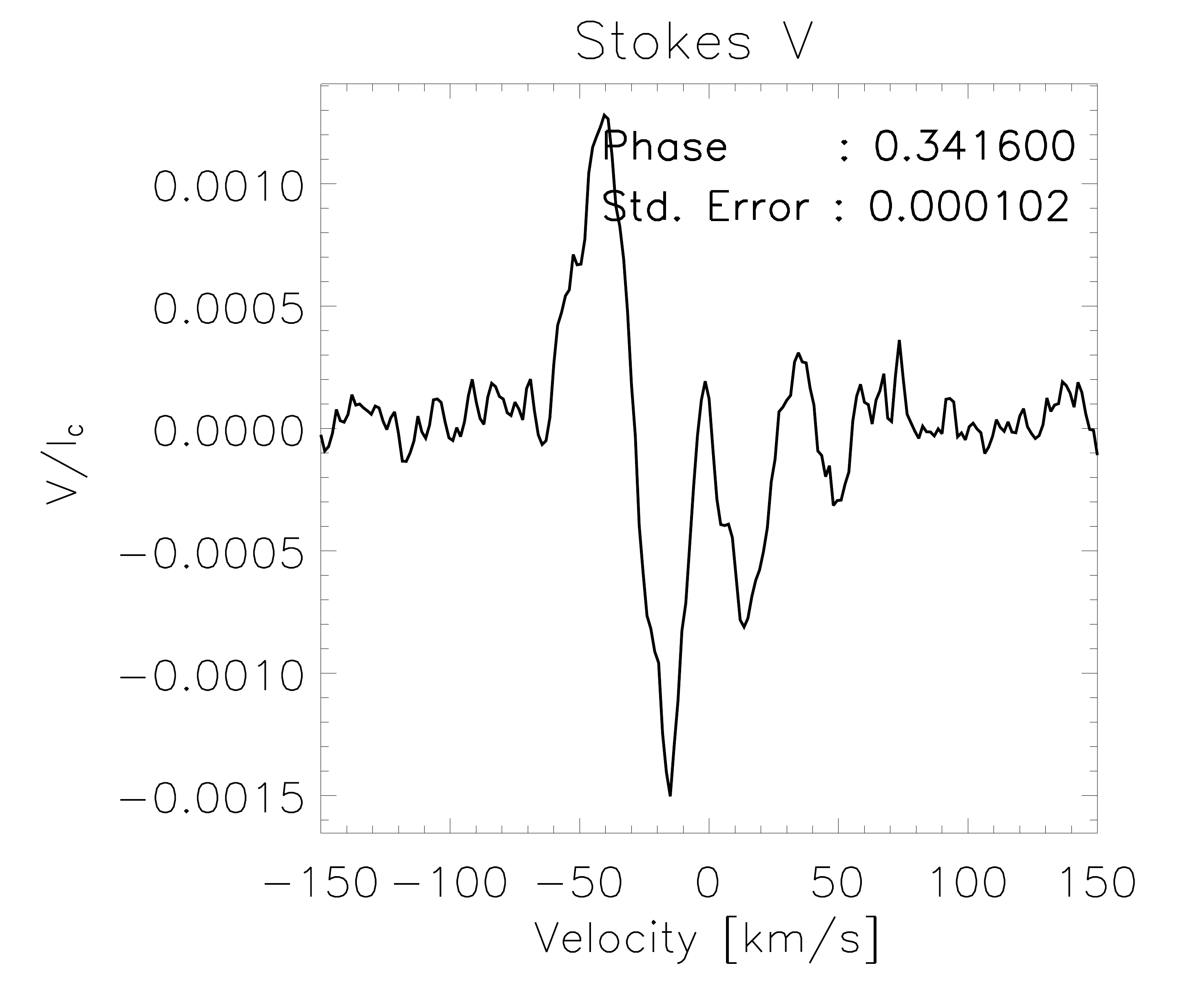}
\includegraphics[width=4.25cm]{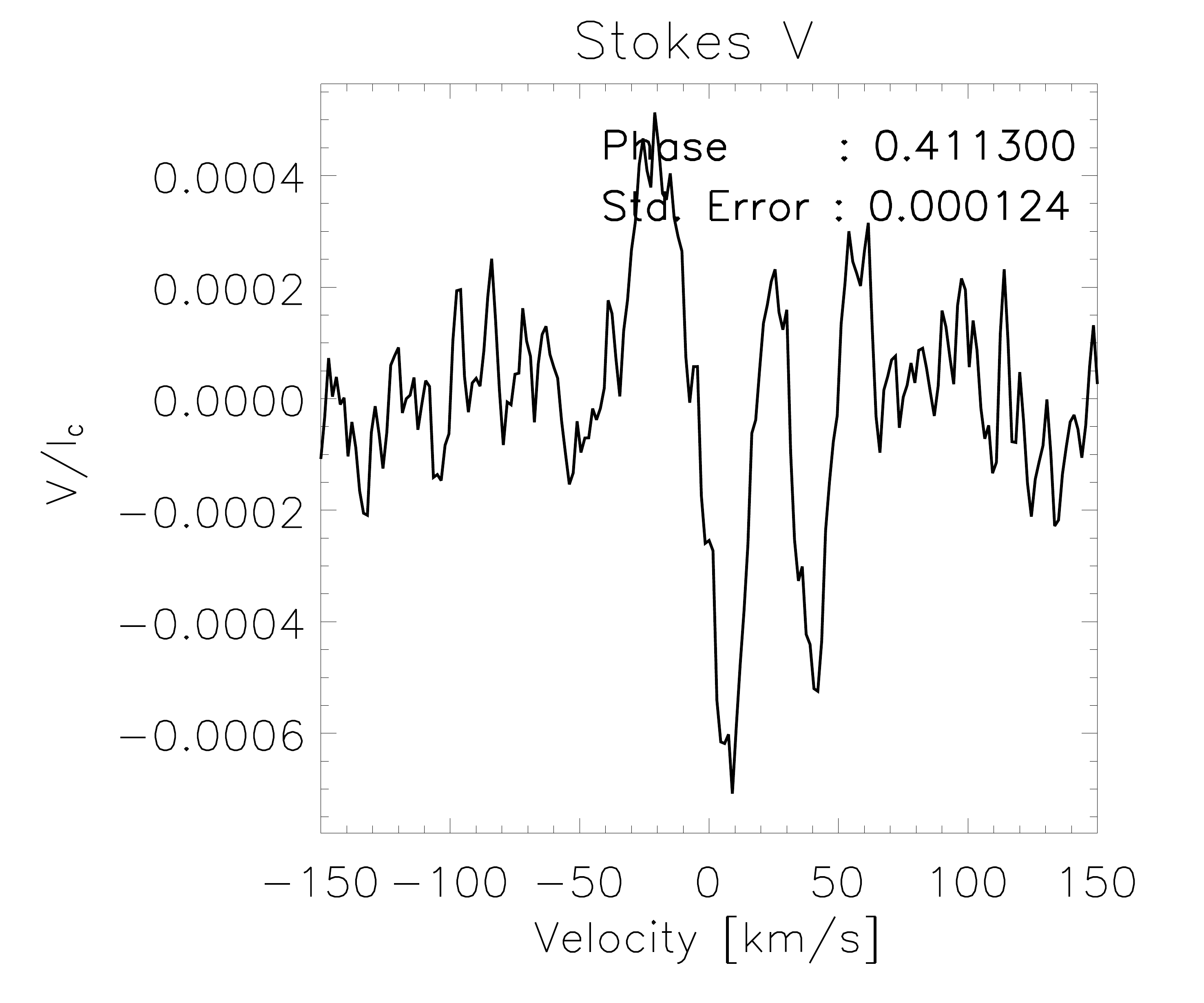}
\includegraphics[width=4.25cm]{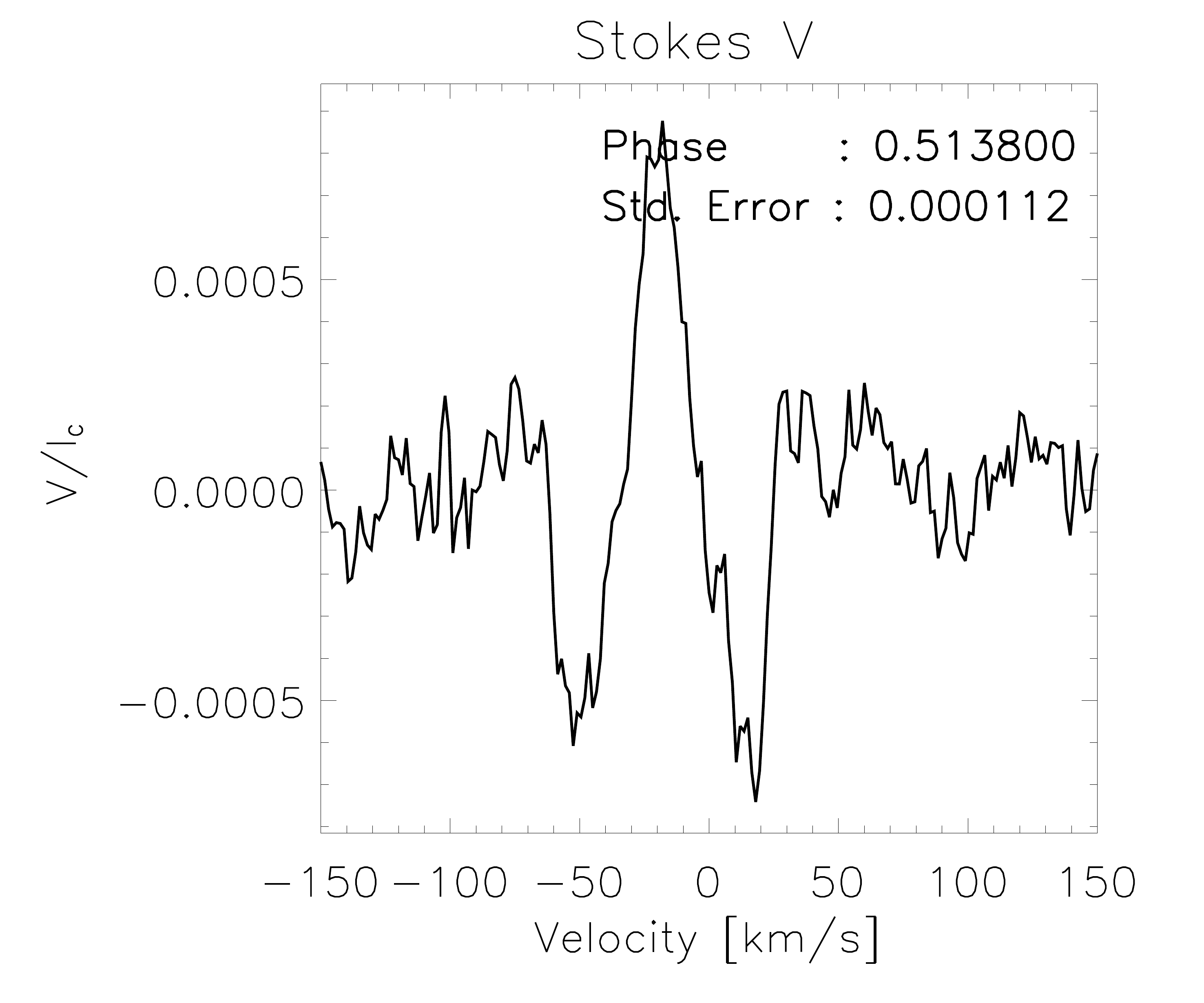}
\includegraphics[width=4.25cm]{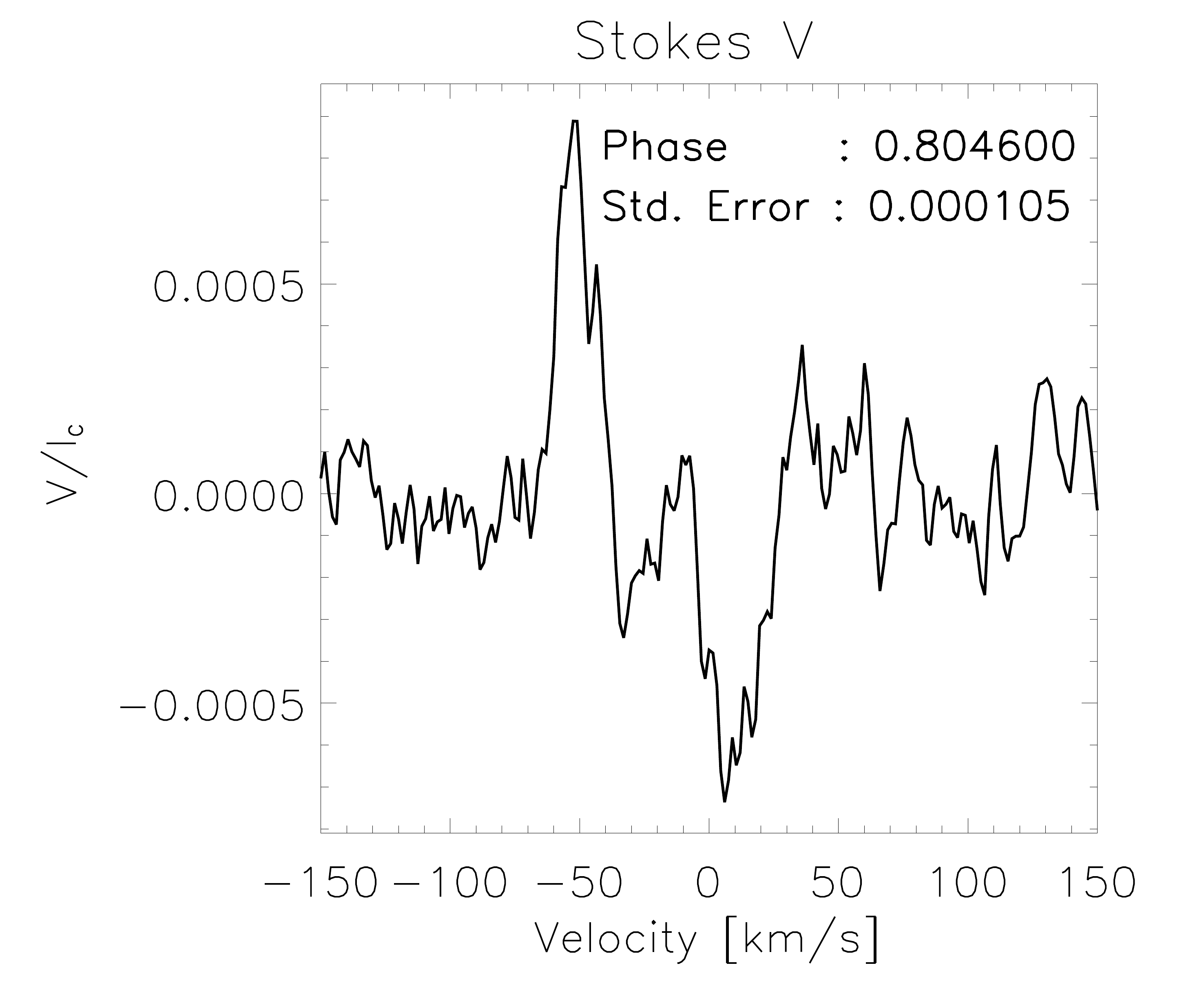}
\includegraphics[width=4.25cm]{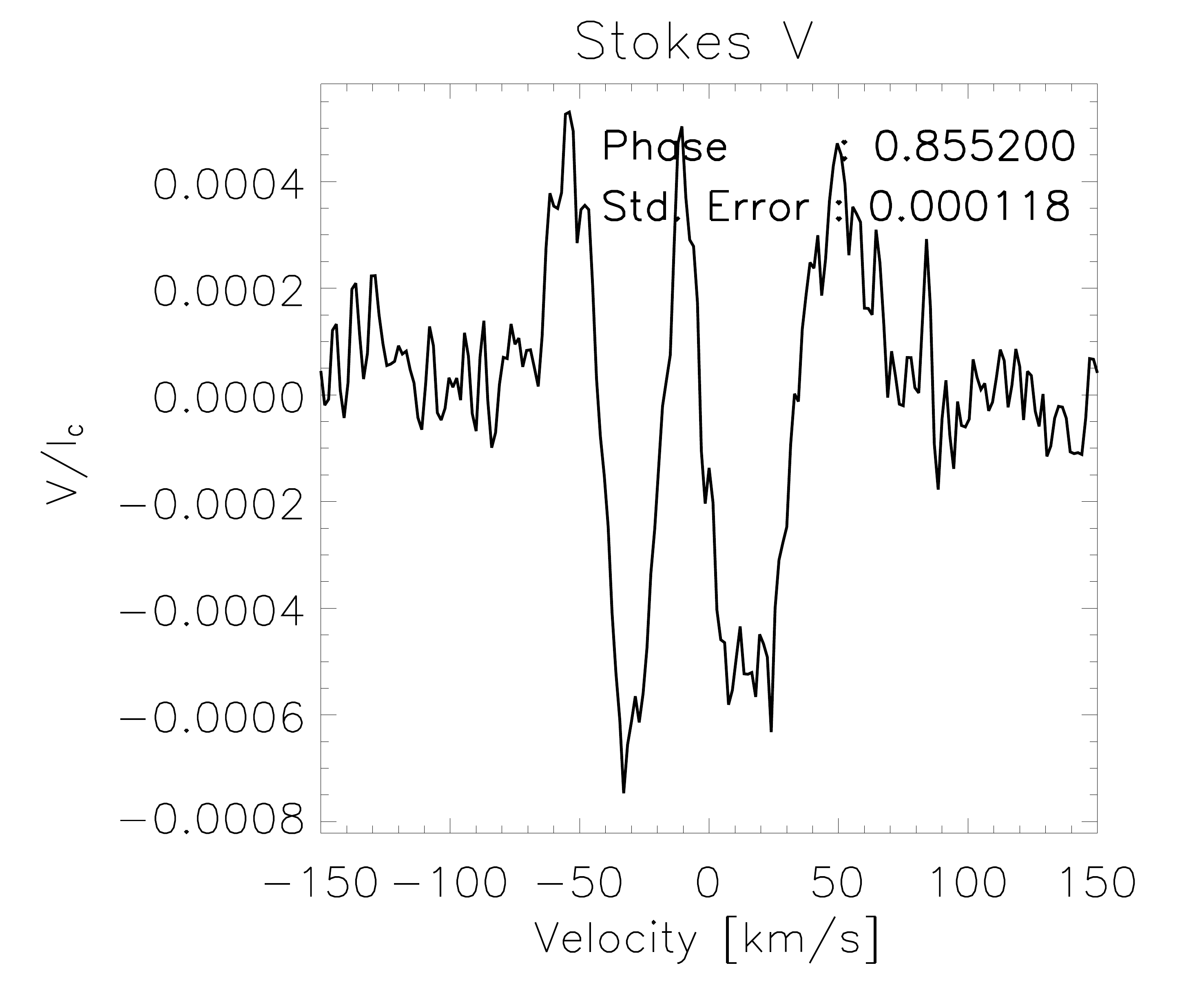}
\includegraphics[width=4.25cm]{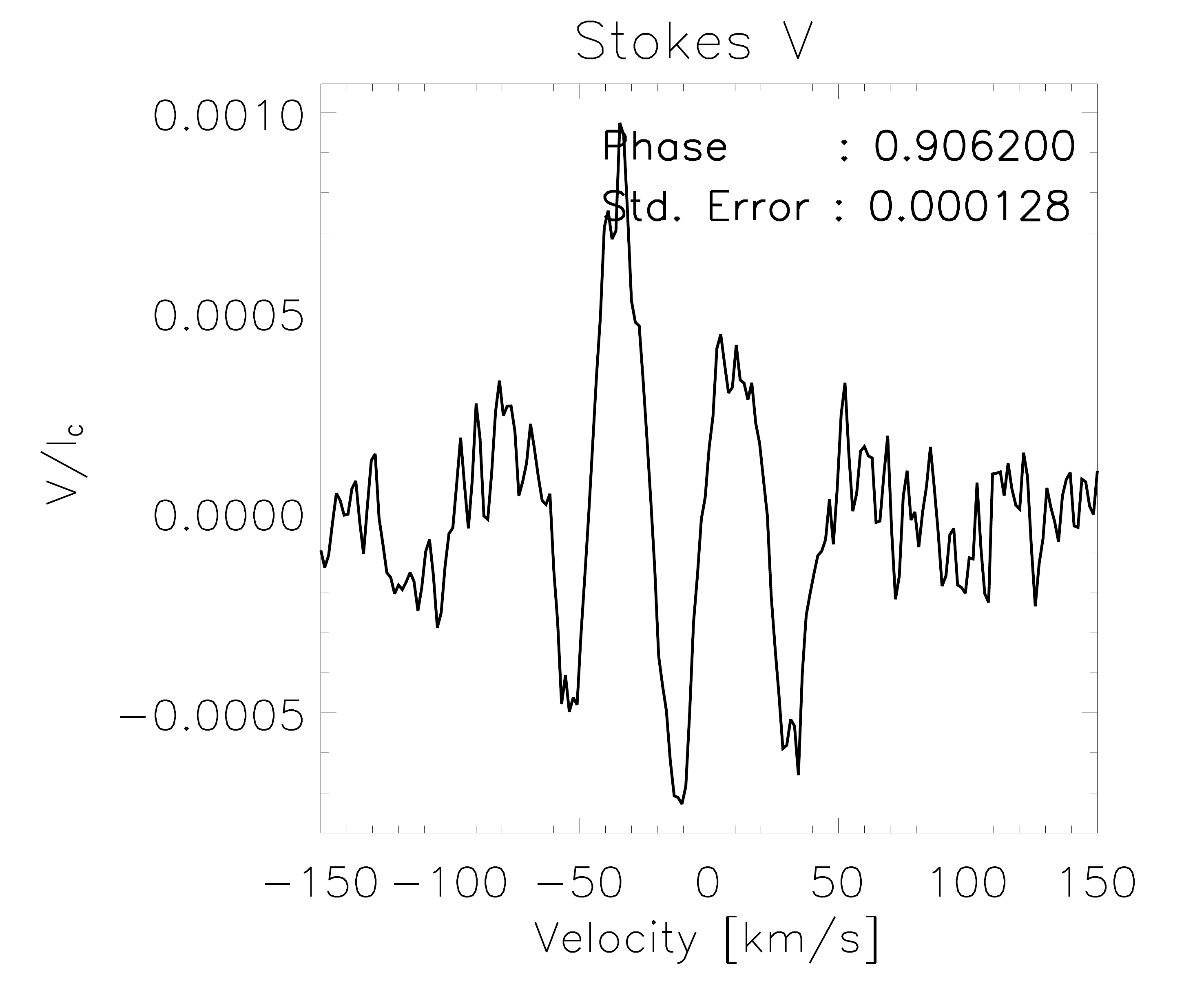}
\caption{Reconstructed SVD Stokes~V profiles for each
rotational phase of V410~Tau. Each profile is reconstructed from
an observation  matrix with a sample size of 929
spectral lines . The averaged standard error is given together
with the corresponding rotational phase.} 
\label{Fig:3}
\end{minipage}
\begin{minipage}{\textwidth}
\centering
\includegraphics[width=4.25cm]{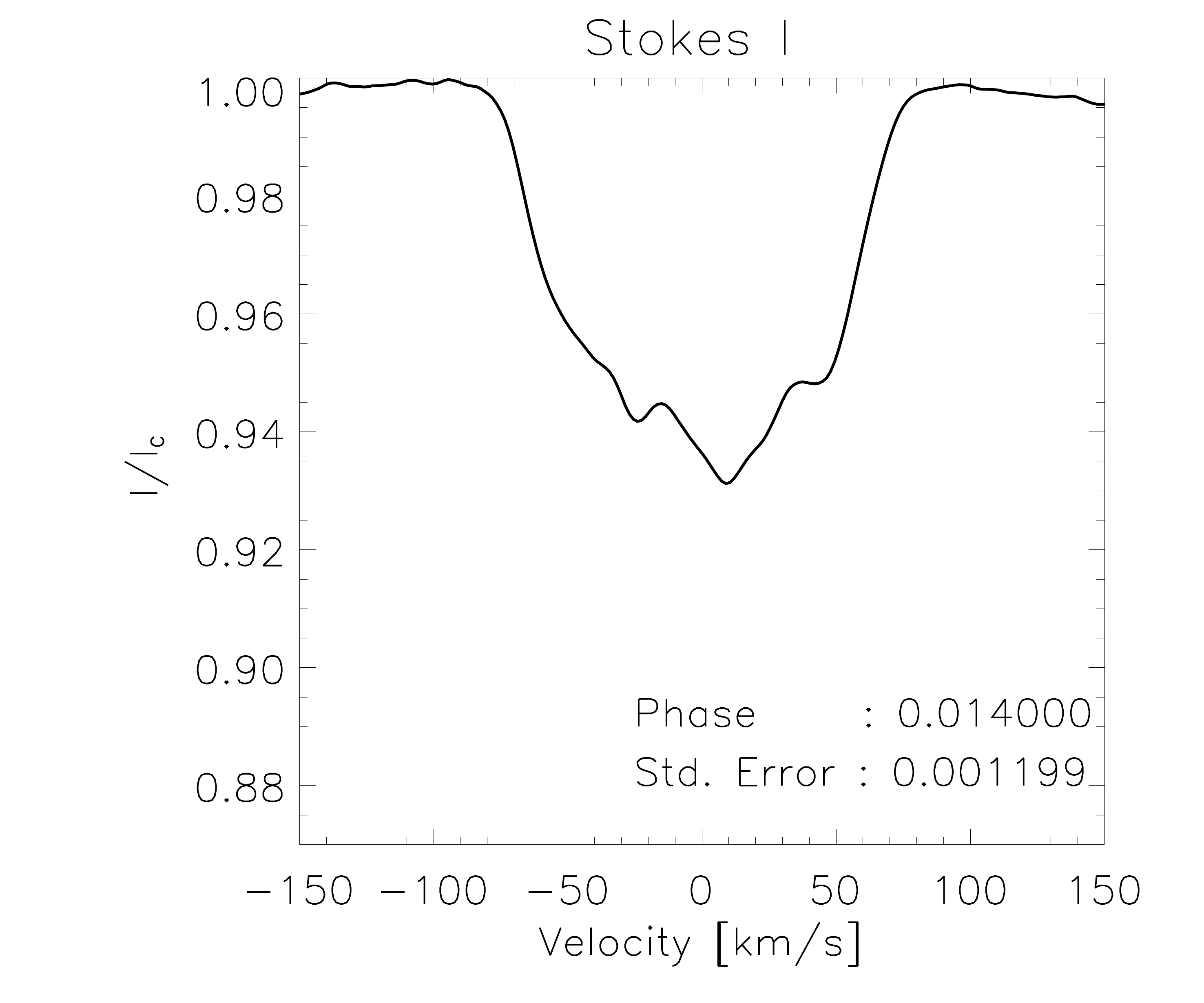}
\includegraphics[width=4.25cm]{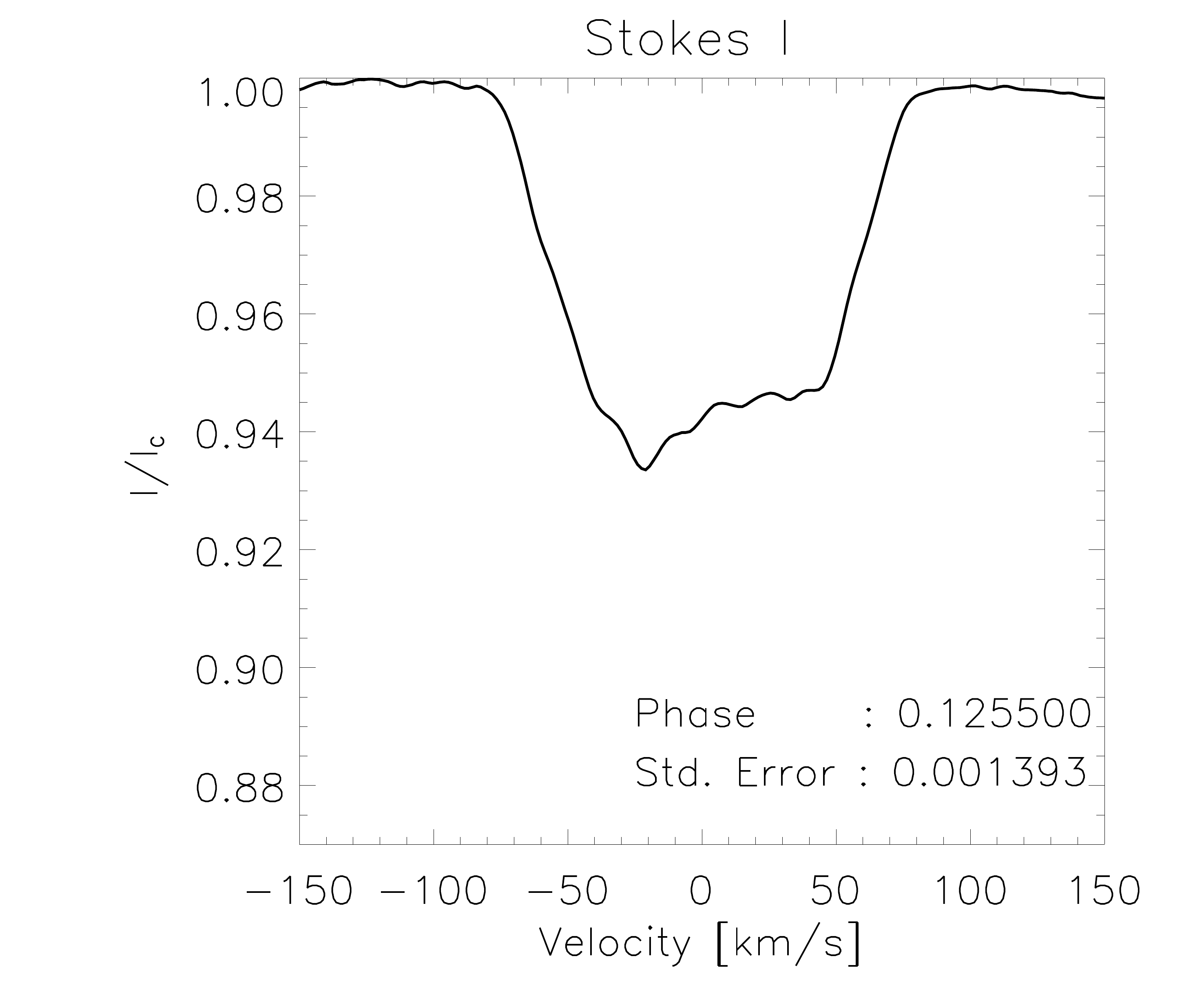}
\includegraphics[width=4.25cm]{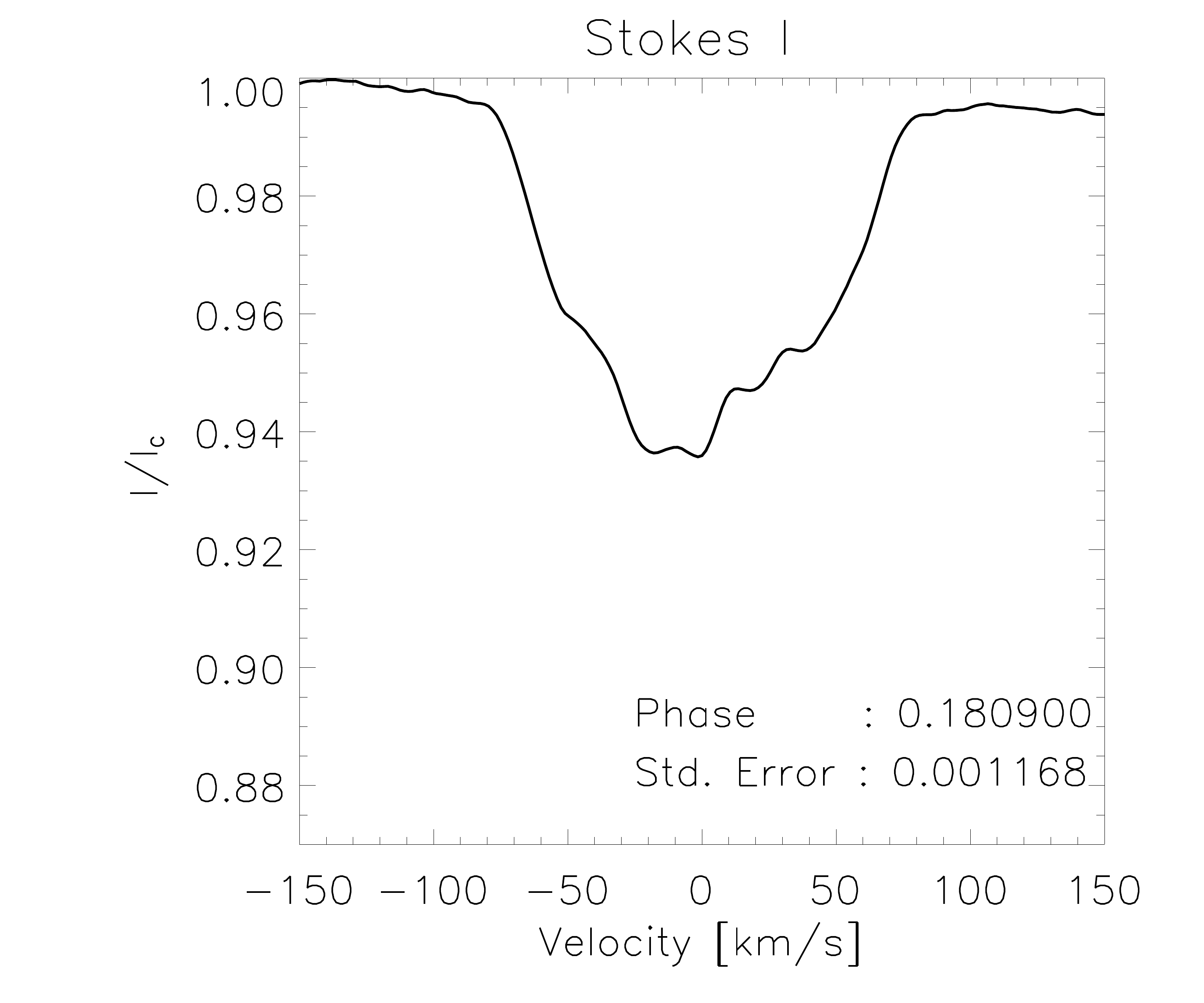}
\includegraphics[width=4.25cm]{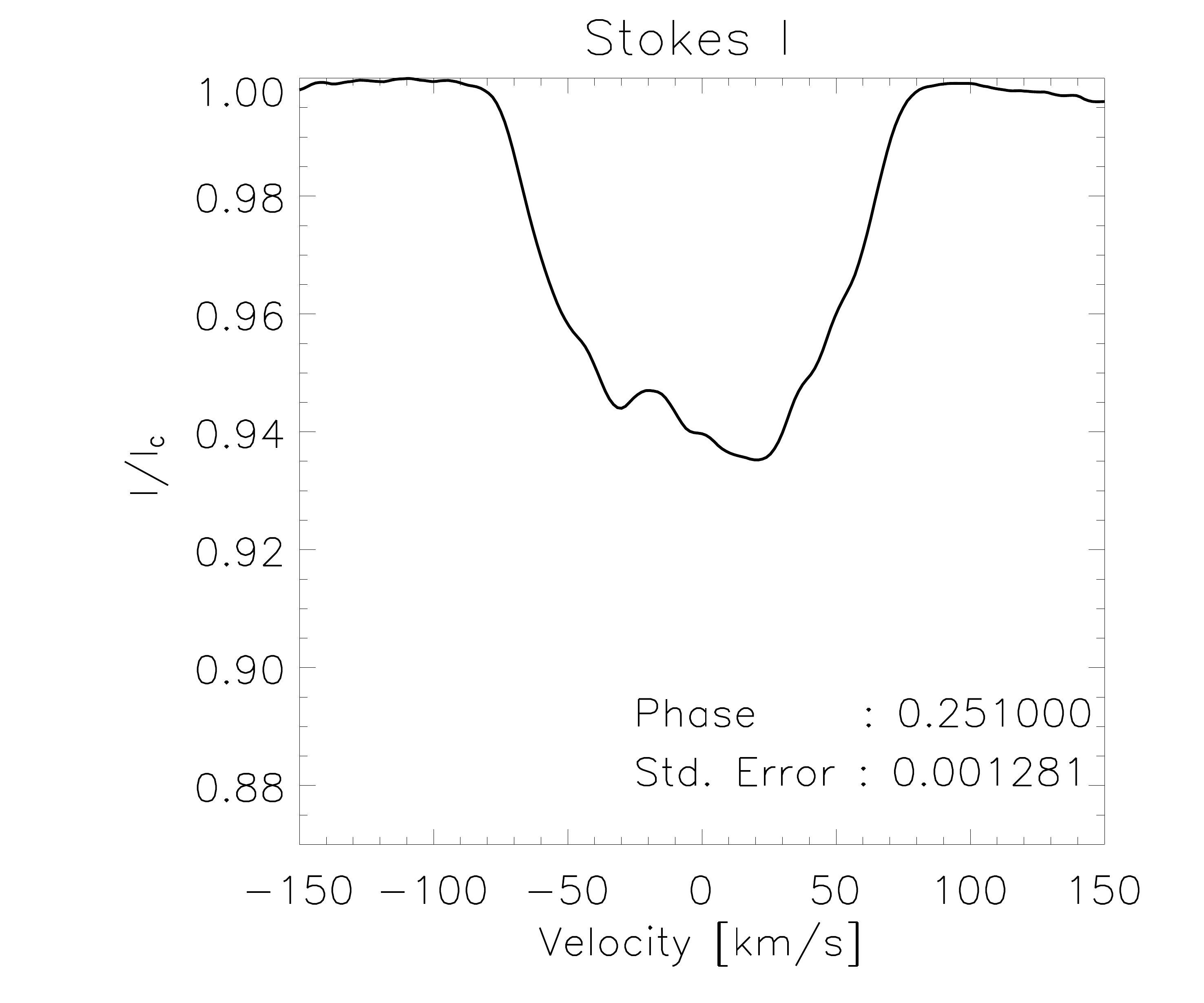}
\includegraphics[width=4.25cm]{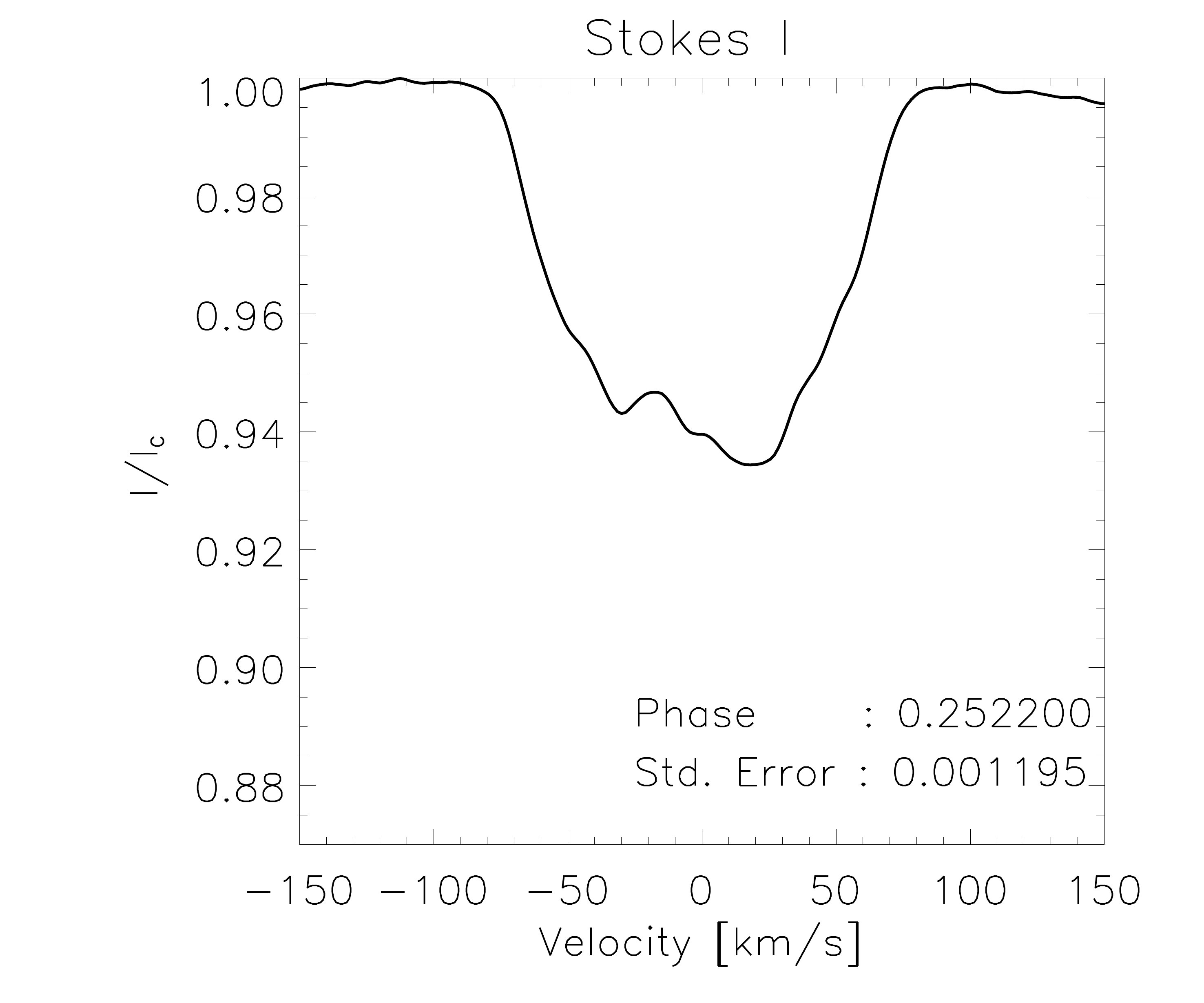}
\includegraphics[width=4.25cm]{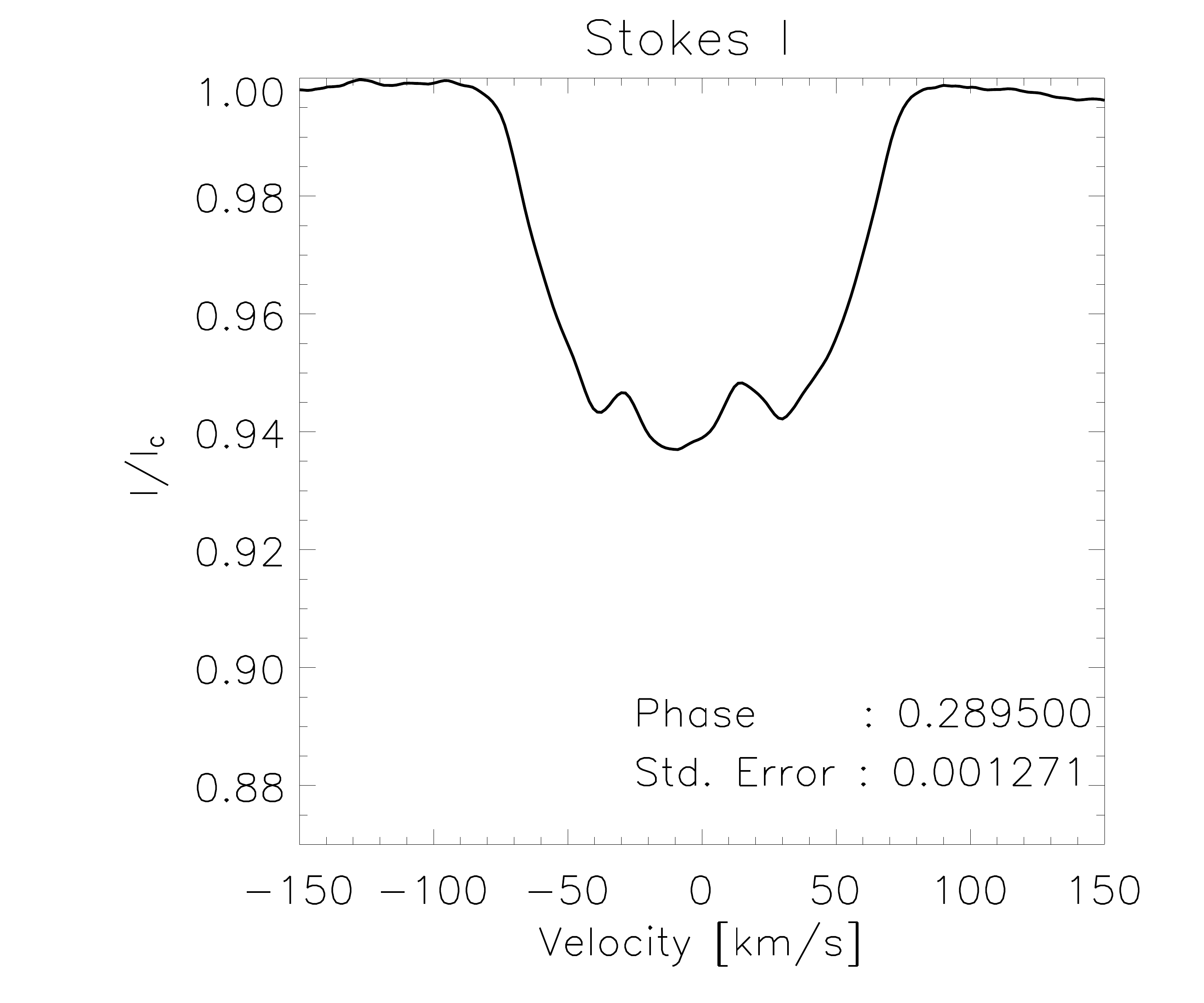}
\includegraphics[width=4.25cm]{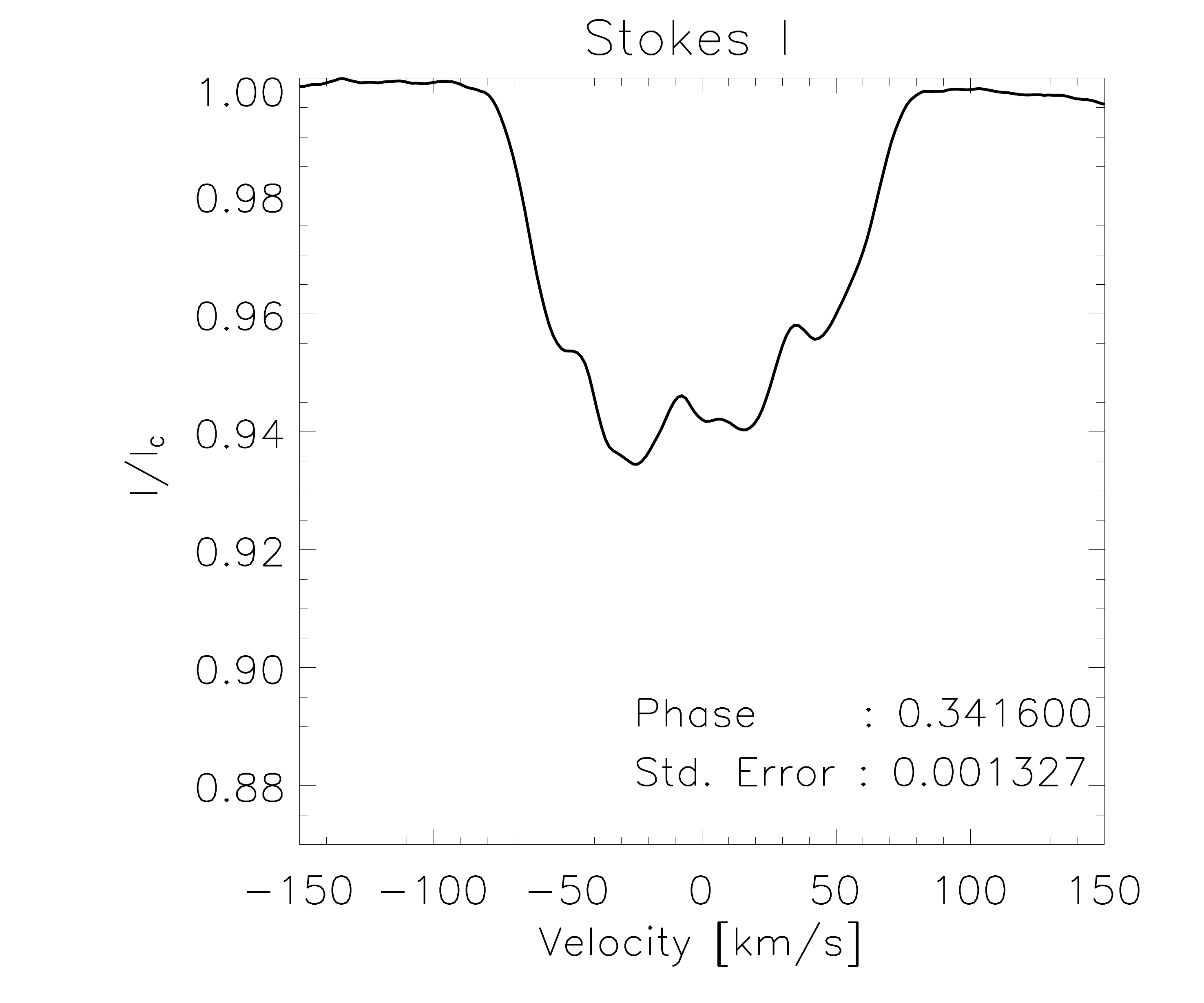}
\includegraphics[width=4.25cm]{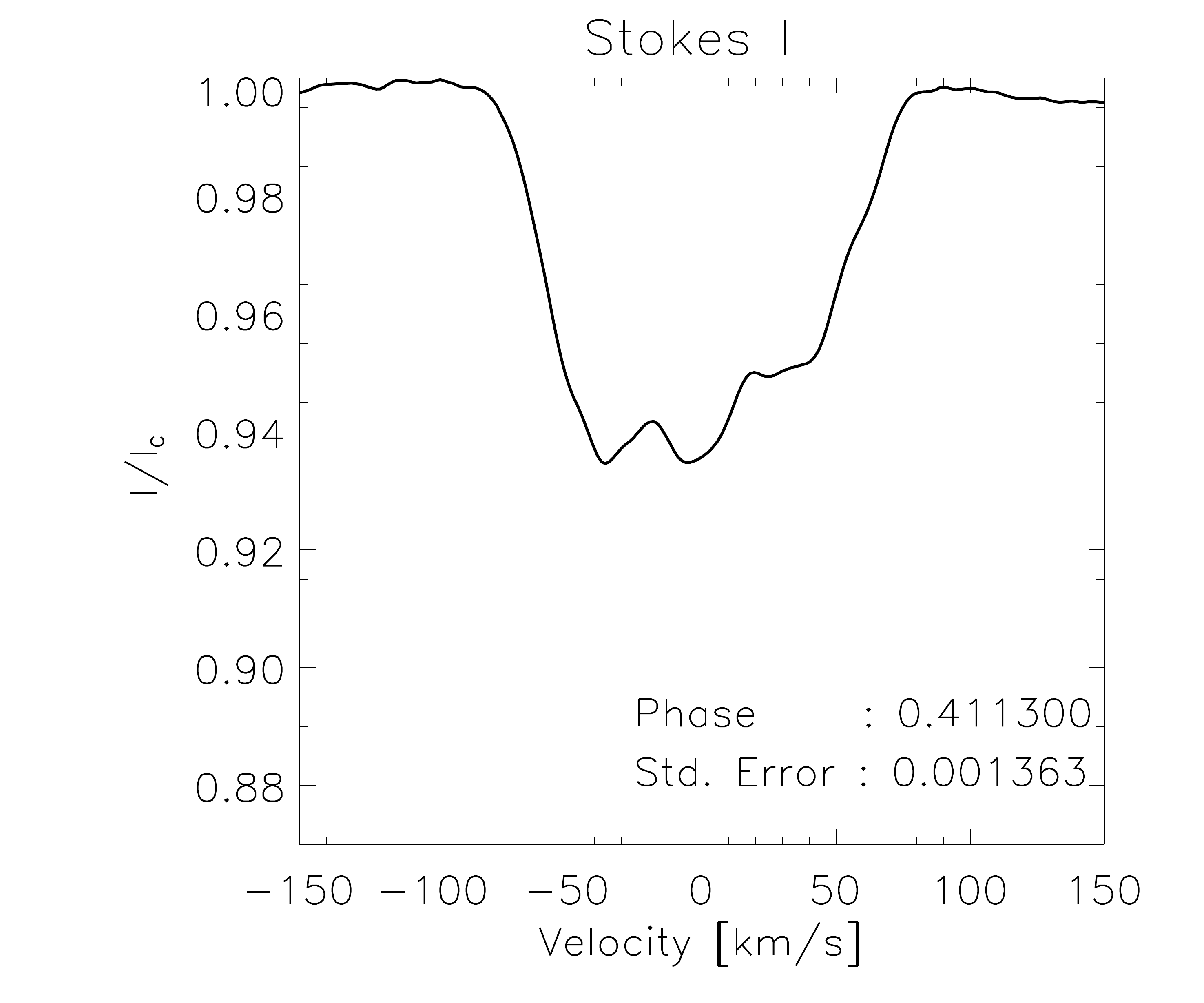}
\includegraphics[width=4.25cm]{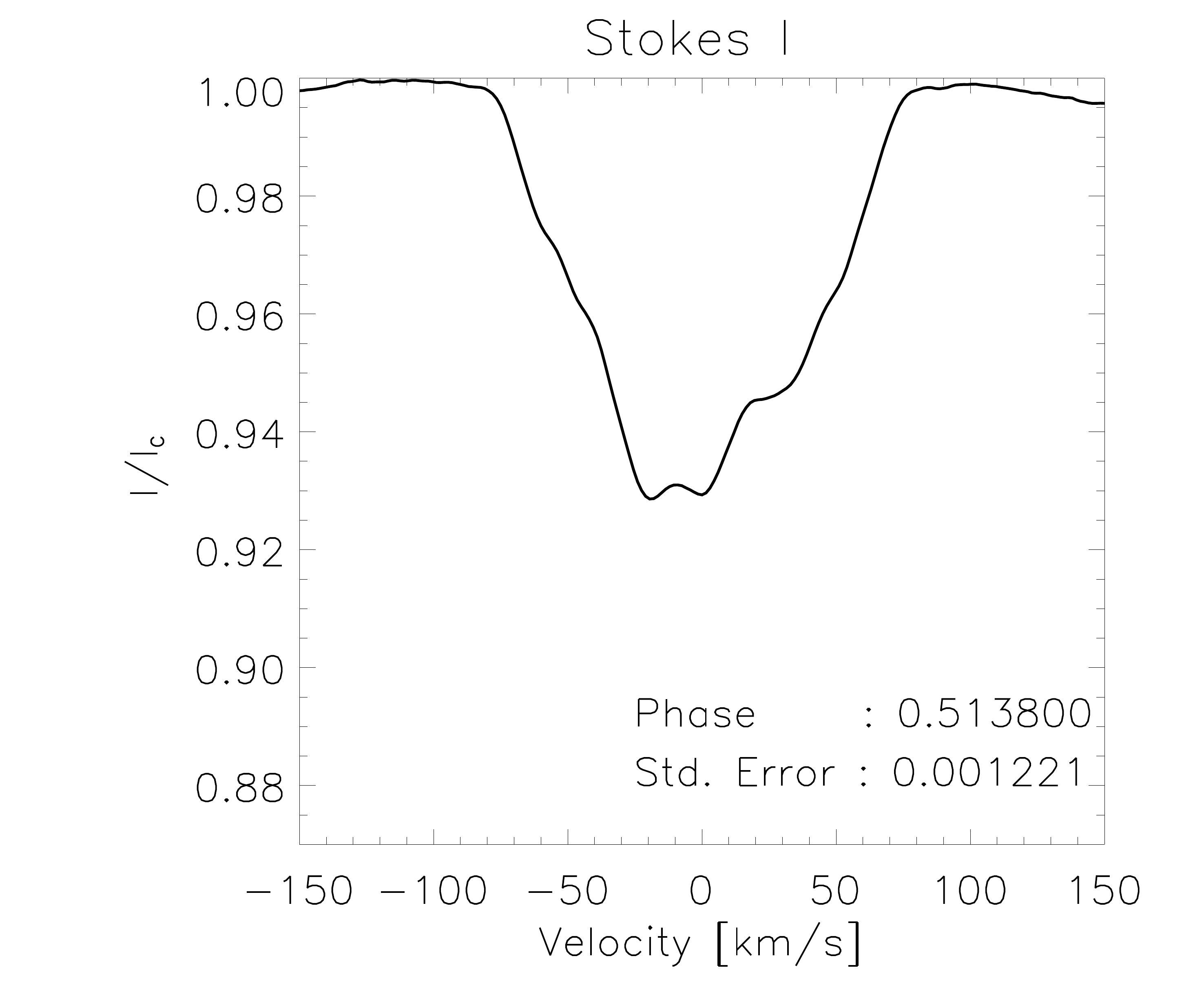}
\includegraphics[width=4.25cm]{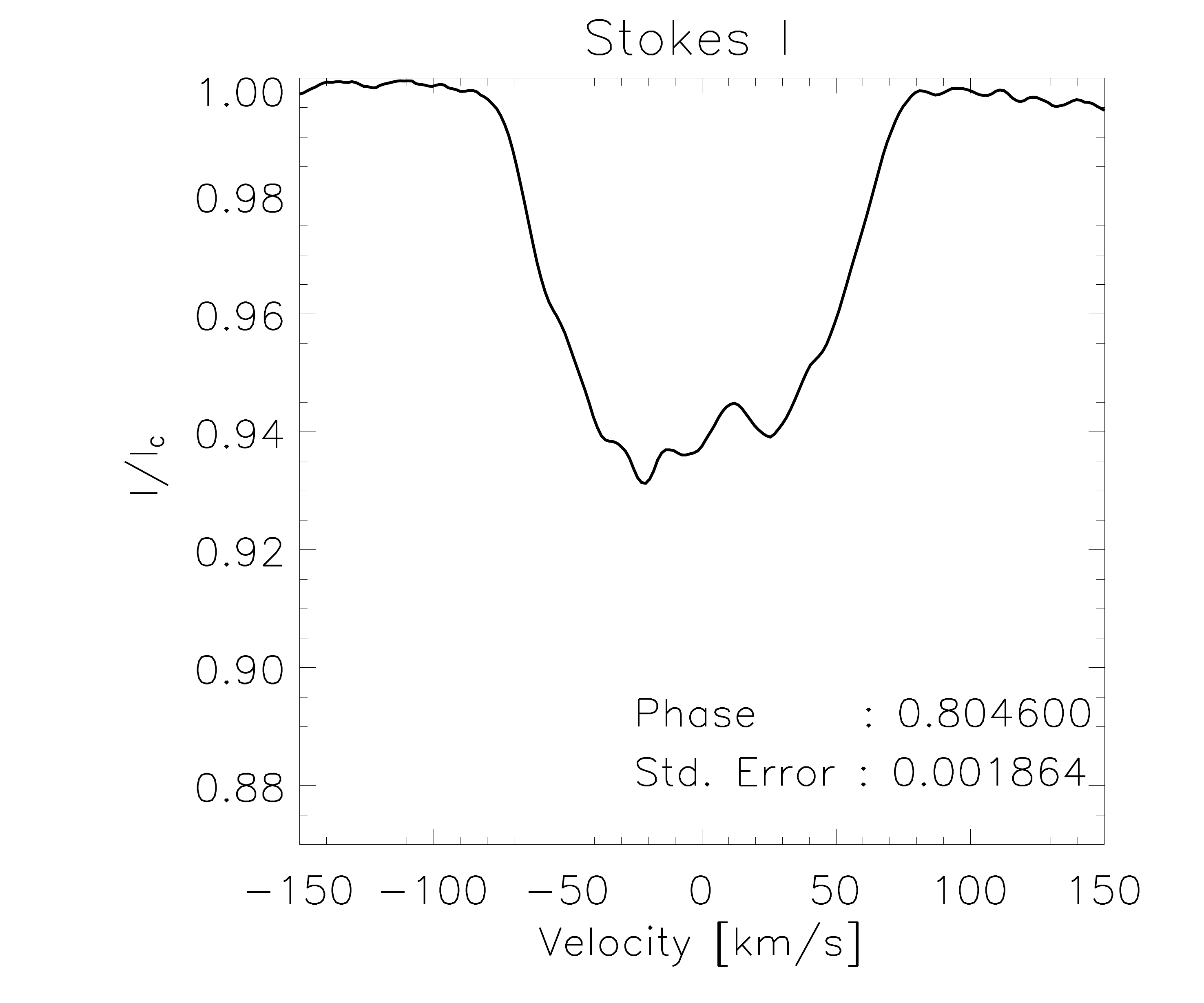}
\includegraphics[width=4.25cm]{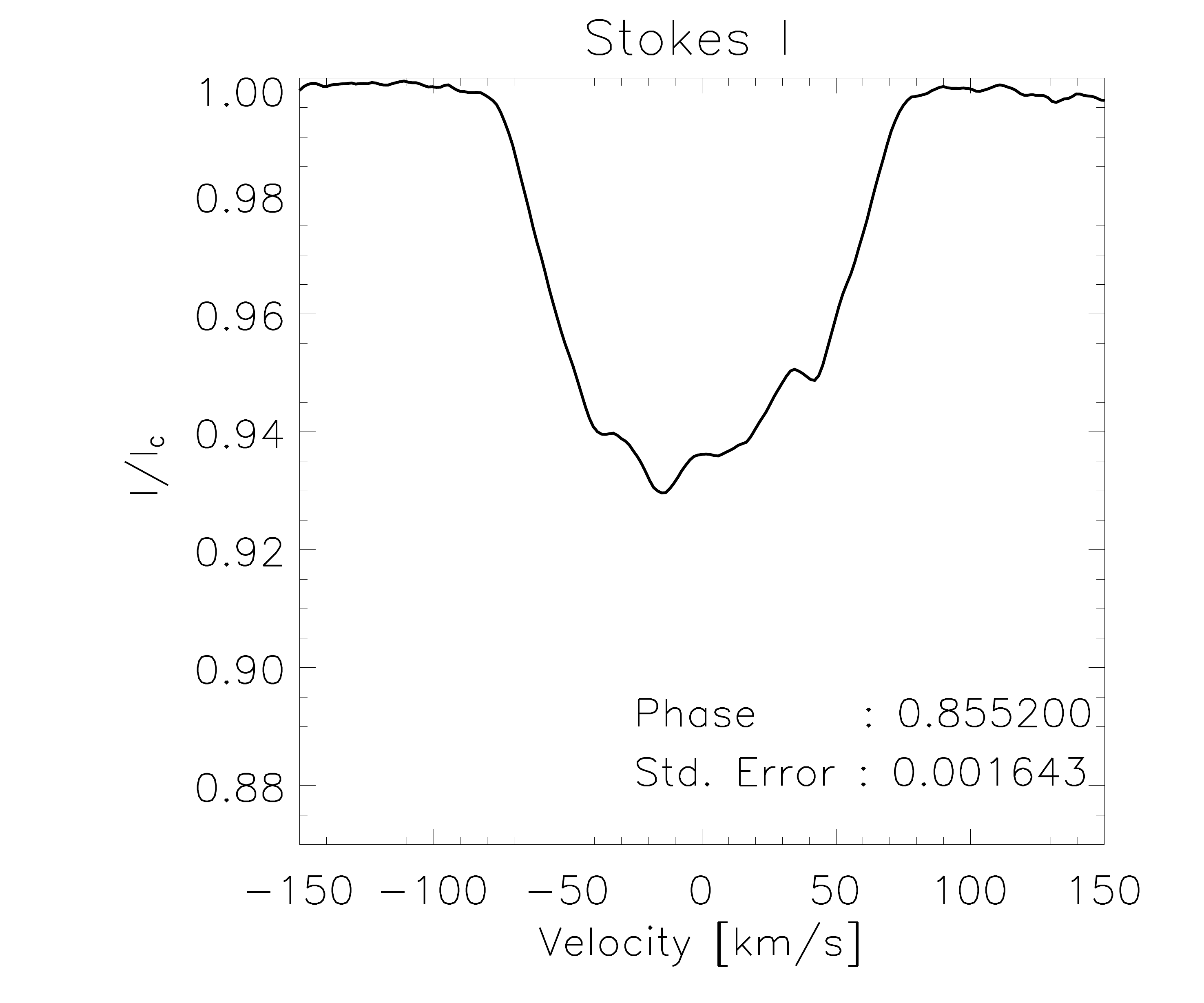}
\includegraphics[width=4.25cm]{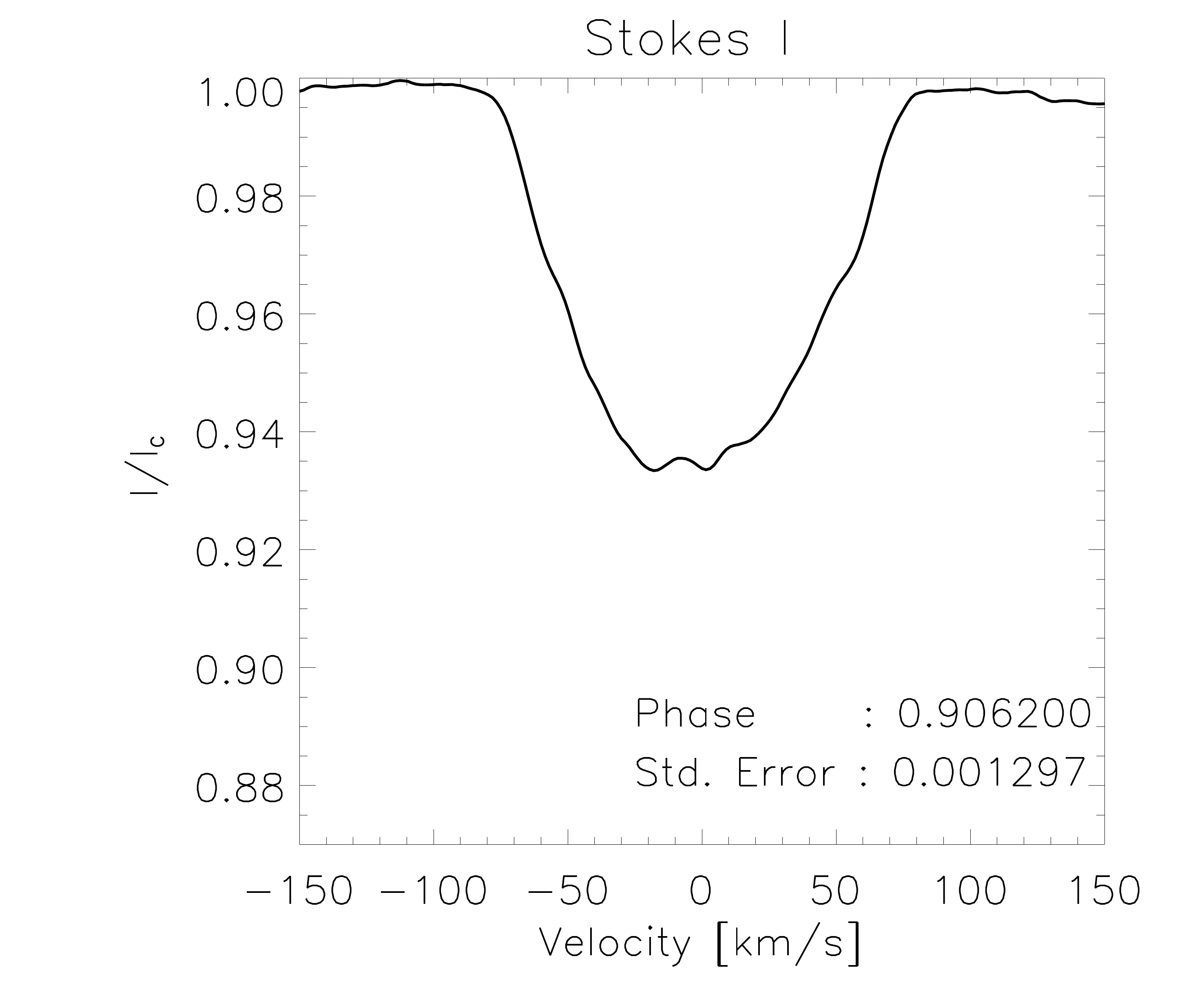}
\caption{Reconstructed SVD Stokes~I profiles for each
rotational phase. For all reconstructed profiles
an observation  matrix with 56 spectral
lines is used. The averaged standard error is given together with the
corresponding rotational phase.} 
\label{Fig:4}
\end{minipage}
\end{figure*}

Note, that all matrices, the original observation matrix and each
of the randomized matrices, carry the same amount of total energy
(variance), i.e. the total sum over all of the eigenvalues are
equal. After sorting the eigenprofiles for all covariance matrices
according to the magnitude of their respective eigenvalues, we can
finally identify the eigenprofiles of the observation covariance
matrix that contain significant profile information. 
This means we determine all those
eigenprofiles whose eigenvalues $\tilde{\lambda}_i$ are larger
then the corresponding mean eigenvalues $\langle \lambda_i^* \rangle$ of the
randomized matrices. This provides a noise threshold value
according to Eq.~(\ref{Eq:3_12}) such that $\tilde{\lambda}_i \; >
\; \langle \lambda_i^* \rangle \; = \; \sigma^2$ which in turn allows us to 
determine the effective rank of the signal subspace.

Fig.~\ref{Fig:1} shows the eigenvalue spectrum of the
observation matrix and the mean eigenvalues from all the
randomized matrices for one observational phase (0.014). The
magnitude of the leading eigenvalues (solid line) obtained
from the observation matrix provides a first evidence that the
observation matrix contains significant information 
different from pure noise or random fluctuations 
which is indicated by the randomized eigenvalue spectrum (dashed line). 
The crossing point at which the eigenvalues from the observation
covariance matrix drop below the eigenvalues of the randomized
matrices marks the critical dimension in the subspace. Up to
eigenvalue 27 the respective eigenprofiles contain significant
information above the noise level. The estimated reduced rank is
thus 27. This estimate allows us to proceed and use Eq.~(\ref{Eq:3_15})
to calculate the rank-reduced (subspace) averaged Stokes~V profile. This
procedure is then performed for all rotation phases to reconstruct
the entire set of Stokes~V profiles used in the following DI/ZDI
inversion. Note that we omitted the error bars for both eigenvalue
spectra in Fig.~\ref{Fig:1} because the standard deviation for all
values is lower then 2\% , which hardly exceeds the thickness of
the plotted lines.
For the Stokes~I profiles, we have a significant larger
\emph{relative} signal-to-noise ratio (it is in fact the S/N
estimate from \ref{Sect:2}). Therefore, we do not start from a negative
hypothesis that no signal or coherent information is present in
the data set but instead directly apply Eq.~(\ref{Eq:3_12}) to
determine the signal subspace. Fig.~\ref{Fig:2} shows the
normalized eigenspectrum for the first 20 eigenvalues of the
observation covariance matrix (solid line) and the error variance
$\sigma_{obs}$ as determined from the observation (dashed
line). It is interesting to see that almost the entire energy
(variance) of the observation covariance matrix can be described
by a very small number of eigenprofiles. The crossing point
indicates that the subspace dimension is five. The first five
eigenprofiles contain already 99.98\% of the total energy of all
line profiles within the observation matrix. In the following
step, we again use Eq.~(\ref{Eq:3_15}) to calculate the
subspace averaged Stokes~I profiles.

Finally, Fig.~\ref{Fig:3} and Fig.~\ref{Fig:4} show the
reconstructed set of Stokes~V and Stokes~I profiles for all
rotational phases. The standard errors are provided again by a
Bootstrap method with a re-sampling number of 1,000. For each
velocity bin, we thus obtained an estimate of the standard
error which is condensed in Fig.~\ref{Fig:3} and Fig.~\ref{Fig:4}
to a mean standard error value averaged over the velocity domain.
Its typical value is 1 $\times$ 10$^{-4}$ for Stokes~V and 
1 $\times$ 10$^{-3}$ for Stokes~I which corresponds to a 
signal-to-noise level of 10,000 or 1,000 respectively


\section{Stellar surface reconstruction with \emph{iMap}}
\label{Sect:4}

\subsection{A brief description of iMap}
\label{Sect:4.1}

In this study, we use the \emph{iMap} code \citep{Carroll07b,Carroll09}
to simultaneously reproduce the temperature and
magnetic vector field distribution from a sequence of observed
Stokes~I and V~profiles. The forward modeling in \emph{iMap} is based on
polarized radiative transfer to allow for the best possible
accuracy in line profile modeling \citep{Carroll09}. 
All line parameters necessary for the synthetic
calculation are taken from the Vienna atomic line list (VALD)
\citep{Kupka99}. The underlying model atmospheres are based on
Kurucz ATLAS-9 atmospheres \citep{Castelli04}. The actual calculation
and integration of the polarized radiative transfer equation in
LTE can be performed either by a numerical integration or by a
number of trained artificial neural networks (ANN) \citep{Carroll08}.

The \emph{iMap} code is equipped with a new inversion module. While
the former versions relied on a conjugate gradient method \citep{Press92} 
with a local entropy regularization \citep{Carroll07b}, 
the current version of \emph{iMap} uses an iteratively
regularized Landweber method \citep{Engl96}. Many different regularization
methods exist for linear and non-linear inverse problems but
interestingly only two of the most restrictive ones received
attention in the field of ZDI and DI namely the Tikhonov
regularization \citep[see e.g.][]{Piskunov02} and the
maximum entropy method \citep[e.g.][]{Vogt87}. There is
also a long list of comparisons between these two methods and
their benefits and respective shortcomings 
\citep[see, e.g.][]{Rice02}. The three major issues that constitute an
ill-posed, inverse problem are : (i) a solution may not exist,
(ii) the problem may not be unique, (iii) the
solution does not depend continuously on the data. We will not
go into problem (i) which would require a discussion about the adequacy 
of the underlying ZDI/DI model assumptions
\citep[see][for a discussion]{Carroll09}, but instead we
assume (as usual) that our model admits a solution. Then, we are
left with the uniqueness problem (ii) and the numerical stability 
problem (iii) which both depend also on the noise level of the data. 
Problems (ii) as well as (iii) are closely related and 
can effectively be addressed by regularization methods that put additional 
constraints on the solution. Please also see in this context the paper of \citep{Kochukhov02}
who describes the degeneracy caused by using just the 
information of the Stokes~I and Stokes~V profiles.
In our iterative approach, implemented in \emph{iMap}, problem (ii) and (iii) will be addressed 
by an iteration process where the step size and an appropriate
\emph{stopping rule} provide the regularization of the inverse problem \citep{Engl96}.

Iterative regularization for inverse problems has been the subject of
various theoretical investigations over the recent years 
\citep{Hanke97,Engl04,Egger05,Kaltenbacher08}. The Landweber
iteration, which is used here, rests on the idea of a simple
fixed-point iteration, derived from minimizing the sum of the
squared errors. Our new inversion routine follows exactly this
line and can be described as follows: Written in a concise vector
notation the problem setting is
\begin{equation}
\min_{\vec{x}} \: \frac{1}{2} \| \vec{I}(\vec{x})  - \vec{O} \|^2  \: ,
\label{Eq:4_1}
\end{equation}
where $\| \|$ is the $L_2$ norm and $\vec{I}$ is the synthetic model profile over all spectral
lines, wavelengths or velocities, and rotational phases, $\vec{O}$ is the corresponding observation. 
The vector $\vec{x}$ contains all our
free parameters of the model, i.e. the temperature and the
magnetic field vector for each surface element. The iteration now
proceeds along the negative gradient direction and updates the
current estimate of the solution vector, $\vec{x}_k$, in the
following manner
\begin{equation}
\vec{x}_{k+1} \; = \; \vec{x}_k + w_k \vec{I'}(\vec{x}_k) \left (
\vec{O} - \vec{I}(\vec{x}_k) \right ) \: . 
\label{Eq:4_2}
\end{equation}
Here, $\vec{I'}$ is the gradient vector with respect to all
surface element values and $w_k$ is the weight factor that can
adaptively accelerate the iteration process. In the conventional
Landweber iteration process, $w_k$ is set to unity. To accelerate
the procedure we use a variant of the steepest descent \citep{Kaltenbacher08}
and set $w_k$ to
\begin{equation}
w_k \; = \; \frac{\|u_k\|^2}{\| \vec{I'}(\vec{x}_k) u_k \|^2} \; ,
\label{Eq:4_2b}
\end{equation}
where $u_k = \vec{I'} * (\vec{O} - \vec{I}(\vec{x}_k))$.

The semi-convergence \citep{Hanke95} of the method requires a stopping rule 
before it enters into the noise level of the data
to regularize the procedure and to avoid overfitting. 
One common, and well studied criterion for the stopping condition is the Morozov discrepancy 
principle which can be written for the iterative approach as
\begin{equation}
\| \vec{I}(\vec{x}_{k^*})  - \vec{O} \|  \; \leq \; \tau \delta <
\;  \| \vec{I}(\vec{x}_k)  - \vec{O} \| \; ; \;  0 \leq k < k^* \; ,
\label{Eq:4_3}
\end{equation}
where $\delta$ is an upper bound for the data error (i.e. noise), $\tau$ a positive number and $k^*$ the maximum iteration
number. In terms of stability and convergence it can be shown that $\tau$ has to satisfy
$\tau \geq 1$  \citep{Engl96}. 
Formally the noise estimate for the ZDI/DI can be derived from the observations. Given a noise contribution $\sigma_i$ for 
each velocity $i$ we can combine the individual errors to form a vector $\vec{\sigma}$ such as,
$\vec{\sigma}^{T} = (\sigma_1,\sigma_2,...,\sigma_n)$ with $n$ the number of velocity points. 
The error can then be expressed as
$\delta =  \| \vec{\sigma} \|$. If the noise estimate is homogeneous (i.e. equal for all velocities) or
does not vary much over the velocity domain and time we may use the maximum of $\vec{\sigma}$ to write for $\delta$ the relation 
\begin{equation}
\delta = \max(\vec{\sigma}) \sqrt{n} \; .
\label{Eq:4_3b}
\end{equation}
If we assume that Eq.~(\ref{Eq:4_1}) follows a $\chi^2$ distribution we may 
use the number of degrees-of-freedom
of the problem to determine the error resulting from the limited degree of freedom of the model
to write
\begin{equation}
\delta = \max(\vec{\sigma}) \sqrt{n-p}. \; ,
\label{Eq:4_3c}
\end{equation}
where $p$ is the number of parameters in the model. If the inversion reduces
the error function Eq.~(\ref{Eq:4_1}) down to the threshold $\delta$ we would ensure that
the reduced $\chi^2$ is close to one.
But let us emphasize here that DI as well as ZDI (given we have correctly modeled the problem) 
are non-linear problems \citep[see][]{Carroll09} and moreover the problem is generally ill-posed, 
which makes it by no means a simple task 
to adequately determine the real degree of freedom of the problem.
The reduced $\chi^2$ as a measure of the \emph{goodness-of-fit} may therefore 
only of limited use.

If the data are pre-processed like in this work
the noise estimate for the SVD reconstructed data can be directly obtained 
from the error estimates provided by the bootstrap procedure given in the last section.

The simultaneous DI and ZDI procedure for each surface element, is
implemented in the following way. Instead of using one long vector
$\vec{x}$ that incorporates the entire information of the surface
temperature and the magnetic field, we proceed with an
alternate minimization \citep{Byrne04} of Eq.~(\ref{Eq:4_1}).
Alternate minimization means here that after each iteration for
the temperature $\vec{x}_{k+1}^T$ a parallel inversion of the
magnetic field vector $\vec{x}^B$ is performed, which uses the
information of the $k+1$ iteration of the temperature
minimization. The alternate iteration at step $k+1$ for
$\vec{x}_{k+1}^T$ and $\vec{x}_{k+1}^B$ can then be written as
\begin{eqnarray}
\vec{x}_{k+1}^T \; =  \; \vec{x}_k^T + w_k
\vec{I'}(\vec{x}_k^T,\vec{x}_k^B) \left ( \vec{O} -
\vec{I}(\vec{x}_k^T,\vec{x}_k^B) \right )
\: , \hspace{0.4cm} \nonumber \\
\vec{x}_{k+1}^B \; =  \; \vec{x}_k^B + w_k
\vec{I'}(\vec{x}_{k+1}^T,\vec{x}_k^B) \left ( \vec{O} -
\vec{I}(\vec{x}_{k+1}^T,\vec{x}_k^B) \right ) \: . 
\label{Eq:4_4}
\end{eqnarray}
Note that both $\vec{x}_{k+1}^T$ and $\vec{x}_{k+1}^B$ depend on
each other due to radiative transfer effects.

We want to note here that after a careful analysis of the
convergence and stability behavior of the inversion,
the strict simultaneity of the temperature and magnetic inversion 
is not necessarily required whereas the order of the inversion
is of importance. 
The dependence of the DI on the magnetic inversion is not strong,
this is mainly due to the weakness of the effect that the 
magnetic field introduces via the Zeeman effect on the intensity 
profile (Stokes~I).  
For rapid rotating cool stars the temperature information encoded in line profile bumps 
of the Stokes~I profiles is therefore only weakly affected.
The dependence of ZDI on the temperature inversion however is strong, which means 
the temperature information 
obtained from the DI must be part of the magnetic inversion as the temperature
leads to a non-trivial scaling of the Stokes~V profile amplitude and width. 
A \emph{consecutive} DI-ZDI approach is therefore equivalent to the 
simultaneous inversion as done in this work.

\subsection{Multi-line inversion}
\label{Sect:4.2}

For DI as well as for ZDI we could apply the least-squares
minimization of Eq.~(\ref{Eq:4_1}) to all of our available and reconstructed
spectral lines simultaneously, i.e. using a combined long spectral
line vector such that the error function reads
\begin{equation}
E \; = \; \frac{1}{2} \; \sum_{p=1}^{N_p} \; \sum_{k=1}^{N_k} \;
\sum_{m=1}^{N_m}  \; (O_{p,k,m} - I(\vec{x})_{p,k,m})^2 \: ,
\label{Eq:4_5}
\end{equation}
where $\vec{x}$ is the model parameter vector, $p$ the
rotational phase, $k$ the spectral line, $m$ the wavelength or the
velocity index respectively,
and $N_p$, $N_k$, $N_m$ their respective total numbers. This is
the approach usually pursued in DI with {\sc TempMap} \citep[e.g.][]{Rice00,Rice11}
but fails for ZDI since the
polarimetric signals are usually buried deep within the noise and therefore require a 
signal extraction which provides a reconstructed composite line profile. 
The SVD reconstruction process 
extracts a single (weighted) mean SVD line profile that comprises the information
of all the contributing lines of the observation matrix and therefore the error function
can no longer be applied to the sum of squared differences of the individual lines
but needs to be applied on the squared differences between the reconstructed SVD profile 
and a proper synthetic equivalent.
The strategy here is in fact to compute a synthetic mean profile for all
contributing spectral lines used in the SVD reconstruction process by accounting for 
all radiative transfer effects and line modeling characteristics (e.g. model atmospheres, 
spectral line parameters and line blends). 

We briefly show that the 
result of this strategy provides an error which is on average
at least as low as in the conventional case, where the sum of differences from each 
contributing line is computed according Eq.~(\ref{Eq:4_1}). 
Let us begin with the conventional case of Eq.~(\ref{Eq:4_5}). For brevity we
incorporate the rotational phases and the wavelengths or velocities into the vector
notation such that $\vec{I}_k$ and $\vec{O}_k$ represent the Stokes
profile of one particular spectral line $k$ taken over the entire
velocity range and over all available rotational phases. If we correctly 
model each individual spectral line that contributes to the averaging
process, then we can express the difference between the observed Stokes profile
$\vec{O}_k$ and the synthetic profile $\vec{I}_k$ as
being exclusively due to the inherent noise $\vec{\epsilon}_k$
such that $\vec{I}_k$ corresponds to the observation $\vec{O}_k$
within the limits of $\vec{\epsilon}_k$;
\begin{equation}
\vec{I}_k \; = \; \vec{O}_k + \vec{\epsilon}_k \: .
\label{Eq:4_6}
\end{equation}
The average sum-of-squares error for a particular line profile can
then be written as
\begin{equation}
E_k \; = \; \mathcal{E} \left[ (\vec{I}_{k} - \vec{O}_{k})^2
\right] \; =  \; \mathcal{E} \left[\vec{\epsilon}_k^2 \right] \: ,
\label{Eq:4_7}
\end{equation}
The mean error $\tilde{E}$ of
all individual spectral lines is then given as
\begin{equation}
\tilde{E} \; = \; \frac{1}{N_k} \; \sum_{k=1}^{N_k} E_k \; =  \;
\frac{1}{N_k} \; \sum_{k=1}^{N_k} \mathcal{E} \left[
\vec{\epsilon}_k^2 \right] \: . \label{Eq:4_8}
\end{equation}
On the other hand, if we perform the averaging process before
minimizing the error function, as we did in our line profile
extractions, we can write for the mean observed profile pattern
\begin{equation}
<\vec{O}> = \frac{1}{N_k} \; \sum_{k=1}^{N_k} \vec{O}_k \: ,
\label{Eq:4_9}
\end{equation}
and the same holds for the mean synthetically calculated profile
\begin{equation}
<\vec{I}> = \frac{1}{N_k} \; \sum_{k=1}^{N_k} \vec{I}_k\: .
\label{Eq:4_10}
\end{equation}
The mean error $\bar{E}$ for the pre-averaged line profile can then be written as
\begin{equation}
\bar{E} \; = \; \mathcal{E} \left[ \left( \frac{1}{N_k} \;
\sum_{k=1}^{N_k} ( \vec{I}_k - \vec{O}_k ) \right)^2 \right] \: .
\label{Eq:4_11}
\end{equation}
This is equivalent to
\begin{equation}
\bar{E} \; = \; \mathcal{E} \left[ \left( \frac{1}{N_k} \; \sum_{k=1}^{N_k} \vec{\epsilon}_k \right)^2 \right] \: .
\label{Eq:4_12}
\end{equation}
We are assuming again that the noise $\vec{\epsilon}_k$ is
uncorrelated and of zero mean, so that all covariance terms
disappear. Then we may write
\begin{equation}
\bar{E} \; = \;  \frac{1}{N_k^2} \; \sum_{k=1}^{N_k} \; \mathcal{E} \left[ \vec{\epsilon}_k^2 \right] \: .
\label{Eq:4_13}
\end{equation}
Comparing this with Eq.~(\ref{Eq:4_8}) shows immediately that the
mean error for the averaged line profile is lower than for the
individual profiles,
\begin{equation}
\bar{E} \; =  \;   \frac{1}{N_k^2} \; \sum_{k=1}^{N_k} \;
\mathcal{E} \left[ \vec{\epsilon}_k^2 \right] \; \leq \;
\frac{1}{N_k} \; \sum_{k=1}^{N_k} \mathcal{E} \left[
\vec{\epsilon}_k^2 \right] \; = \; \tilde{E} \: . 
\label{Eq:4_14}
\end{equation}
Most likely, the idealized assumption of completely uncorrelated
noise is not valid but still, the error of the averaged modeling
$\bar{E}$ is not expected to be larger than that of a conventional
minimization of $\tilde{E}$. Note, that an important aspect here is
the correct synthetic modeling of the averaged profile.
The extension to a weighted averaging is straight forward.

\begin{figure*}[!t]
\begin{minipage}{9.25cm}
\centering
\includegraphics[width=7.5cm]{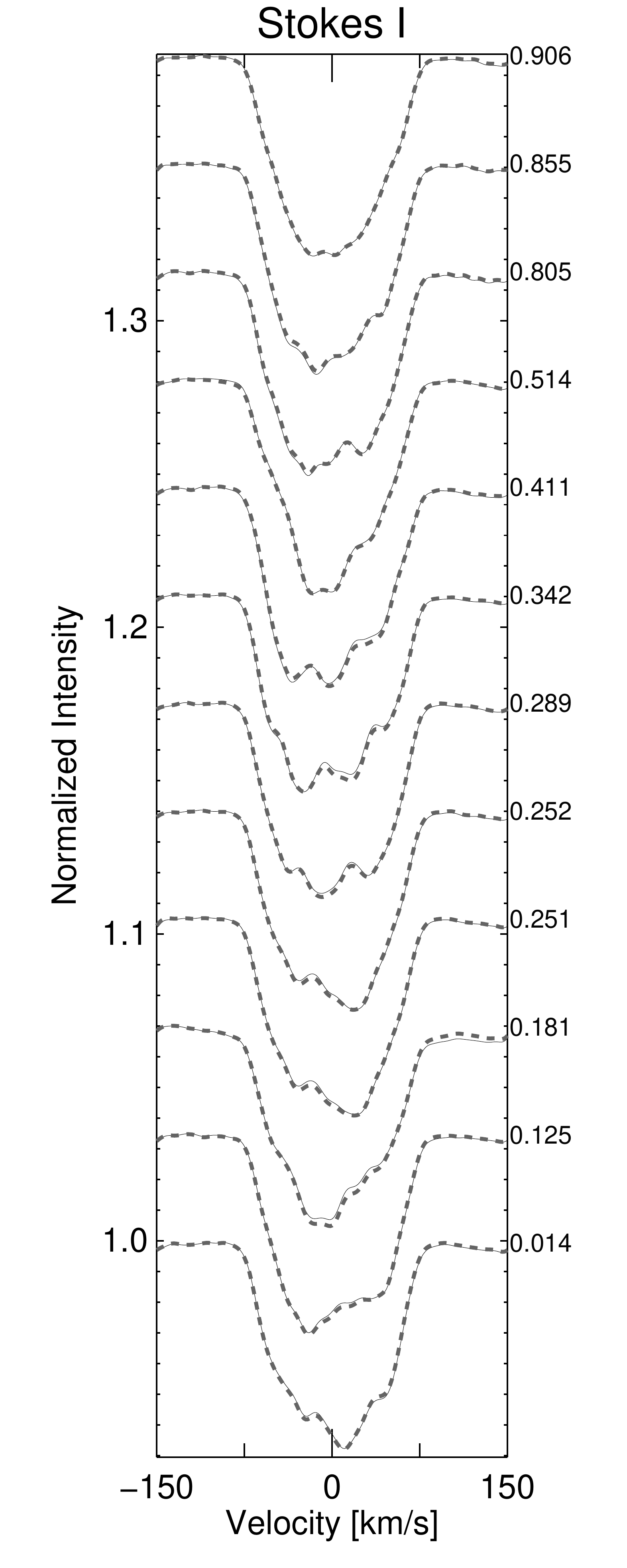}
\end{minipage}
\begin{minipage}{9.25cm}
\includegraphics[width=7.5cm]{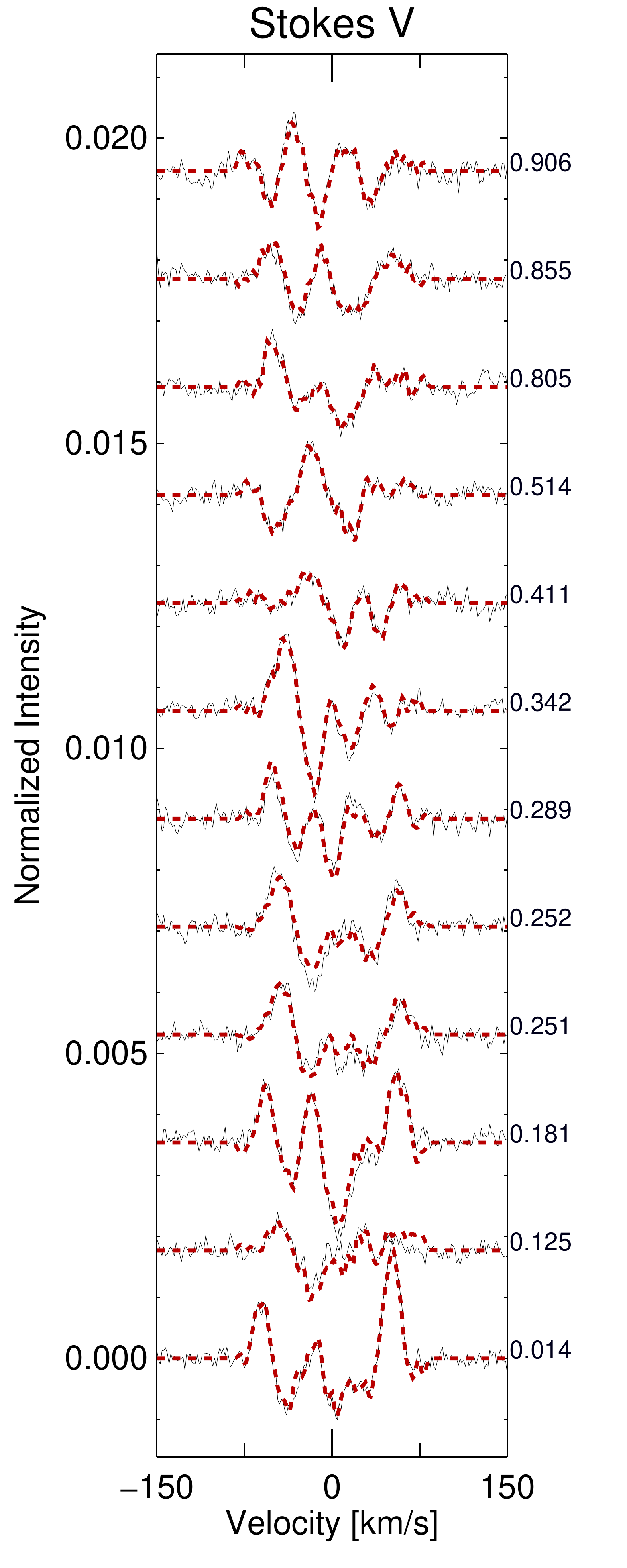}
\end{minipage}
\caption{Stacked plot of the synthetic fits (thick dashed lines)
to the observed Stokes profiles (thin solid lines), 
on the left side Stokes~I and on the right Stokes~V. All spectral line profiles are 
normalized by the continuum intensity.}
\label{Fig:5}
\end{figure*}

\subsection{Differential rotation inversion}
\label{Sect:Differential}

In this section, we introduce our methodology for retrieving a
surface differential rotation signal, if present. Our technique
follows closely that introduced by \citet{Petit02}. Given that
the observations span over several rotational phases and that the
stellar surface obeys a solar-type differential rotation law, the
underlying idea is that the closer the modeled differential
rotation is to the true value the lesser the resulting
reconstructed surface image is smeared out. In other words, not
accounting for differential rotation introduces a blurring effect
in the resulting DI or ZDI. Differential rotation provides
not only constraints on the image space in terms of the resulting
\emph{sharpness} of the image but also on the data space because
the differential rotation alters the velocity distribution of the
individual local line profiles over the visible surface and thus produces 
a modification of the disk-integrated flux profile \citep[see, e.g.][]{Gray05}. 
The data space approach has been extensively used in the past for
determining the differential rotation of a number of stars \citep[e.g.][]{Reiners03,Reiners06}. 
Compared to the approach chosen by \citet{Petit02} and \citet{Donati03} there is
one major difference we apply here. Instead of repeating the
inversion many times with various differential rotation parameters to
sample the $\chi^2$ landscape, we directly incorporate the
differential rotation law into the inversion.
All other restrictions given in Sect.~5.1. of \citet{Petit02}
also hold for our approach with particular emphasis on the
required rotational coverage, note that the
requirements for the phase coverage is less restrictive.

\subsubsection{Assumed differential-rotation law}
\label{Sect:DiffLaw}

The assumption here is that the differential rotation on V410~Tau
can be described, as for so many cool and late-type stars, with a
simplified ``solar-type'' differential rotation law;
\begin{equation}
\Omega(l) \; = \; \Omega_{eq} \; - \; d\Omega \; \sin^2 l \: ,
\label{Eq:4_15}
\end{equation}
where $l$ is the surface latitude, $\Omega(l)$ the respective
rotation rate at a particular latitude, $\Omega_{eq}$ the rotation
rate at the equator, and $d\Omega$ the difference in rotation rate
between the pole and the equator.

\subsubsection{Implementation as a sheared-image analysis within \emph{iMap}}
\label{Sect:Sheared}

The search for the least blurred image consistent with the minimum
of the objective Eq.~(\ref{Eq:4_1}) can also be expressed as an
image deconvolution problem with a latitudinal smearing operator
or point-spread function $P_{\rm lat}$. This concept will be investigated 
and developed in a future paper, instead we concentrate here on describing the
problem in the same way as in \citet{Petit02}. Providing
sharpness as an additional constraint in image space may indeed
provide a means of finding a solution which best matches with a
differential rotation law and the data error (fit) at the same
time. This extra constraint has been provided in terms of
minimizing the information content (maximizing the image entropy)
of the image by \citet{Donati00} and \citet{Petit02}.
However, because maximum entropy is entirely invariant under a
random permutation of the image data (pixel) and does not account
for any spatial coherence in the image domain, we believe this
constraint is not an appropriate measure to quantify distortions
due to surface differential rotation. Moreover differential rotation may, 
despite the distortions, retain to a large extent the morphological 
coherency of surface structures over several rotation periods. Instead of 
adding an additional constraint to the error function, we want to 
completely rely on the achieved quality of the data fit and implement the
optimization for the two parameters $\Omega_{eq}$ and $ d\Omega$
within the \emph{iMap} code. This extra flexibility amounts to an
additional degree of freedom in the time domain to account for the
time-dependent rotational evolution of the image. The profiles and
the corresponding part of the visible stellar disk are no longer
only determined just by the spectral information of the respective
phase alone but also by the time information \citep{Petit02}.
This minimization process can conveniently be incorporated into
our Landweber iteration in Eq.~(\ref{Eq:4_4}) to yield an
alternating minimization scheme.
If we define the vector $\vec{\Omega}$ which  comprises our two differential
rotation parameter $\Omega_{eq}$ and $ d\Omega$ the process can be described as follows:
After proceeding along the
negative gradient direction in image space (temperature and/or
magnetic field) the process switches to the omega space to perform a minimization
of the squared error, Eq.~(\ref{Eq:4_1}) with respect to the differential parameter vector $\vec{\Omega}$.
This provides a new current estimate of the vector $\vec{\Omega}$.
which then enters again in the next iteration cycle of Eq.~(\ref{Eq:4_4})
This process is repeated until the stopping condition is reached (see Sect.~\ref{Sect:4}). 
We can formally introduce an image-shear operator 
$\mathbb{P}_d$ which differentially rotates a current estimate $\vec{x}_k(\vec{\Omega})$ of the image
(temperature or magnetic field) to write the process as
\begin{equation}
\hat{\vec{\Omega}} \; = \; \arg\min_{\vec{\Omega}} \;\mathbb{P}_d \; \left [
\vec{x}_k(\vec{\Omega}) + w_k \vec{I'}(\vec{x}_k(\vec{\Omega})) \left ( \vec{O} -
\vec{I}(\vec{x}_k(\vec{\Omega})) \right ) \right ] \: , 
\label{Eq:4_16}
\end{equation}
The minimization of Eq.~(\ref{Eq:4_16}) is performed by a gradient descent
method similar to the one used for the Landweber iteration.


\subsection{Fixed stellar parameters}
\label{Sect:Parameters}

The adopted stellar parameters are the same as those used and
partly derived in Paper~I. The effective temperature is 4,400~K,
the logarithmic gravity is 4.0, the projected rotational
velocity is 77.7~\kms , and the inclination of the stellar
rotation axis is $62^\circ$ as determined from DI in Paper~I. The
micro- and macro-turbulence is set to 2.0~\kms\ and 1.5~\kms ,
respectively, solar chemical abundances are assumed for all
elements. As also described in detail in Paper~I, an average
rotational period of 1.871948d was adopted. Rotational phase is
also kept the same as in Paper~I.

\subsection{iMap setup}
\label{Sect:Setup}

For the inversion with \emph{iMap}, we use all 12 available Stokes~I and
Stokes~V SVD profiles from Fig.~\ref{Fig:3} and Fig.~\ref{Fig:4}.
We utilize the multi-line approach described in
Sect.~\ref{Sect:4.2} which requires to synthesize all 929 spectral
line profiles of the observation matrix to finally obtain the mean
synthesized profile. The enormous amount of computation is performed
by our ANN approach \citep{Carroll08}. As mentioned in
Sect.~\ref{Sect:4.1} all line parameters were taken from the VALD
line list and model atmospheres are produced by ATLAS-9 \citep{Castelli04}. 
The stellar surface is partitioned into $6^\circ \times 6^\circ$
segments, whereas the polarized radiative transfer is calculated
on a subpartition of $3^\circ \times 3^\circ$. The initialization
temperature was homogeneously set to 4,400~K. The initial magnetic
field is homogeneously set to zero for all three magnetic
components (radial, azimuthal and meridional). For $v\sin i$,
stellar inclination, gravity, and abundances, the values given in
Sect.~\ref{Sect:Parameters} are used.

The stopping rule according to the discrepancy principle for the
Landweber iteration is set according to the standard errors of the
reconstructed line profiles. As described in Sect.\ref{Sect:4.1} this means that we 
use for the magnetic inversion the largest (maximum) standard error of the reconstructed
Stokes~V profiles and for the temperature inversion the largest
standard error of the reconstructed Stokes~I profiles. The
individual values are shown in Fig.~\ref{Fig:3} and
Fig.~\ref{Fig:4}.

\section{Application to V410~Tau}
\label{Sect:5}
\subsection{Temperature distribution in 2008/09}
\label{Sect:T-map}

\begin{figure*}[!t]
\begin{minipage}{\textwidth}
\centering
\subfigure{\includegraphics[width=\textwidth]{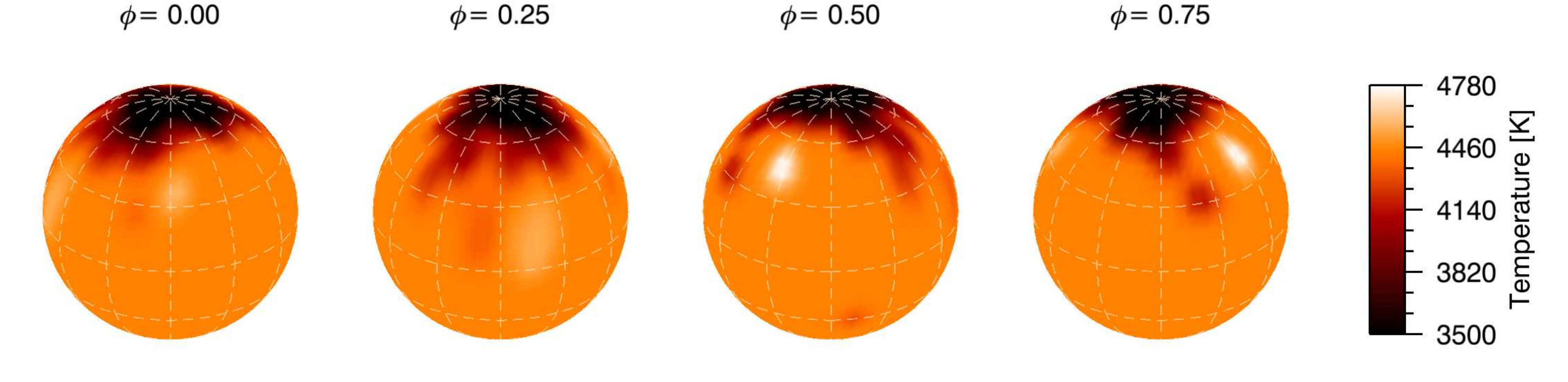}}
\caption{Orthographic maps of the temperature distribution of
V410~Tau at four different rotational phases $\phi$.}
\label{Fig:6}
\vspace{0.75cm}
\end{minipage}
\begin{minipage}{\textwidth}
\centering
\subfigure{\includegraphics[width=\textwidth]{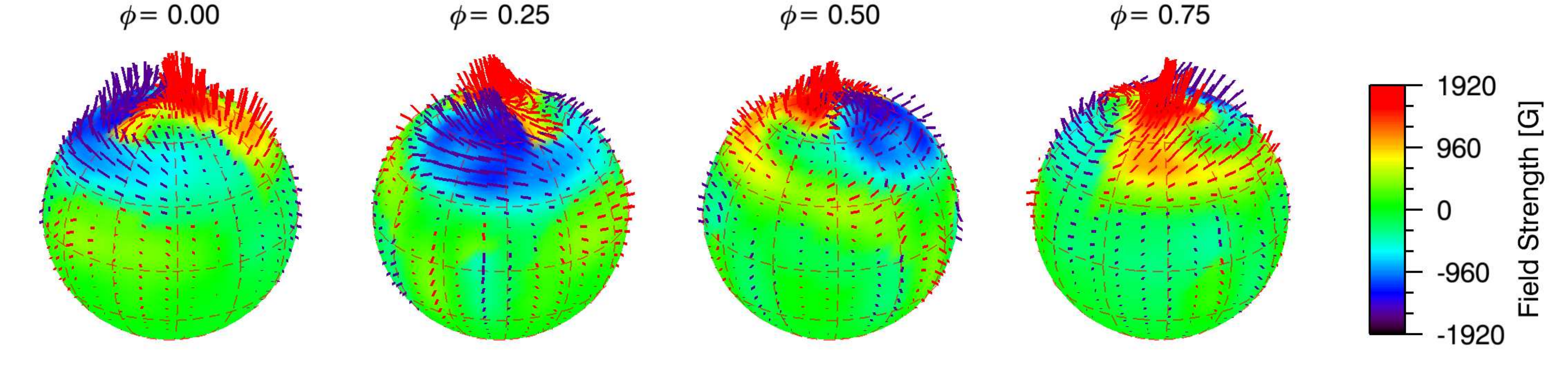}}
\caption{Orthographic maps of the magnetic field distribution of
V410~Tau at four different rotational phases $\phi$. The magnetic
field is color coded and the field lines are proportional to the field
strength (blue negative and red positive polarity). } 
\label{Fig:7}
\vspace{0.75cm}
\end{minipage}
\begin{minipage}{\textwidth}
\centering
\subfigure{\includegraphics[width=\textwidth]{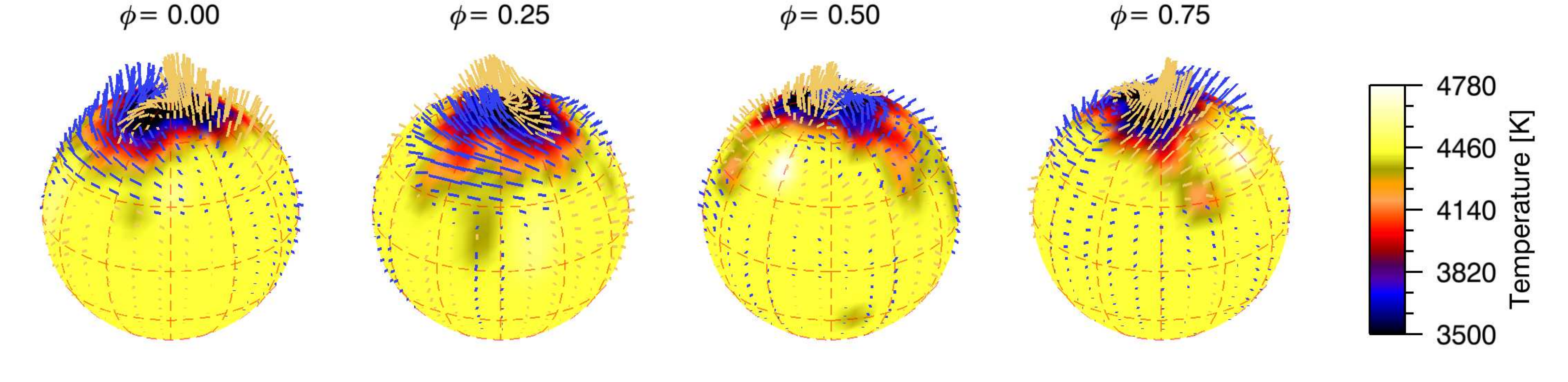}}
\caption{Orthographic maps of the temperature overplotted by the magnetic field lines
at four different rotational phases $\phi$. For a better visibility the positive fields
are color coded in yellow while the negative fields are again in blue. 
As above the field lines are proportional to the strength of the field.} 
\label{Fig:8}
\vspace{0.20cm}
\end{minipage}
\end{figure*}

The synthetic fits relative to the observed Stokes~I profiles are shown
in a stacked plot on the left side of Fig.~\ref{Fig:5}. 
The final temperature map in
orthographic projection is shown in Fig.~\ref{Fig:6}. It is
dominated by a structured, cool, polar cap with a temperature as cool
as $\approx$3,500~K, i.e. $\approx$900~K below the effective
temperature. Several appendages reach down to lower latitudes of
around 30$^\circ$ with decreasing temperature contrast the further
they are away from the cap. The total spot filling factor is 11.9~\%.
From previous investigations, we knew that
V410~Tau shows a polar spot and a number of smaller cool spots
together with localized hot features (see the review in Paper~I).
However, the simultaneous reconstruction of the surface
temperature and the magnetic-field distribution provides
unprecedented accurate surface details.


\begin{figure*}[t]
\centering
\includegraphics[width=\textwidth]{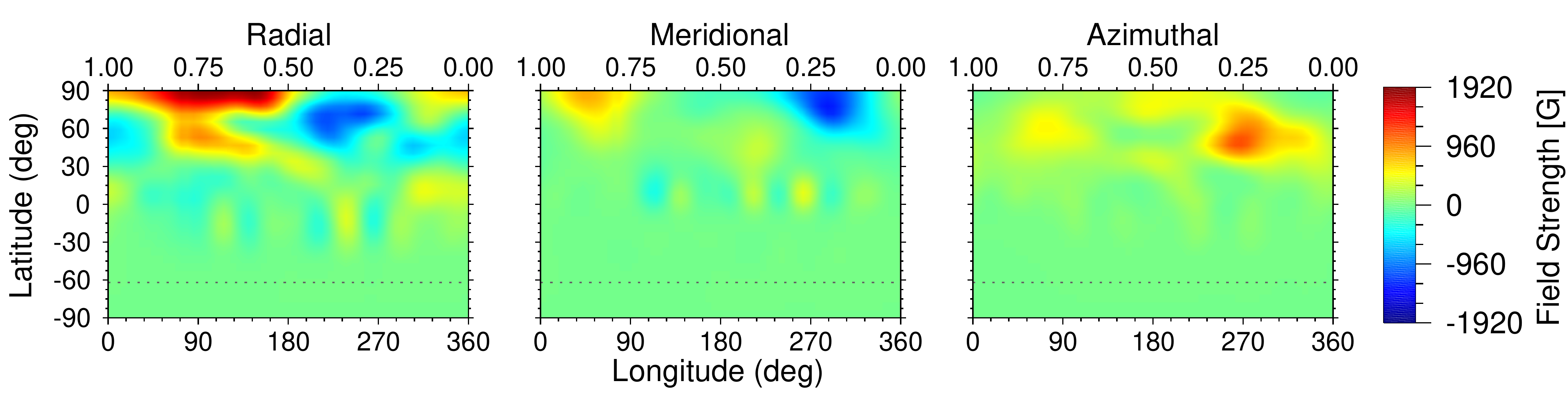}
\caption{Pseudo-Mercator projection of the three magnetic-field
components; left radial, middle meridional, and right azimuthal.
Note, that the lower x-axis is the longitudinal angle, while the upper
x-axis gives the rotational phase.
The dotted line indicates the limit for the visibility due to the
stellar inclination.} 
\label{Fig:9}
\end{figure*}

An interesting feature of the temperature map in Fig.~\ref{Fig:6} is the pair
of isolated spots consisting of a cool spot at phase $\phi\approx
0.7$ and of a hot spot at phase $\phi\approx 0.6$ at intermediate
latitudes. The hot spot's temperature is $\approx$4,800~K, i.e.
$\approx$400~K above the effective temperature while the cool
spot's temperature is $\approx$4,000~K, i.e. $\approx$400~K below
the effective temperature. A similar pair of spots, although
larger in size and with much smaller contrast, is located on the
equator at phase $\phi\approx 0.25$. This pair's temperature
difference is $\approx$400~K in total. Yet a third such pair of
just moderate warm and cool spots may exist around $\phi\approx
0.0$ at latitude $\approx 30^\circ$.

All of above features were independently reconstructed with {\sc
TempMap} and \emph{iMap} in Paper~I; some with slightly
different contrast but the overall similarity is very good.

\subsection{Magnetic field distribution in 2008/09}

The magnetic field reconstruction found by \emph{iMap} provides a solution compatible with the 
observed data and the noise level.
The synthetic fits for the individual Stokes~V profiles are shown
in a stacked plot in Fig.~\ref{Fig:5}. The ZDI map is shown as an
orthographic plot in Fig.~\ref{Fig:7}. The distribution of the
individual magnetic components (radial, azimuthal and meridional)
can be best seen in the pseudo Mercator-projection plot in
Fig.~\ref{Fig:9}. The orientations of the field within the Mercator
plots is defined such that a positive direction in the radial field
points outward while a negative polarity points inward, the meridional
component is positive if it points toward the upper north-pole
and negative in the direction of the south-pole. For the azimuthal component
a positive field is defined in westward direction 
and negative for an eastward direction.

The magnetic topology on V410~Tau is dominated by a bipolar
structure around the visible pole with peak field strengths of up
to $\pm$1.9~kG.
Figure~\ref{Fig:8} emphasizes the relatively good correlation 
of the temperature and magnetic field and shows that the strongest
fields are located in the darkest regions around the pole.
The two polarities appear separated by a sharp,
pole-crossing, neutral sheet. In total the field forms an S-shaped
structure centered at the rotation pole. The topology is
predominantly radial but 1~kG meridional components coexist very
close to the pole. An azimuthal component, on the other hand, is
restricted to a region at $\approx$60$^\circ$-latitude and around
270$^\circ$ longitude, i.e. 180$^\circ$ away from the strongest radial
component, close in distance but on the opposite hemisphere and
always of positive polarity. Its peak field strength reaches 1~kG
around 270$^\circ$ longitude while the rest of it mostly remains near
500~G.One can express the field topology also in terms of the poloidal and
toroidal components which reveals in our case that the majority, 73~\%,
of the surface field of V410~Tau is given in the form of poloidal
fields and just 27~\% being in form of a toroidal component.

The dominant of the two isolated warm-cool spot pairs around
$\phi\approx 0.6$ and $\phi\approx 0.7$ has no recognizable
distinct magnetic-field structure. Both, the cool and the warm
spot, appear with the same general mix of field components as
their immediate surrounding and at least this particular spot pair
is reconstructed with the same (positive) polarity. On the
contrary, the second such spot pairs (around $\phi\approx 0.25$;
see Sect.~\ref{Sect:T-map} above) indicates a mixed polarity,
negative for the warm feature and positive for the cool feature.

It is interesting to note that the strong \Halpha\ flare reported
in Paper~I (located at $\phi\approx 0.18$) coincides with the best
visibility of the sharp neutral sheet of the bi-polar radial
field. It is at least tempting to state that, if truly coincident,
it would very well agree with solar active-region physics where
flares are usually related to the current sheets separating
opposite polarities \citep[e.g.][]{Schrijver05}.

An error analysis of the here presented results is given in the 
appendices \ref{AppError}, \ref{AppTest}, and \ref{AppBlends}.

\subsection{Differential rotation of V410~Tau}

As described in Sect.~\ref{Sect:Differential} a solar like differential
rotation parameterized by $\Omega_{eq}$ and $d\Omega$ is part of
the inversion process. In the current version of \emph{iMap} the sheared
image analysis is applied to the temperature structures only. For
the initialization, $\Omega_{eq}$ is derived from the rotational
period of 1.871948~d (see Paper I). Only a very weak solar-like
differential rotation is present, if at all. The formal value for
the angular velocity at the equator is $\Omega_{eq}$ = $3.356\pm
0.005$, well within the errors of the photometrically determined
rotation period, and for the differential rotation rate between
the equator and the pole we have $d\Omega = 0.007\pm 0.009$. We can not
safely regard this as a detection of a differential rotation on
V410~Tau but conclude that it must be very small, likely smaller
than a factor 30 compared to the Sun.

\subsection{The magnetic nature of V410~Tau}

The probable absence of a radiative core and the non-detection of differential rotation 
in this work and other investigations \citep{Strassmeier94,Skelly10} leaves not much room
for speculation about the nature of the 
dynamo operating in the interior of V410~Tau -- at least in the framework of classical mean 
field dynamos.
As has been shown by \citet{Kueker99} a likely explanation for the generation of the 
magnetic field in weak-lined T Tauri stars can be given in terms of a $\alpha^2$-dynamo. 

\begin{figure*}[!t]
\begin{minipage}{9cm}
\includegraphics[width=9cm]{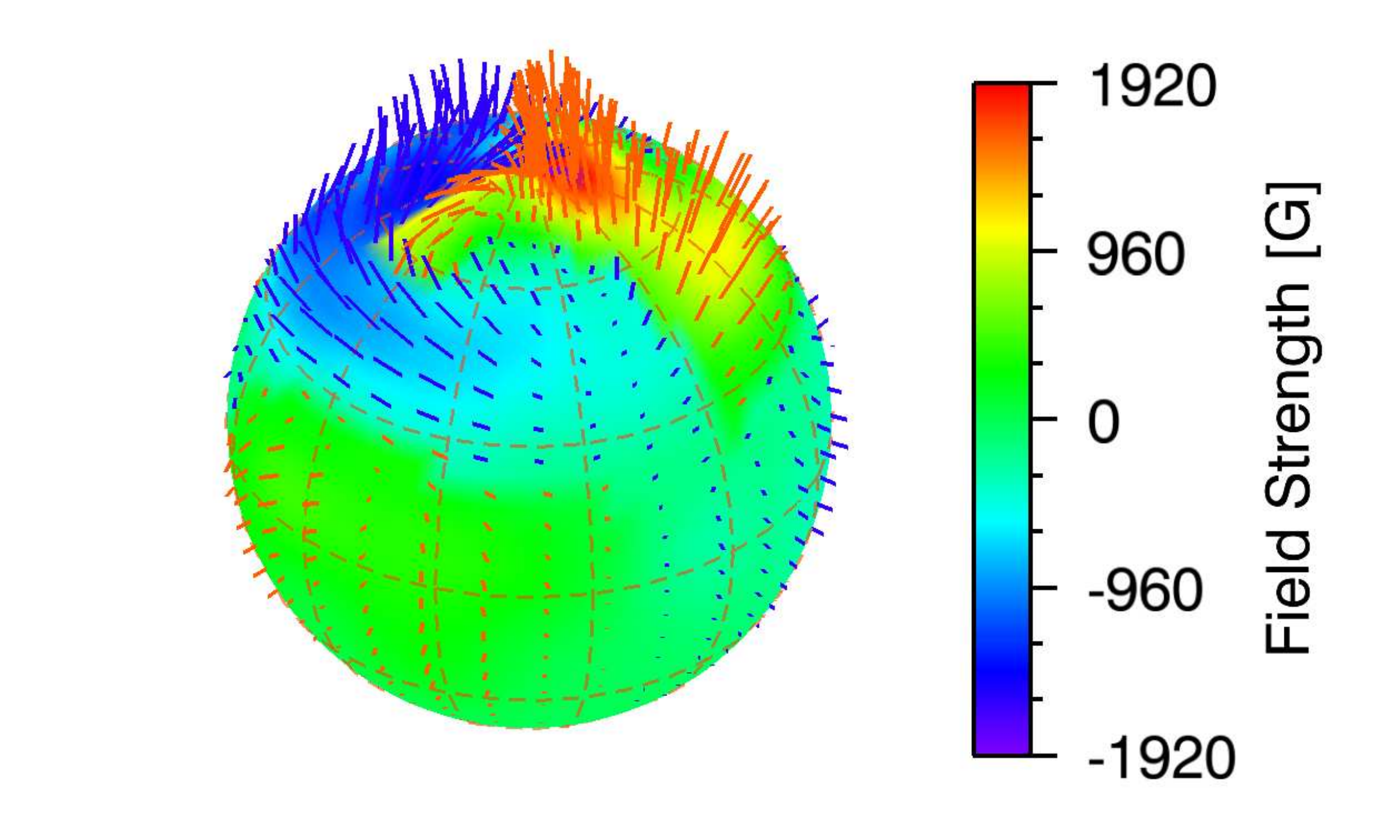}
\end{minipage}
\begin{minipage}{9cm}
\includegraphics[width=9cm]{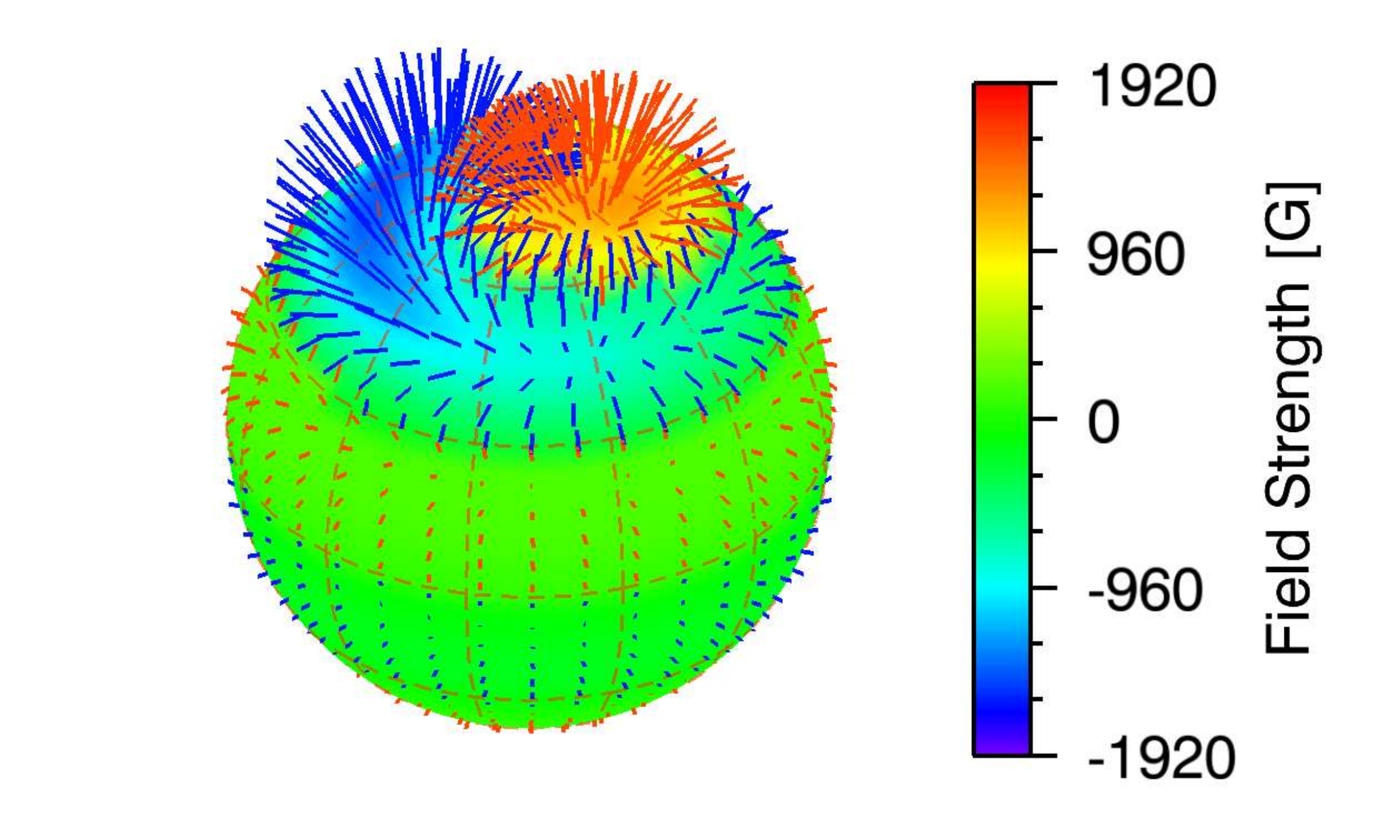}
\end{minipage}
\caption{Intriguing structural similarity between the result of our magnetic surface reconstruction (left) and 
an $\alpha^2\Omega$-dynamo simulation which is scaled to a surface field strength of 1.9 kG (right).} 
\label{Fig:10}
\end{figure*}

From the observational point of view
we can characterize and quantify the surface topology with various degrees of detail but
providing a quantitative link between observation and theory is anything but straightforward. 
In the moment we feel that dynamo theory does not give enough reliable information
to provide such a quantitative link between the surface appearance of magnetic fields 
and their internal generation process.
But this shall not keep us from giving qualitative arguments in favor of an $\alpha^2$-dynamo here. 
Following \citet{Kueker99} for an $\alpha^2$-dynamo 
the field topology exhibit a distinct non-axisymmetric geometry
with S1- or A1-type modes \citep{Ruediger04}.
The pronounced bipolar magnetic field at the pole (Fig. \ref{Fig:7}) can easily be identified with such
a non-axisymmetric geometry.
However the fact that the non-axisymmetric mode of such an
$\alpha^2$-dynamo solution can only rotate with the star itself (except for a slight
drift) may not really fit to recent observations that V410~Tau exhibit a periodic change of its 
activity in the range between 4.8 to 5.5 years \citep{Stelzer03,Savanov12}. But a possible 
explanation can still be given in terms of an $\alpha^2\Omega$-dynamo \citep{Elstner05}, where
the non-axisymmetric solution is modulated by an oscillatory axisymmetric mode.
The excitation of such an $\alpha^2\Omega$-dynamo can already occur at
a very low differential rotation rate \citep{Elstner05}. To qualitatively compare this dynamo model 
with our reconstructions we have taken a snapshot of the non-axisymmetric mode of these simulations 
at a time when both magnetic polarities are of approximately the same strength. The surface field strength 
was scaled to be comparable to that of the reconstructed surface field of V410~Tau and the
symmetric field (relative to the equator) at the south pole has been damped.
The similarity between our reconstruction and the $\alpha^2\Omega$-dynamo simulation 
shown in Fig.~\ref{Fig:10} is quite remarkable and may in fact strengthen the hypothesis that the field 
of weak-lined T Tauri stars is generated by an $\alpha^2$-type dynamo or in cases where a weak differential rotation 
has already developed by an $\alpha^2\Omega$-dynamo.

An interesting model which accounts for the evolution
of pre-main sequence dynamos were presented by \citet{Kitchatinov01} and may also
be applicable to our findings. According to their simulations
the dynamo during the early phases of the pre-main sequence stage, when the core development 
is still in progress and the rotational velocity is large, gives rise to a strong 
non-axisymmetric mode. In the course of the evolution when differential rotation sets in 
the star develops an increasing axisymmetric dynamo mode which can lead to an 
oscillating behavior \citep{Ruediger94} before the field is eventually completely 
dominated by the axisymmetric mode at a later stage. 

How does V410~Tau compares with other T Tauri stars that have been subject to magnetic investigations.
Zeeman-Doppler-Imaging has been applied to a small number of classical T Tauri stars and have been
compared by \citet{Hussain12} to find a relation between mass, period, magnetic field strength
and field complexity. Although no clear relation could be found from the limited sample, there seems to be
a tendency that fully convective T Tauri stars may harbor simpler surface fields (i.e. mainly a dipole pole
component) while T Tauri stars which have developed a radiative core have the tendency to more
complex fields \citet{Hussain12}.
It is not clear if V410~Tau as a weak-lined T Tauri fit into this picture. Given its evolutionary stage
it may be reasonable to assume, in the light of the above trend, that V410~Tau exhibit a more complex
surface field than just a dipole (spherical harmonic numbers $l=1$ and $m=0$). 
This is in fact true, but our magnetic reconstruction shows just a little more additional complexity
with a clear dominance of the S1 mode  ($l=1,m=1$)  and the A1 mode ($l=2,m=1$).
The apparent diversity of the magnetic field of T Tauri stars 
certainly requires a greater sample in order to provide a more general picture of the magnetic 
evolution and the underlying dynamo for these pre-main sequence stars.

At the end we also want to comment on the apparent differences between 
the reconstructed surface fields of \citet{Skelly10} and our magnetic and temperature maps. 
In a recent magnetic investigation of V410~Tau \citet{Skelly10} found 
that 50~\% of the total field has a toroidal component and the other 50~\% are poloidal 
and that the majority of the field  shows strongly inclined large-scale azimuthal fields which are 
distributed over one entire hemisphere, whereas our reconstructed topology has only 27~\% toroidal fields
and 73~\% poloidal fields with a strong radial and bipolar component around the polar region.
Moreover, in contrast to our reconstructed maps where the strong magnetic fields are well correlated 
with the temperature, see Fig.~\ref{Fig:8}, the fields of \citet{Skelly10} seem to show almost no correlation with 
their brightness maps (i.e. a proxy of the temperature distribution).
The fact that these data were observed 
at almost the same time as ours (January 2 to January 17, 2009) at the 
Telescope Bernard Lyotl (TBL) with the NARVAL spectropolarimeter at the Pic di Midi, is 
certainly a reason for a closer inspection and comparison of the two results.
Without having the possibility for doing such a detailed comparison of the data as well as of the  inversion codes
we can only speculate about the reasons for the disparity of the two reconstructions. 
But we want to emphasize that in contrast to \citet{Skelly10} our approach makes no assumptions on the
surface topology in terms of a spherical harmonic decomposition or potential field structure. It is furthermore 
fully based on polarized radiative transfer and line profile modeling and it pursues a strategy which
simultaneously invert the temperature and magnetic field. This give us enough confidence to believe in the 
validity of our results.


\section{Summary and conclusions}

Despite the relative high levels of magnetic activity of many rapidly rotating late-type stars 
the typical spectropolarimetric signals that these stars produce are mostly below the noise level 
of a typical spectropolarimetric observation and require therefore specialized techniques to extract 
and reconstruct the line profiles.  
To address this problem we have developed a new SVD based spectral-line 
extraction technique. For Stokes~V this approach is a full reconstruction
method which allows us in this work to reach an improvement in the signal-to-noise ratio 
by a factor of 40.
For Stokes~I where the actual signals are orders of magnitudes larger then for Stokes~V 
the SVD method reduces to a denoising technique but still allows to increase the 
signal-to-noise ratio by a factor of 5. 

The interpretation of polarimetric data in terms of ZDI always require
the introduction of additional information. This is true for the underlying 
model used in the inverse problem as well as for the regularization of the problem itself.
To pose a minimum number of surface constraints on the solution we have chosen a new regularization 
approach for DI and ZDI in terms of an iterative regularization of the problem.
The new variant implemented in the inverse module of \emph{iMap} shows a good and stable behavior and 
converges within 400 iterations.  

A common problem for ZDI is that the real strong magnetic fields which are
associated with cool surface regions produce only a fraction of the photon flux
compared to the unaffected quiet or even hot surface regions.
To prevent our code from these ill-defined flux-weighting and the mutual effects between the 
temperature and magnetic fields we have pursued a strategy that performs a simultaneous 
DI and ZDI to retrieve both the temperature and the magnetic field distribution at the same time.

The reconstruction of the temperature and magnetic field of V410~Tau 
reveals that the majority of the strong fields of 2~kG are located in the cool spots, 
in particular within the large polar spot. The
reconstruction clearly show that both polarities coexist within
the large polar spot and that the entire polar-field topology
appears to be dominated by a twisted bi-polar structure separated by a
relatively sharp neutral line. It is reminiscent of an
over-dimensioned solar active region. Even the time of a strong
flare coincides with the best visibility of this neutral line,
just like for solar active regions. However, due to its shear size
of approximately 11\% of a hemisphere, the bi-polar feature appears S-shaped
as if it had been dragged around the pole due to the rapid stellar
rotation. The absence of a detectable differential rotation and the pronounced
non-axisymmetry of the field may suggest 
an $\alpha^2$-dynamo operating in the interior of V410~Tau. 
This is also supported by the intriguing similarity between our reconstructed 
surface fields and the dynamo simulation of \citet{Elstner05},
and may support the theory that the underlying mechanism responsible for the magnetic
field generation in weak-line T Tauri stars is an $\alpha^2$-type dynamo.


\begin{acknowledgements}
The authors would like to thank Nadine Manset and the other CFHT
staff for their support of the observations. KGS is grateful to
the German Science Foundation (DFG) for support under grant
STR645/1. JBR acknowledges support from the Natural Science and
Engineering Research Council of Canada (NSERC).
\end{acknowledgements}


\vspace{2cm}

\appendix

\section{Error Analysis}
\label{AppError}  

In this section we want to investigate the significance of our reconstruction results, in particular
we want to quantify how sensitive the solution of the inverse problem is with respect to the initial conditions.
In our DI and ZDI setup we deal with four parameter spaces
(temperature, radial magnetic field, azimuthal magnetic field and meridional magnetic field) each of which with a 
dimension equal to the number of surface segments. The inversion algorithm has to navigate through the 
combined parameter spaces to find a solution that is compatible with the data.
In order to study the stability of the inversion relative to the initial starting conditions
(different positions in the parameter spaces) we use a simulation that runs the inversion with the original 
data set but from randomly chosen starting positions.
Each parameter space is independently initialized by choosing a random value for each surface segment. 
Though we may create a considerable spread among the individual surface segments we can only consider a small 
fraction of the overall parameter space. Any exhaustive study to sample the parameter space 
is way out of reach for this high-dimensional setting and it is not the purpose here, instead we want to gain a 
quantitative measure how the result varies when the initial conditions are changed.


\begin{figure*}[h]
\begin{minipage}{\textwidth}
\centering
\includegraphics[width=\textwidth]{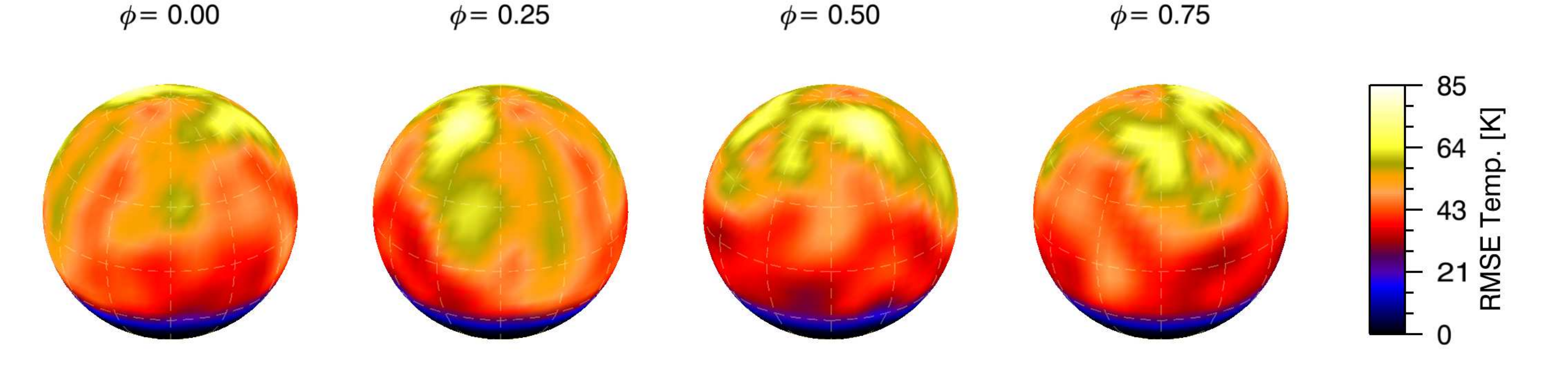}
\caption{Temperature error map in orthographic projection at four different rotational phases.
The maximum RMS error value is 80 K and the mean RMS error value is 50 K. Regions below a latitude of -30$^{\circ}$ are set
to a constant value and do not participate in the error estimation.}
\label{Fig:A1}
\vspace{0.75cm}
\end{minipage}
\begin{minipage}{\textwidth}
\centering
\includegraphics[width=\textwidth]{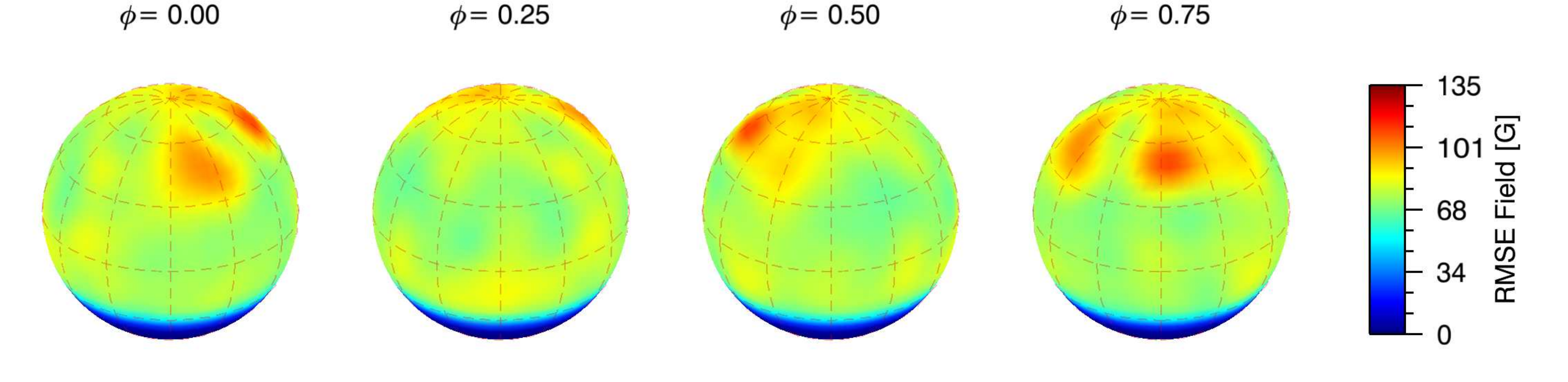}
\caption{Radial magnetic field error map in orthographic projection at four rotational phases, 
the maximum RMS error value is 125 G and the mean RMS error value is 64 G.}
\label{Fig:A2}
\vspace{0.75cm}
\end{minipage}
\begin{minipage}{\textwidth}
\centering
\includegraphics[width=\textwidth]{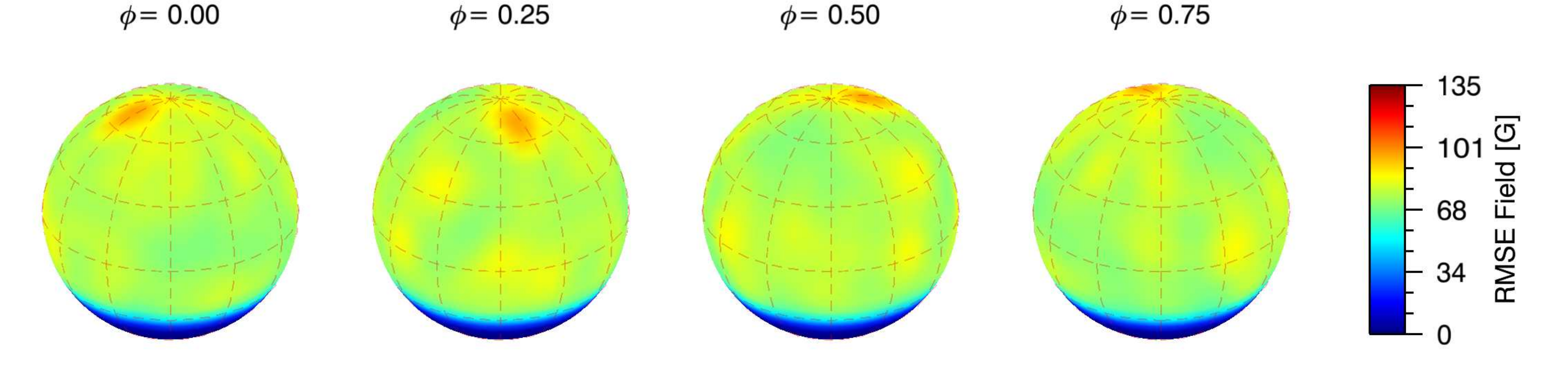}
\caption{Meridional magnetic field error in orthographic projection, 
the maximum RMS error value here is 101 G and the mean RMS error value is 51 G.} 
\label{Fig:A3}
\vspace{0.75cm}
\end{minipage}
\begin{minipage}{\textwidth}
\centering
\includegraphics[width=\textwidth]{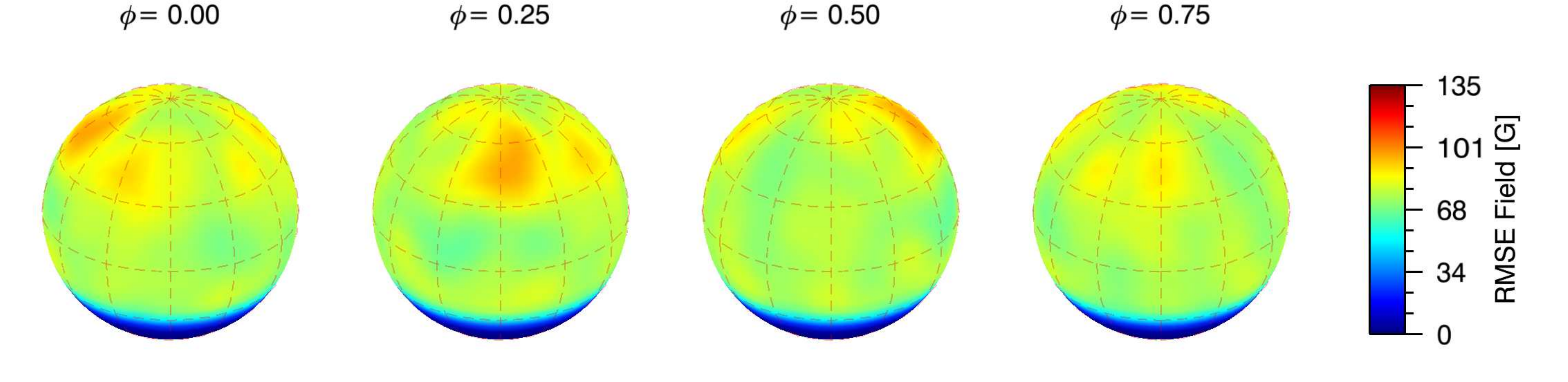}
\caption{Azimuthal magnetic field error in orthographic projection with 
a maximum RMS error value of 106 G and a mean value of 52 G.} 
\label{Fig:A4}
\end{minipage}
\end{figure*}

A random generator on the basis of a normal distribution provides the values for each parameter space and 
surface segment. The underlying normal distribution has a mean value which is set to the effective temperature for 
the temperature component and to zero Gauss for all three magnetic components.
The standard deviation is 250 K for the temperature and 250 G for all the magnetic components 
to provide a large spread among the individual surface segment values.
The simulation (inversion with different initializations for all parameters and segments) 
is run for 100 times on the original data set. For the so obtained 100 DI and ZDI maps we
calculated the root-mean-square (RMS) error to finally get RMS error maps for each parameter.
The individual inversions are all run to the same accuracy level given by the noise of the reconstructed
data. Since the derivatives with respect to the magnetic parameters as well as with respect to the temperature
become to weak below a latitude of -30 degrees to provide any substantial changes during the inversion we fix 
these segment values to zero for the magnetic maps and to the effective temperature for the temperature maps. 

The RMS error map for the temperature is shown in the orthographic plot Fig.~\ref{Fig:A1}. It shows a peak value of
80 K at positions associated with the coolest regions (polar spot). Other regions 
hardly exceed a RMS error of 50 K. The RMS error map for the three magnetic components is shown in 
Fig.~\ref{Fig:A2}, Fig.~\ref{Fig:A3}, and Fig.~\ref{Fig:A4}. 
The error map of the radial magnetic field has a peak value of 125 G, the average 
RMS error value is 64 G. The RMS error values for the meridional and azimuthal field exhibit 
peak values of 101 G and 106 G respectively and both have a mean RMS error value around 60 G.
What can be seen from these maps is that the error values are correlated with the 
absolute field strength as well as with the temperature. This emphasizes once more
the influence of the temperature on the magnetic field determination.
For the given photospheric conditions of V410~Tau a temperature change of just 80 K in a spot like
region causes a change in the amplitude of the Stokes~V signal of 4\%. In the strong field regime
of the polar spot this difference in amplitude is equivalent to a magnetic field of 60 G !
Given this temperature dependence, the noise level in the data as well as
the fact the a random small scale magnetic field on the surface produces local 
Stokes V signals that effectively cancel each other out,
the obtained uncertainties from the error simulation are surprisingly low and show that the 
inversion always settle in the proximity of the same (local) minimum and that the solution is 
robust against perturbations of the initial conditions.

\section{Test Inversion with a dynamo model}
\label{AppTest}

\begin{figure}
\centering
\includegraphics[width=9cm]{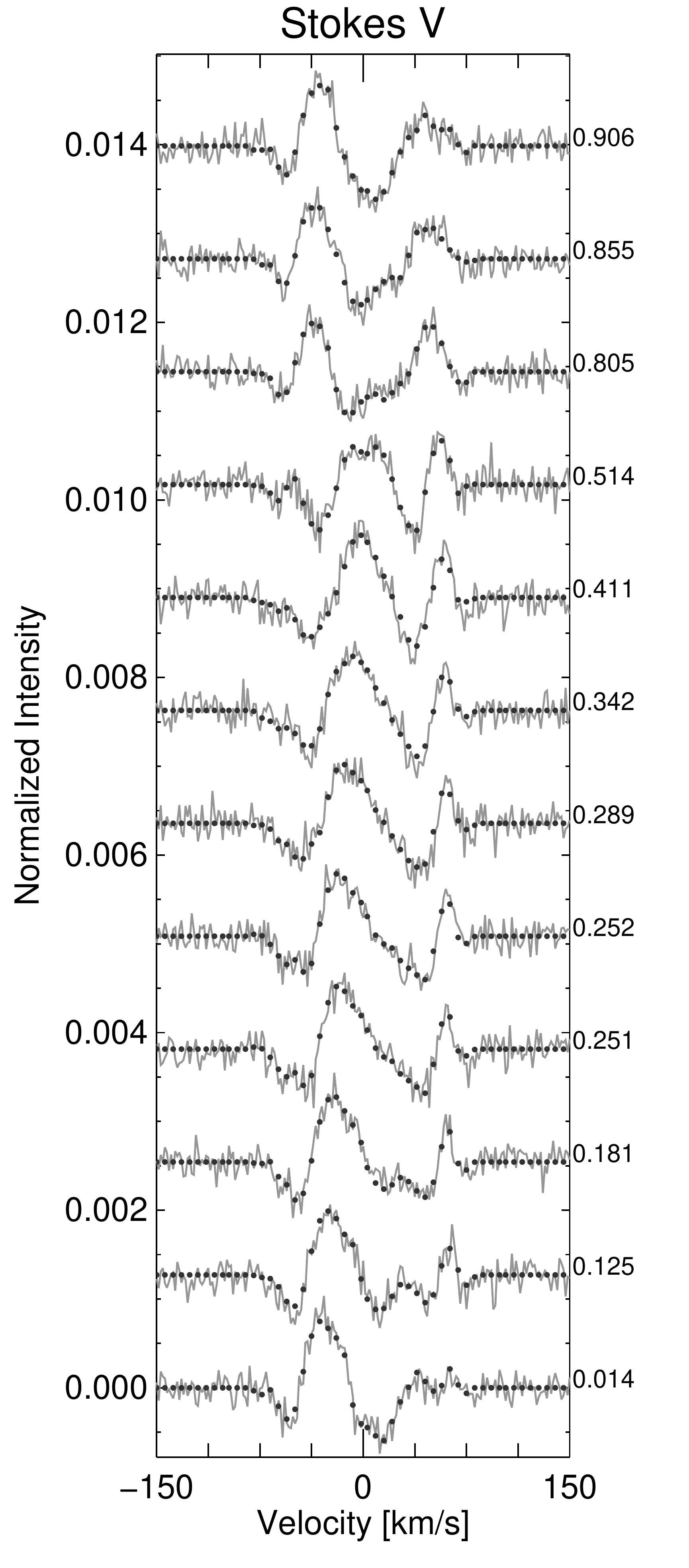}
\caption{Stacked plot of the synthetic fits (dotted points)
and the synthetic Stokes~V profiles (grey solid lines) obtained from the
dynamo model.}
\label{Fig:B1}
\end{figure}

In this section we want to examine if the inversion is able to retrieve a surface field topology 
as complicated as that of a simulated $\alpha^2\Omega$-type dynamo with the given data set.
The question that we want to investigate here goes beyond the testing of an inversion code
with simplified surface topologies, what we want to do here is to seek a more direct
answer to our problem, can we trust the hypothesis of our work that the reconstructed 
surface topology of V410~Tau is in fact similar to that of a given
$\alpha^2\Omega$-type dynamo simulation.


\begin{figure*}[h]
\vspace{0.25cm}
\begin{minipage}{\textwidth}
\centering
\includegraphics[width=\textwidth]{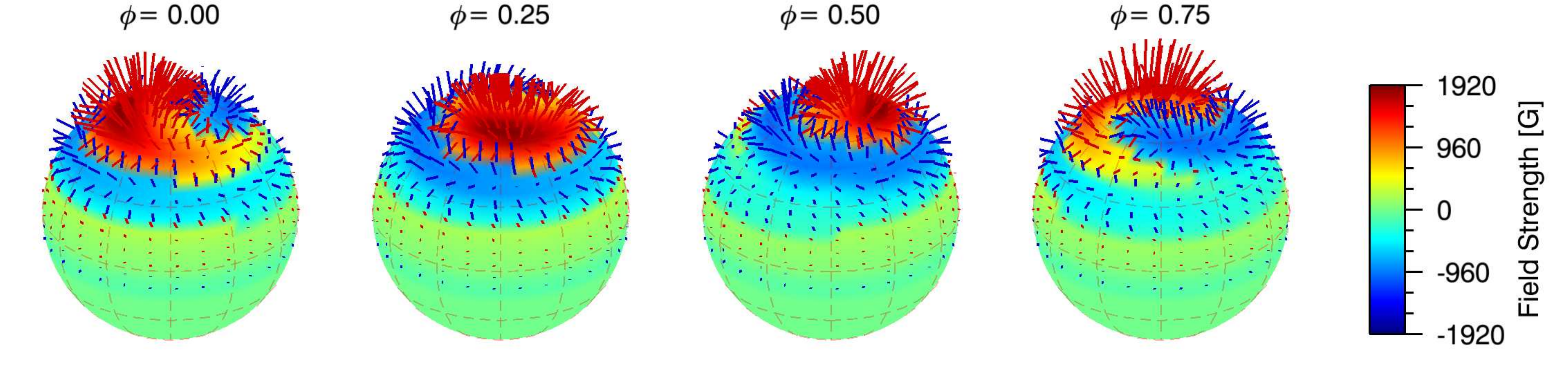}
\caption{Orthographic maps of the original dynamo simulation which serves as a
synthetic model for the test inversion, The model is shown at four different rotational phases 
$\phi$.}
\label{Fig:B2}
\vspace{0.75cm}
\end{minipage}
\begin{minipage}{\textwidth}
\centering
\includegraphics[width=\textwidth]{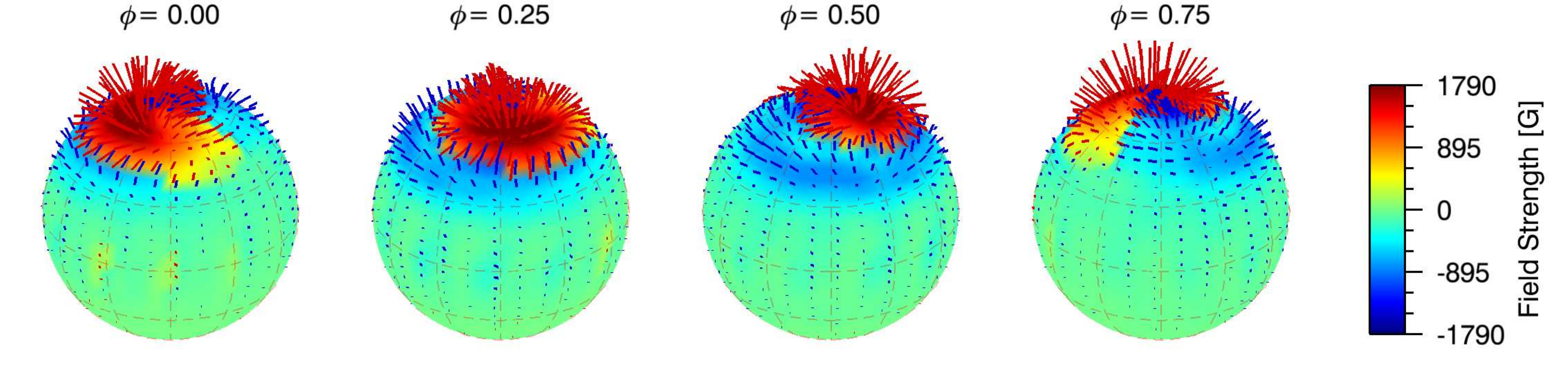}
\caption{Reconstructed surface field from 12 phase modulated Stokes V profiles.
The number and order of the phases as well as the introduced noise correspond to the real 
observations of V410~Tau. Given the observational constraints imposed on this test inversion the dynamo 
model could be very well reproduced.} 
\label{Fig:B3}
\end{minipage}
\end{figure*}
For that reason we synthesized from the $\alpha^2\Omega$-dynamo simulation of \citet{Elstner05},
12 phase resolved spectra at phases which correspond to our observations. The dynamo model was
scaled to have a peak field strength of 1920 G. The stellar parameter are 
those from V410~Tau given in Sect.~\ref{Sect:Parameters}. Noise is added according to
the Bootstrap analysis in Sect.~\ref{Sect:3.2}. The inversion with \emph{iMap} is run with the same 
setup as the original inversion for V410~Tau, which also means that the inversion 
is stopped at the same noise level. Only Stokes~V profiles are used for this magnetic inversion.
The result is illustrated in the orthographic plots Fig.~\ref{Fig:B2} and Fig.~\ref{Fig:B3} as well as 
in the profile plot Fig.~\ref{Fig:B1}. Given the number of available phases and the noise level in the data, 
the original magnetic field structure is remarkably good reproduced. 
Although the maximum field strength is not quite reached in the reconstruction 
shown in Fig.~\ref{Fig:B3} one can see
that the intertwined nature of the field structure and the rapid variation of the field lines around the
polar region is well reproduced. The quality of the fit of the inversion relative to the original synthesized
Stokes~V profiles is shown in  Fig.~\ref{Fig:B1}. This demonstrates that with
the given data set the inversion code can in fact reconstruct a surface field 
with a topology as complex as a $\alpha^2\Omega$-type dynamo.

\section{The influence of line blends}
\label{AppBlends}

Here we want to investigate how much information about the line blends may get lost
within the rank estimation of the truncated SVD procedure.
Note that in our analysis we compare the SVD reconstructed line profile with the 
synthetic (weighted) mean of all contributing lines (929 lines for Stokes V 
and 56 for Stokes I) where all significant line blends are accounted for.
One might suspect that if information about the 
line blends is leaked into the noise subspace it may lead to a mismatch between
the synthesized line profiles and the reconstructed profile which may cause
problems in the subsequent inversion.

\begin{figure*}[!t]
\begin{minipage}{9cm}
\includegraphics[width=8cm]{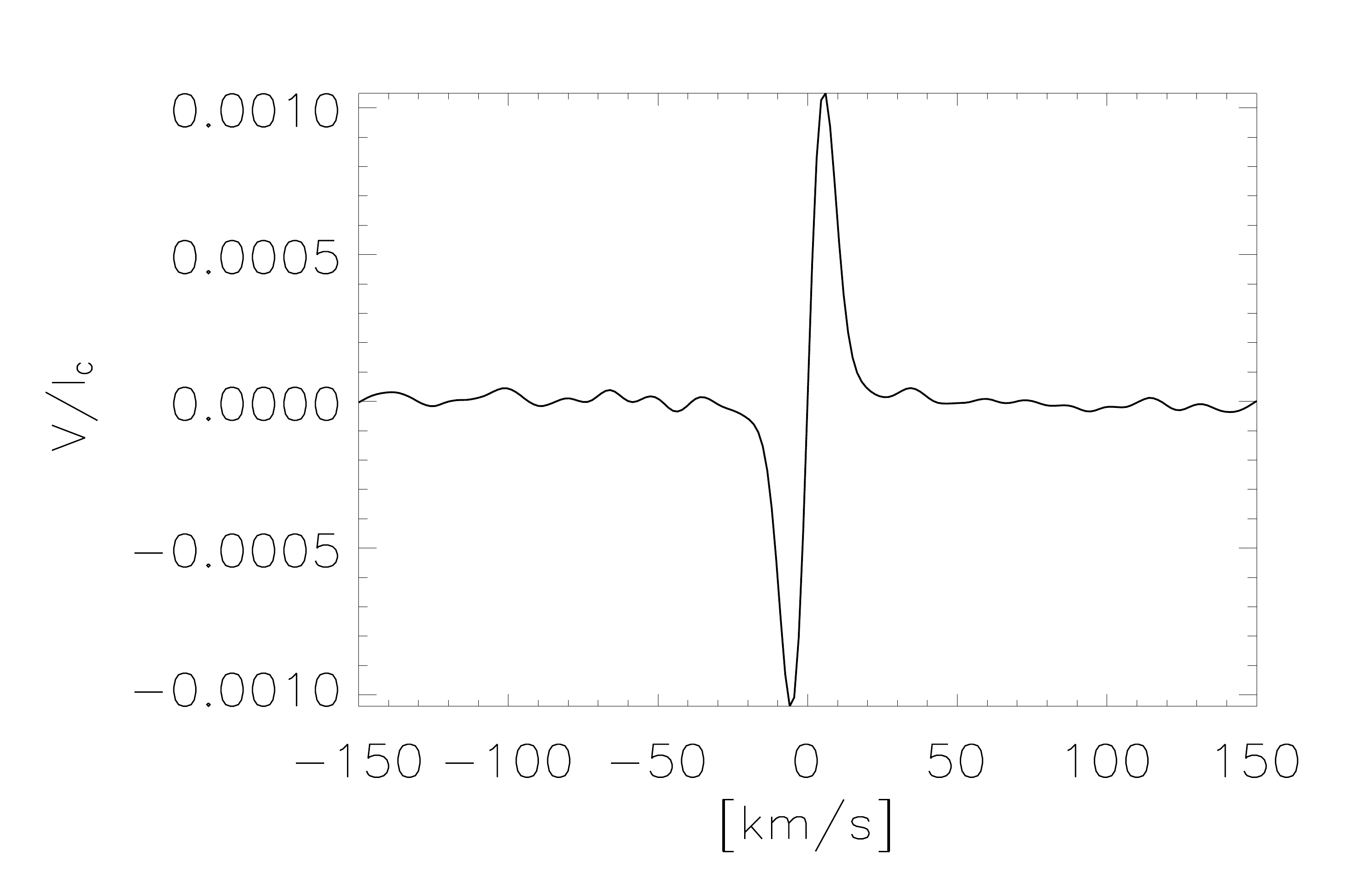}
\end{minipage}
\begin{minipage}{9cm}
\includegraphics[width=8cm]{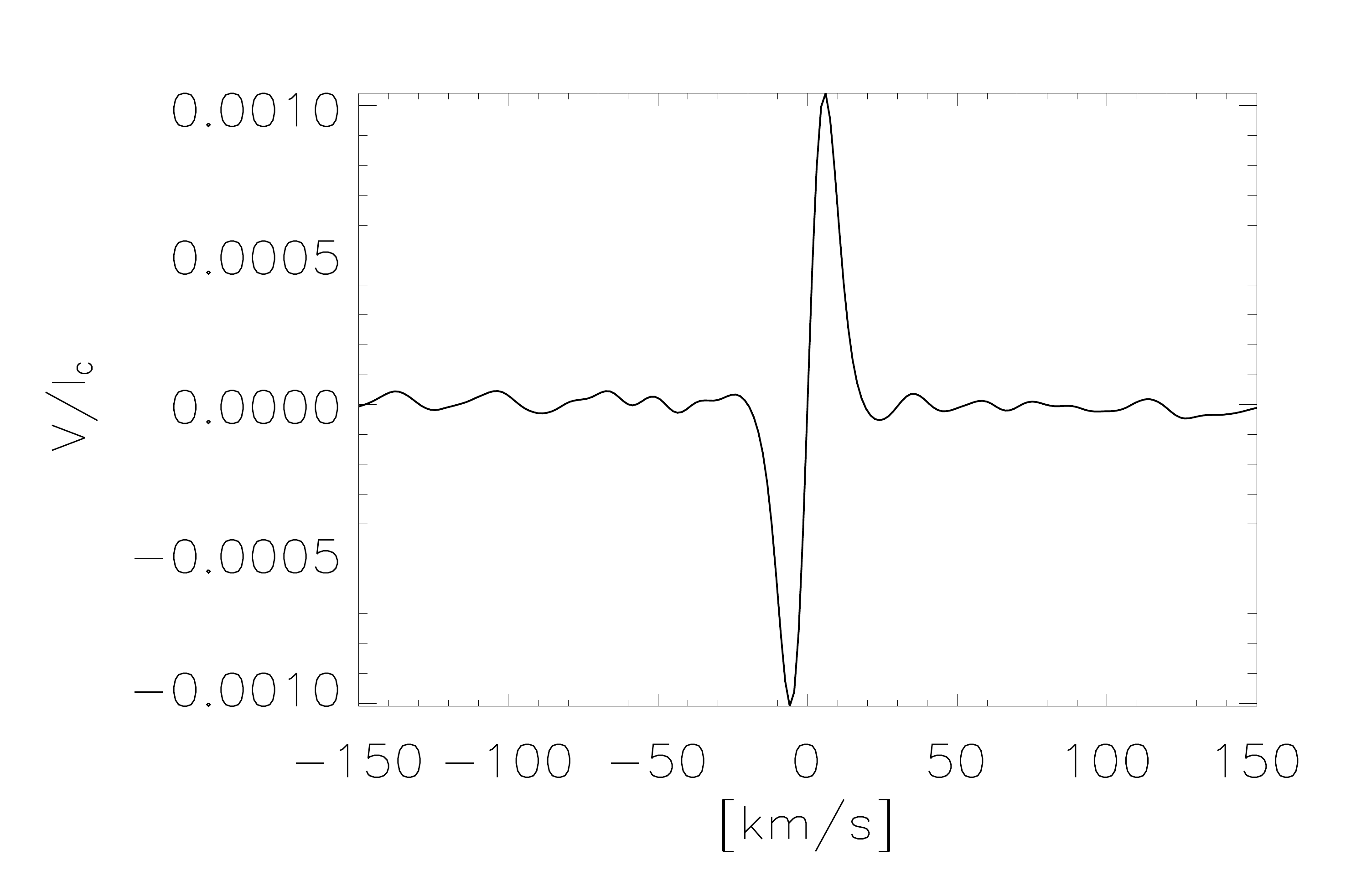}
\end{minipage}
\caption{Reconstructed SVD profile originating from all 929 spectral lines (left). 
The reconstruction is performed with all eigenprofiles. The ''quasi-noise'' level introduced by
the line blends has a median absolute deviation (MAD) of $1.56 \times 10^{-5}$.
On the right, reconstructed SVD profile with only one eigenprofile corresponding to the largest eigenvalue.
The difference between the reconstruction using the full set of eigenprofiles is hardly visible and the RMS
between them is $2.22 \times 10^{-5}$.
} 
\label{Fig:C1}
\end{figure*}

Let us concentrate on the Stokes V profiles here and have a look on
how line blends influence the resulting profile from the two-stage SVD reconstruction.
Line blends affect the individual lines used in the observation matrix in different ways,
they occur at different positions within the line profiles, and are of different strength 
and number. So it will be of interest to see how much of the blends are finally
recognized as systematic and correlated effects within the SVD reconstruction.
For that purpose we have modeled a simple monopole star with a homogeneous magnetic field
of 10 G and a small rotational velocity of 5 km s$^{-1}$. All other stellar parameters are assumed
to be the same as that for V410~Tau. We synthesize the line profiles in the velocity domain
and use a value of $\pm$ 150 km s$^{-1}$, around the respective line center of each contributing line.
For such a small rotational velocity the Stokes V signals are therefore much thinner than the used 
velocity range. This wide span of the velocity domain will give us additional information 
about the systematics introduced by the line blends relative to the base level (i.e. zero polarized continuum). 
The 929 lines are calculated without any noise. After 
calculating the line profiles and their respective line blends we applied the SVD analysis and 
used all available eigenprofiles for the reconstruction. The result of the two stage process is an 
average Stokes~V profile that is the same as a regular average since we have used the entire set of 
eigenprofiles.
In  Fig.~\ref{Fig:C1} on the right side we see the result of the SVD reconstruction. It is clearly seen that the
Stokes V profile has its regular shape in spite of the many contributing line blends. The \emph{wiggling} is
the result of the line blends. Note again no noise is present in that synthetic test case.
This wiggling of the base line (the continuum) shows no dramatic variation within the velocity range.
We may quantify this wiggling in terms of a noise measure, and use the
median absolute deviation (MAD) for that purpose which is defined as, 
MAD$(\vec{V}) = median(|V_1-median(\vec{V})|,...,|V_n-median(\vec{V})|)$, 
where $median(\vec{V})$ is the median of the vector components of $\vec{V}$.
The MAD give us a value of
$1.56 \times 10^{-5}$. Given the strength of the reconstructed Stokes V signal, which has an amplitude 
comparable to that observed for V410~Tau, we can see that the blends introduce
an effect that is below the noise level of the real reconstructed observations.   
How much information is lost if we would just use the eigenprofile corresponding to the largest eigenvalue?
In that case we would pretend that our signal subspace is of dimension one and all significant signal information is 
comprised within the first eigenprofile.
Performing the two-stage SVD reconstruction under this assumption results in a 
Stokes V profile shown on the right side in Fig.~\ref{Fig:C1}. The difference between the reconstruction with the full set of eigenvectors
and that using only a single eigenprofile is very small. We calculated the RMS between the two reconstructed profiles 
which gives a value of $2.22 \times 10^{-5}$.
The contribution of the line blends relative to the Stokes signal for the set of 929 spectral lines used
in this work is apparently very small, i.e. ten times smaller than the noise level deduced in Sect.~\ref{Sect:3.2}.
The remaining systematic effects introduced by the blending are essentially 
captured by the first eigenprofile which is also the signal carrying component. This demonstrates that the influence
of line blends for the set of spectral lines used in this work is small and that the leakage of information 
carried by eigenprofiles belonging to smaller eigenvalues (i.e. noise space) is negligible.


\end{document}